\documentclass[a4paper,14pt,english]{extreport}

\usepackage{amsfonts,amsmath,amssymb,amscd} 

\usepackage{lscape}		
\usepackage{geometry}	

\usepackage{cmap}						
\usepackage[utf8]{inputenc}				
\usepackage[T2A, T1]{fontenc}				
\usepackage[main=english, russian, english]{babel}	
\usepackage{tempora}     
\usepackage{newtxmath}   
\usepackage{fix-cm}

\usepackage{indentfirst} 

\usepackage[usenames]{color}
\usepackage{color}
\usepackage{colortbl}

\usepackage{longtable}					
\usepackage{multirow,makecell,array}	

\usepackage[singlelinecheck=off,center]{caption}	
\usepackage{soul}									

\usepackage{cite} 

\usepackage[linktocpage=true,plainpages=false,pdfpagelabels=false, unicode]{hyperref}

\usepackage{graphicx} 

\usepackage{tocloft}

\usepackage{alltt}
\usepackage{listings}

\usepackage{bmpsize}
\makeatletter
\renewcommand{\@biblabel}[1]{#1.}	
\makeatother

\graphicspath{{images/}} 

\definecolor{linkcolor}{rgb}{0.9,0,0}
\definecolor{citecolor}{rgb}{0,0.6,0}
\definecolor{urlcolor}{rgb}{0,0,1}
\hypersetup{
    colorlinks, linkcolor={linkcolor},
    citecolor={citecolor}, urlcolor={urlcolor}
}

\newtheorem{definition}{Definition}
\newtheorem{lemma}{Lemma}
\newtheorem{theorem}{Theorem}
\newtheorem{corollary}{Corollary}

\def\Re{\mathop{\rm Re}\nolimits}
\def\diag{\mathop{\rm diag}\nolimits}

\def\col{\mathop{\rm col}\nolimits}

\def\proof{{\sf Proof.} }

\def\ac{\mathop{\alpha_{\mathbb C}}\nolimits}
\def\mc{\mathop{\mu_{\mathbb C}}\nolimits}
\def\ar{\mathop{\alpha_{\mathbb R}}\nolimits}
\def\mr{\mathop{\mu_{\mathbb R}}\nolimits}

\makeatletter
\DeclareSymbolFont{AMSa}{U}{msa}{m}{n}
\DeclareMathSymbol{\AMSSquare}{\mathord}{AMSa}{"03}
\DeclareRobustCommand{\square}{\ensuremath{\AMSSquare}}
\makeatother			

\begin{document}

\renewcommand{\abstractname}{Abstract}
\renewcommand{\alsoname}{see also}
\renewcommand{\appendixname}{Appendix}
\renewcommand{\bibname}{Bibliography}
\renewcommand{\ccname}{outgoing}
\renewcommand{\chaptername}{Chapter}
\renewcommand{\contentsname}{Contents}
\renewcommand{\enclname}{encl.}
\renewcommand{\figurename}{Figure}
\renewcommand{\headtoname}{incoming}
\renewcommand{\indexname}{Index}
\renewcommand{\listfigurename}{List of Figures}
\renewcommand{\listtablename}{List of Tables}
\renewcommand{\pagename}{Page}
\renewcommand{\partname}{Part}
\renewcommand{\refname}{References}
\renewcommand{\seename}{see}
\renewcommand{\tablename}{Table}			

\selectlanguage{english}

\thispagestyle{empty}

\begin{center}
Saint Petersburg State University\par
\par
\end{center}

\vspace{20mm}
\begin{flushright}
Manuscript copy

Translated from Russian
\end{flushright}

\vspace{20mm}
\begin{center}
{\large SALISHCHEV SERGEY IGOREVICH}
\end{center}

\vspace{5mm}
\begin{center}
{\bf \large Synthesis of signal processing algorithms with constraints on minimal parallelism and memory space
\par}

\vspace{10mm}
{
Specialty 01.01.09 --- discrete mathematics and 
\par mathematical cybernetics
}

\vspace{10mm}
Dissertation for the degree

of Candidate of Physical and Mathematical Sciences
\end{center}

\vspace{20mm}
\begin{flushright}
Scientific advisor:

Doctor of Physical and Mathematical Sciences, Professor at SPbU

Barabanov Andrey Evgenyevich 
\end{flushright}

\vspace{10mm}
\begin{center}
{Saint Petersburg -- 2016}
\end{center}

\newpage			
\tableofcontents
\clearpage		

\chapter*{Introduction} 
\addcontentsline{toc}{chapter}{Introduction} 

Energy efficiency is one of the key characteristics of semiconductor wireless
devices, as it determines the battery life of the device and its thermal regime.
Battery life and operating temperature define the device's usage scenarios and,
in many cases, its overall applicability.

Modern wireless low-power devices face increasing demands for computational power
while preserving high autonomy and compactness. This is caused by a shift in the
way the device is used toward constant interactive engagement with the
surrounding environment and the information space. Devices now support
low-energy wireless data transfer protocols with high throughput, such as
ZigBee, WiFi, and Bluetooth. In addition, continuous processing of audio and video
data becomes essential for implementing augmented reality and voice control in a
``hands-free'' mode. Consequently, new requirements arise for device energy
efficiency, since the sensor data preprocessing chain and the wireless connection
have to remain continuously active.

Preprocessing may include components such as adaptive beamforming in a microphone
array, adaptive noise suppression, adaptive suppression of far-end and near-end
echo, voice activity detection, recognition of keywords and key events from
other sensors, signal packaging for wireless transmission for downstream cloud
processing, feature extraction (cepstral coefficients, linear prediction
coefficients, feature descriptors for video images, etc.), speaker
identification by voice, and recognition of basic device control commands,
fingerprint identification, and other biometric identification of the user.

Subsequent data processing may include computationally complex algorithms such
as speech recognition, object recognition, and context-aware semantic analysis
(previous user actions, geographic location, audio-visual information, other
sensor data). However, these algorithms start operating only after speech or an
object of interest is detected and the user is authenticated at the preprocessing
stage. In the current generation of devices they are implemented in the cloud.
Therefore, their contribution to device power consumption at the current stage
of technological development is relatively small. Nevertheless, there is a
demand to migrate these algorithms onto the device without reducing battery life
in order to lessen dependence on wireless data connectivity and increase user
privacy. This necessitates optimizing the algorithms and their hardware
implementation for power consumption.

Standard power consumption models for semiconductor circuits take into account
only the active power expended on switching logic elements. Thus, the energy
estimate for an algorithm is proportional to its complexity in elementary
operations, and the power optimization problem has no independent meaning.

Such models are consistent with practical reality for circuits with lithographic
nodes larger than 45 nm. For smaller lithographic nodes, leakage current losses
come to the forefront; they are proportional to the area of the circuit
connected to the power supply and do not depend on the algorithm's computational
complexity. The circuit area is composed of the memory size and the number of
parallel computational elements. Therefore, a more complex energy consumption
model is required, taking into account computational parallelism and the size of
the memory used. The problem of choosing optimal parallelism to minimize power
consumption when modeling speedup using Amdahl's law was considered by Wu and
Lee~\cite{Amdahl-Power} for multicore superscalar processors. Unlike our task,
processors operate at high frequency and high supply voltage, which allows one
to ignore memory power consumption.

Chapter 1 describes a new power consumption model and examines the optimization
of its parameters. The chapter also considers the impact on power consumption of
other factors such as memory architecture, numerical data representation,
processor and operating system architecture, the use of managed runtime
languages (Java, C\#, etc.), parallel computing models, and automatic
verification tools.

As a cross-cutting example of practical significance that illustrates the
proposed approach to optimizing energy efficiency, we consider the problem of
adaptive far-end echo cancellation in conferencing systems using linear
filtering with a long impulse response. To solve this problem we employ the
superfast Schur algorithm for factoring Toeplitz matrices, based on the FFT.

The basic algorithmic blocks for digital signal processing algorithms are
adders, multipliers, and storage devices. 

Further up the hierarchy of complexity are procedures for computing elementary
functions such as $\ln,\exp,\sin,\cos,\sqrt{x}, 1/x$. These functions occur
frequently in digital signal processing algorithms, and the energy efficiency
and speed of their computation can significantly affect the overall energy
efficiency of the device. Computation may use iterative methods, tabulation,
function approximation by polynomials, or their combinations. Various kinds of
splines can be regarded as a combination of tabulation and polynomial
approximation. Computation using programmable processor instructions is also
possible. This method is not the most energy efficient for specialized hardware
and can be used only for one-off computations that do not significantly affect
overall algorithmic complexity. The slowest hardware computation method is the
CORDIC algorithm~\cite{CORDIC}, which computes one bit of the value per cycle
using only addition and shift operations. The next group of methods is based on
bipartite tables~\cite{Bipartite-table} and computes a group of bits per cycle.
As in CORDIC, only addition and shift operations are used. However, the table
size grows rapidly with increasing required precision. For computations at
higher precision, piecewise polynomial approximation is usually employed. The
challenge in hardware implementation of the piecewise polynomial approximation
method is achieving an optimal balance between accuracy, table size, and
polynomial degree, which determines energy expenditure. For elementary
functions, the tables are redundant and can be significantly reduced by exploiting
the smoothness of the functions. Strollo's work~\cite{10.1109/TC.2010.127}
proposed reducing table sizes by 40\% without a significant increase in
computation using a two-link smooth spline; however, accuracy is determined
empirically by testing over all admissible data, which makes the method
inapplicable in its present form for high precisions.

Chapter 2 describes methods of reducing the tables required to compute
elementary functions at a given precision, significantly improving on Strollo's
result.

Among more complex basic signal processing algorithms, the Fast Fourier Transform
is the most commonly used. FFT refers to a family of algorithms for computing
the discrete Fourier transform with computational complexity of ${\mathcal O}(n
\ln n)$, where $n$ is the FFT length. Major areas of research in FFT optimization
include minimizing the total number of operations, optimizing execution on
general-purpose processors with vector instruction support, and optimizing
implementations for specialized semiconductor circuits. The fastest algorithm in
terms of operation count to date is Johnson's algorithm
\cite{journals/tsp/JohnsonF07}, based on the split-radix approach. For short
FFTs, the nonrecursive Sorensen algorithm of the same type is often used
\cite{1164804}. The main difference between parallel specialized circuits and
general-purpose processors lies in memory access discipline. Processor-oriented
algorithms assume random access to both data and code memory, which requires
complex and power-hungry hardware mechanisms, whereas algorithms for specialized
circuits employ memory of specific structure, optimized for power consumption
and area, and a fixed compute block for butterfly operations. Thus, under
parallelism requirements, specialized circuits are more energy efficient because
they do not require special hardware modifications to provide random memory
access and instruction fetch and memory addressing. Existing FFT implementation
algorithms on specialized hardware can be categorized by computation time
asymptotics ${\mathcal O}(1)$, ${\mathcal O}(\ln n)$, ${\mathcal O}(n)$,
${\mathcal O}(n \ln n)$. The algorithm of choice depends on how the FFT compute
block is applied. Under the additional conditions of resource reuse and
controllable transform length, often encountered in real-world tasks with long
FFTs, algorithms with ${\mathcal O}(n \ln n)$ runtime and random-access memory
turn out to be more efficient because they provide flexibility in FFT length and
use library memory. Such algorithms include Johnson's streaming FFT
\cite{johnson1992conflict}. When parallelism is increased by enlarging the FFT
radix beyond 2, pure-radix FFT algorithms lose efficiency, and mixed-radix
computation becomes necessary.

Johnson's algorithm applies only to pure radices or, with minor modifications,
to mixed radices without parallel butterfly execution. The modification of
Johnson's algorithm proposed by Jo and Sunwoo~\cite{jo2005new} is specialized
for radices 2/4. Applying it to other mixed radices requires adapting the
algorithm.

In many tasks, self-sorting FFT algorithms are more efficient; these include the
Johnson-Burrus and Templeton implementations
\cite{Hegland:1994:SIF:201897.201901}. The Johnson-Burrus algorithm is
formulated for single-bank random-access memory; using it with other memory
types requires adaptation. The Templeton algorithm is strictly sequential and
formulated for scalar processors.

Algorithmic adaptation is also required for the most efficient memory
architecture provided by the component library, for example, single-port memory.
Such adaptations for in-place FFT algorithms have yet to be considered in the
literature.

Developing streaming FFT algorithms with memory access constraints essentially
amounts to scheduling a synchronous data flow graph. The results of each
butterfly are written back to the same locations from which their inputs were
read. In multibank memory, all wings of each butterfly must land in different
banks; otherwise conflicts arise. At the same time, the total number of memory
cells is minimal-equal to the FFT length. The combinatorial problems of finding
initial data distributions that guarantee conflict-free stages for arbitrary
radices constitute the core of Chapter 3.

Another commonly used class of algorithms in digital signal processing is $LU$
factorization of Toeplitz and inverse Toeplitz matrices. This is required when
solving the Yule-Walker equations for constructing optimal noncausal linear
filters. Such problems of large size frequently arise in acoustic echo and noise
cancellation applications and in separating signal sources. This is due to the
low propagation speed of sound and reverberation. Factorization problems are
usually solved using Levinson and Schur algorithms with complexity of ${\mathcal
O}(n^2)$, where $n$ is the autocorrelation vector length. For large-scale tasks,
one can use the fast Schur algorithm proposed by Ammar and Gragg
~\cite{Ammar87thegeneralized} and Voevodin and Tyrtyshnikov~\cite{VoevTyrt} with
complexity ${\mathcal O}(n \ln^2 n)$, or the preconditioned conjugate gradient
method with iteration complexity ${\mathcal O}(n \ln n)$. Both algorithms rely
on the FFT and can serve as examples of energy-efficient implementations of
complex algorithms with FFT.

The fast Schur algorithm is described in numerous papers and appears as a long
sequence of algebraic instructions containing convolutions of matrix polynomials
and recursive function calls. The number of additions and multiplications, as
well as the required memory, has been evaluated. 

In this work we studied the impact of parallelism on the algorithm's
implementation complexity expressed in the energy expenditures described in
Chapter 1. To this end, the algorithm had to be represented as a traversal of a
weighted graph. Based on the study of this graph, Chapter 4 improves the
required memory value and calculates the length of the critical path in terms of
read and write operations. Furthermore, the optimal degree of parallelism for
the fast Schur algorithm on 4096 samples was found for execution on a streaming FFT
accelerator; it turned out to be $p=4$.

The \underline{\textbf{objective}} of this work is to develop algorithms that minimize the
complexity of computations typical of statistical signal processing. Complexity
is measured by energy efficiency, modeled in Chapter 1. It depends, in
particular, on table lengths when computing standard elementary functions, on
memory access in FFT implementations, on self-sorting FFTs, and on parallelism
in the fast inversion algorithm for Toeplitz matrices.

To achieve this \underline{\textbf{objective}}, the following \underline{\textbf{tasks}} were addressed:
\begin{enumerate} 
\item Formulate a complexity functional and other requirements imposed on algorithms for their effective hardware implementation; 
\item Develop efficient integer algorithms for computing elementary functions with a given precision and minimal table length; 
\item Develop a schedule for implementing the FFT data flow graph in multibank memory;
\item Investigate the parallelism of the fast Schur algorithm.
\end{enumerate}

\underline{\textbf{Main propositions submitted for defense:}} \begin{enumerate}
\item A method for qualitative power estimation and selection of optimal
parallelism for energy-efficient specialized CMOS compute blocks.
\item A method of computing elementary functions using an almost smooth
four-link quasi-spline and an accuracy estimate for polynomial approximation
with fixed-point coefficients constrained on a uniform grid.
\item A theorem on FFT data placement in multibank memory when computing with
arbitrary mixed radices.
\item A theorem on data placement and computation order for a self-sorting FFT.
\item A theorem on data placement and computation order for an FFT with single-port memory.
\item An analysis of the energy efficiency of the $LU$ factorization algorithm for
real Toeplitz matrices on a convolutional accelerator for echo cancellation
tasks using the fast Schur algorithm.
\end{enumerate}

\underline{\textbf{Scientific novelty:}} \begin{enumerate} 
\item An energy consumption model has been developed for a low-power digital
circuit executing a known computational algorithm. The model includes dynamic
and static energy losses and idle shutdown. The problem of selecting optimal
parallelism within this model has been solved.

\item The problem of minimizing power consumption when computing values of
standard functions has been reduced to minimizing table lengths. New
quasi-spline approximation methods with nonuniform tabulation have been
developed, convenient for hardware implementation, which made it possible to
shorten the table lengths of all principal standard functions compared with known
analogs, leading to a substantial reduction in power consumption and an increase
in the speed of hardware computational blocks.

\item A theorem has been proven on FFT data placement in multibank memory when
computing with arbitrary mixed radices, guaranteeing homogeneity of the
synchronous data flow graph for computations, which ensures maximum computation
speed at a given parallelism and zero buffer size. Explicit FFT formulas have
been derived as Kronecker products stage by stage for arbitrary orders. 

\item A theorem has been proven on the self-sorting modification of the FFT in
multibank memory for mixed radices, as well as a similar theorem for a computing device with
single-port memory.

\item For the fast Schur algorithm, the minimal memory space has been found,
the length of the critical path has been computed, and the optimal parallelism has
been evaluated.

\end{enumerate}

The \underline{\textbf{practical significance}} of the dissertation is ensured by reducing
the area and power consumption of the components considered and increasing their
versatility, which makes a significant contribution to improving the energy
efficiency of autonomous wireless devices implemented on specialized semiconductor logic circuits.

The \underline{\textbf{reliability}} of the results presented in the work is ensured by the
practical implementation of the proposed schemes and algorithms as semiconductor
circuits. The practical implementation included developing models in SystemC, 
automatic verification using the Aegis for SystemC system,
gate-level logic synthesis of semiconductor circuits for accelerators from
SystemC models, logic simulation of circuit operation, and synthesis of a
virtual topology for a low-power semiconductor chip fabrication process with 22 nm
geometries.

\underline{\textbf{Dissemination of results.}} The main results were presented at the
International Conference of the Audio Engineering Society (AES) (Russia, Saint
Petersburg, 2003), the international conference on Computer Data Analysis and
Modeling (CDAM) (Belarus, Minsk, 2004), the young scientists' conference
"Gyroscopy and Navigation" (Russia, Saint Petersburg, 2004), the seminar of the
Department of Theoretical Cybernetics of Saint Petersburg State University
(Russia, Saint Petersburg, 2015, 2016), and Intel Labs workshops (2013-2015).  

\underline{\textbf{Personal contribution.}} 
The author proposed an energy consumption model for a low-power digital circuit
executing a known computational algorithm and investigated the optimal
parallelism in it. The author proposed modifications for the streaming FFT
accelerator architecture with random-access memory and justified the correctness
of the resulting algorithms. The author proposed a modification of the
piecewise-polynomial approximation architecture with flexible constraints on
four segments and justified the correctness of the resulting algorithms. The
author proposed a modification of the superfast Schur algorithm for
implementation on a convolutional accelerator and investigated its complexity and optimal parallelism.

\underline{\textbf{Publications.}} The main results on the dissertation topic are
presented in 12 printed publications~\cite{salishev2005echo, salishev2005echo2, 
salishev2012poly, salishev2013fft, barabanov2002echo, salishev2005echo3, 
salishev2009mre, salishev2008mre, Frampton:2009:DMH:1508293.1508305, 6089227,
Glukhikh:2013:SAA:2547165.2547276, salishev2014continuous}, including
4~\cite{salishev2005echo, salishev2005echo2, salishev2012poly, salishev2013fft}
in journals recommended by the Higher Attestation Commission,
5~\cite{barabanov2002echo, salishev2005echo3,
Frampton:2009:DMH:1508293.1508305, 6089227, Glukhikh:2013:SAA:2547165.2547276}
in proceedings of international conferences in English, of which
3~\cite{Frampton:2009:DMH:1508293.1508305, 6089227,
Glukhikh:2013:SAA:2547165.2547276} are indexed by Scopus,
1~\cite{salishev2014continuous} is a US patent application.

The works~\cite{barabanov2002echo, salishev2013fft, Frampton:2009:DMH:1508293.1508305, salishev2008mre, salishev2009mre, Glukhikh:2013:SAA:2547165.2547276, 6089227} were co-authored.  
In~\cite{salishev2013fft} the author is responsible for the problem statement,
the formulation of all theorems and their proofs, except for the proof of Theorem 4. 
In~\cite{Frampton:2009:DMH:1508293.1508305, salishev2008mre} the author
contributed the section devoted to practical implementation experience. 
In~\cite{salishev2009mre} the author is responsible for the problem statement, 
the analysis of existing systems to extract common requirements, and the
section on using Java in systems programming. 
In~\cite{Glukhikh:2013:SAA:2547165.2547276, 6089227} the author is responsible
for the problem statement and the development of the error detection algorithm
for synchronization via reachability analysis. 
In~\cite{barabanov2002echo} the author performed mathematical modeling.

\underline{\textbf{Scope and structure of the work.}} The dissertation consists of an
introduction, four chapters, and a conclusion. The total length of the
dissertation is \textbf{191} pages of text. The main text spans \textbf{152}
pages with \textbf{9} figures, \textbf{14} tables, and \textbf{5} appendices.
The bibliography contains \textbf{85} references.

\clearpage	
\chapter{Energy Efficiency Factors} \label{chapt1} 
Energy efficiency is one of the main parameters of wireless 
semiconductor devices.

Energy efficiency optimization is possible at various levels: transistor,
gate, architectural, and system levels. Optimization issues at gate and
transistor levels have been repeatedly considered and lie in the field of
semiconductor physics.

Algorithmic optimizations related to the system level are typically
considered within the framework of existing microprocessor architectures and are aimed
at improving performance, rather than energy efficiency.

Architectural optimizations are closely related to algorithmic ones, since
they determine the relative cost of operations, memory model, and parallelism. However,
they are usually not considered together with algorithmic optimizations for
improving energy consumption, since for a fixed processor
architecture they are predetermined and are constraints, rather than optimization
parameters.

One of the standard approaches to improving energy efficiency is the
development of specialized semiconductor logic circuits with fixed
functionality for implementing a certain set of algorithms. However, this
approach has several serious disadvantages. First, it limits
the flexibility of the resulting system, since reusing it to implement
other algorithms becomes impossible. Algorithms in signal processing
and data analysis are constantly evolving, and the development and
manufacturing time of a semiconductor circuit is at least 6 months. Thus,
the circuit may become obsolete already during preparation for production. 

Algorithm errors
that make it into the hardware implementation of the device cannot be corrected without
releasing a new version of the chip, which makes such errors costly. This
drives conservatism in algorithm selection and additional effort on
verifying their correctness during hardware development
of the device, which further slows down the development process. Moreover, the circuit
cannot be used in other similar devices having slightly
different functionality (for example, cell phone, wireless headset,
head-mounted device, watch), if this variability was not planned during
development. Second, in chips manufactured using modern lithographic
processes, a significant portion of energy losses is created by leakage currents, which
are proportional to the total area of elements connected to power on the chip.
Thus, to minimize leakage currents, large base blocks, such as
multipliers, adder trees, controlled shift registers, register
memory, static memory should be maximally reused between
different parts of algorithms and, if possible, completely disconnected from
power during idle time.

This requirement is conceptually close to the concept of a programmable processor.
By programmability is meant the presence of effective development
tools for programs in high-level languages (C, C++, Fortran, etc.) and
the possibility of efficient compilation of programs implementing a wide class
of digital signal processing algorithms.

Development based on a programmable processor has the following
advantages over a specialized non-programmable device:
\begin{enumerate} 
\item possibility of reusing computational resources
between parts of the algorithm to save chip area and improve
energy efficiency; 
\item reuse of data memory by various parts
of algorithms to reduce its size and reduce the need for copying
data, which leads to improved energy efficiency; 
\item possibility
of correcting errors in the algorithm after chip manufacture, which allows
separating the processes of hardware and software development to reduce overall
development time; 
\item possibility of reusing the block and architecture
of the chip in other types of devices by changing the algorithm to reduce
development time and device production cost; 
\item presence of developed programming tools (compiler,
debugger, operating system, libraries) to reduce development time.
\end{enumerate}

On the other hand, general-purpose signal processors often
turn out to be insufficiently performant or energy efficient for
use in low-power autonomous devices for the following reasons:
\begin{enumerate} 
\item additional blocks for implementing programmability;
\item insufficient parallelism associated with sequential program structure;
\item excessive width of standard data types, requiring wider memory and
more complex computational blocks; 
\item use of part of the memory bandwidth for program loading;
\item time losses on executing branches and loops;
\item small granularity of universal operations, leading to excessive 
memory accesses.
\end{enumerate}

Thus,
development of specialized energy-efficient digital data processing
systems with high performance and resource reuse capability
appears to be the most advantageous strategy both from the point of view of flexibility and
from the point of view of energy efficiency. A general architecture for such computational 
blocks based on RAM was proposed by Hartenstein~\cite{hartenstein1991novel} and is based on 
data flow, in contrast to conventional processors controlled by a program 
(control flow). The architecture is controlled using address generator templates
and data flow configurations, thus it can implement 
arbitrary predetermined parallelism and resource reuse 
without overhead for program execution.

These computational blocks can then be
integrated with an extensible processor core as an instruction set
extension or as accelerators for computing library functions. Thus,
both the flexibility of a programmable processor and high performance
and energy efficiency of specialized semiconductor circuits are preserved. This strategy
underlies this work.

Since development of effective programming tools
is an extremely complex and resource-intensive task, the processor should
be based on a standard microarchitecture and instruction set with a set of
extensions accelerating execution of algorithms specific to this domain.
Such standard extensible microarchitectures are ARM, ARC, MIPS,
Sparc. All of them have excellent support from open-source development
tools such as GNU GCC and Clang/LLVM. Open license
on development tools allows relatively easy adaptation of them to
required sets of extensions.

Within the framework of processor specialization, two tasks are solved. First, a
memory architecture and data representation most convenient for this class
of algorithms is chosen, for example, vector width for vector computations, word width,
use of floating-point numbers, presence of penalty for unaligned access
to memory, width and number of additional vector registers.

Second, a set of additional processor instructions accelerating
program execution, characteristic of this class of algorithms, is chosen, for example, shift
with rounding, addition with saturation, finding the most significant nonzero bit,
FIR filtering, FFT butterfly, computation of elementary functions, bit
permutations, etc.

The task of choosing memory architecture is the most complex and important, since
most algorithms when parallelized and specialized turn out to be
limited precisely by memory bandwidth, rather than computational blocks.
Moreover, memory architecture determines the context of algorithm implementation and can
significantly influence the course of computations. In this case, the preferable
case is when the processor core and specialized accelerator have shared
memory, since overhead for data copying is reduced.

A distinctive feature of programmable processors is the presence
of a program in memory, which increases the necessary memory bandwidth
for instruction fetch. Usually specialized processors have Harvard
architecture, which, unlike the traditional architecture for general-purpose
processors, von Neumann architecture, has additional program memory,
connected to the processor by a separate bus. This eliminates the so-called "von Neumann
bottleneck", when accesses to program and data compete for memory
bus. The next bottleneck becomes the process of decoding and executing
commands. To simplify the instruction decoding block, most
specialized processors decode instructions one at a time in the order of fetch
from memory. Instructions are decoded at the rate of one instruction per synchronization
cycle. Modern processors execute instructions in a pipeline. That is,
one instruction passes through several processing stages: fetch from memory,
decoding, reading data from registers or memory, computing result,
computing address of next instruction, writing result to registers or memory.
If an instruction execution stage takes several synchronization cycles, then
results of sequential instructions can be reordered, if
they are data-independent.

Most algorithms used in signal processing and adaptive
control consist in applying a small computational kernel to
a significant volume of data. In program code this is reflected in loops with
small body and significant number of iterations. When choosing instruction set
extension, overhead for instruction decoding and
loop organization, instruction execution duration and possibility of their
reordering should be taken into account. These overheads can be comparable to costs of
the computations themselves. In case of significant overhead for efficient
algorithm implementation, the computational block should be developed not as
an instruction set extension, but as an accelerator for computing library functions,
since in this case it does not interact with the processor pipeline, and
overheads are minimized.

Implementation of semiconductor circuits can be performed in logical
synthesis languages Verilog, VHDL or using high-level logical
synthesis tools from languages C, C++, SystemC, System Verilog, BlueSpec System Verilog, Haskell.
Use of high-level logical synthesis significantly reduces time
of development and verification of logical circuits. This method was used in this
work.

There are two main approaches to developing energy-efficient circuits from the point
of view of choosing optimal parallelism of computations and clock frequency:
\begin{enumerate} 
\item "uniform work", when performance and
clock frequency are matched according to computational load; 
\item
"burst to sleep", when the device functions at maximum effective
frequency and after completion of work goes to sleep with power disconnection from part of blocks.
\end{enumerate} 

A critical difference in low-power lithographic processes
with small geometric norms (45~nm and less) compared with older ones is
the dominance of power losses due to leakage currents over active power,
associated with switching of logical states of the circuit when operating in
low-energy modes with reduced clock frequency. Leakage currents
are essentially proportional to the area of the circuit connected to power on the chip.
The "race-to-sleep" approach may be more advantageous for such circuits for
several reasons: 
\begin{enumerate} 
\item memory block area, as a rule,
dominates over computational circuit area, while memory size is a
characteristic of the algorithm and weakly depends on computation speed; 
\item during
idle time, complete disconnection from power of blocks including part of memory
is possible to save energy; 
\item dynamic clock frequency management is not required
in low-energy modes, which simplifies the development process; 
\item
the device has a reserve of computational resources, which increases possibilities
of modernization and reuse without changing the architecture.
\end{enumerate}

Basic algorithmic blocks for digital signal processing algorithms
are adders, multipliers, and memory devices. For adders and
multipliers, the main way to reduce energy consumption is to reduce
the number of gates used. For implementing adders, carry-save
arithmetic and tree structure of computations~\cite{carry-save} is used.
This allows avoiding long chains of carries, which leads to path balancing
and reduction of delay on the critical path. For implementing multipliers
a combination of Wallace~\cite{Wallace-Tree} and
Booth~\cite{Booth} methods is used. Multiplication is considered as summation of results
of bit-wise multiplications in carry-save arithmetic. This allows efficiently
combining addition and multiplication in dependency graphs of computations, inserting
full adder with carry only before writing values to register or to
memory. For implementing full adder, ripple carry circuits, carry lookahead
circuit~\cite{carry-lookahead}, and parallel
prefix method~\cite{parallel-prefix-add} are usually used. These methods differ in
ratio of number of gates and critical path length. Logical synthesis
tools perform automatic optimizations of computation graphs from adders and
multipliers to achieve optimal circuit area at a given target
frequency, which is achieved by balancing path lengths in the circuit. Manual
optimizations at the level of basic operations generally lead to worse
results in energy consumption compared to automatic optimizations in
logical synthesis tools.

At the next level are parallel vector operations on small
vectors, vector addition, multiplication, operations on quaternions, linear
filtering. Energy-efficient implementation of these operations does not present
difficulties from the algorithmic point of view, however it requires choosing and accounting for
specifics of memory architecture, to ensure the necessary level of parallelism
without additional energy costs per operation compared to
sequential implementation.

\section{General Model of Energy Consumption of Clocked Logic Circuits}
Consider factors influencing energy efficiency of computational blocks when
constructing low-power semiconductor circuits specialized for
a group of algorithms. We will also take into account two additional criteria:
development speed and convenience of composition of computational blocks for implementing
complex algorithms.

Physical factors include threshold voltage, leakage currents, parasitic
capacitance. Other factors are architecture, type and size of memory, precision
of intermediate computations, logical synthesis methods. Let us analyze
interrelationships of these factors and their influence on energy consumption.

Since the considered circuits operate continuously, power is used to characterize their
energy consumption. Power of a semiconductor circuit
consists of active and static dissipated power. $$P=P_{a}+P_{s}$$
Active power is due to switching of semiconductor circuit state and
is related to computation intensity. Static power does not depend on
computations, but only on the amount of hardware connected to power.

Complex computational blocks are developed in the form of clocked
semiconductor circuits. The circuit consists of registers storing state, and
combinational logic circuits describing transitions between states. Transition
between states occurs on strobe of the clock signal. Such a circuit can
be represented as a finite automaton.
$$
F=(R, \sigma, \Sigma); \Sigma \subset R \subset \{0,1\}^{n};\qquad \sigma: R
\mapsto R
$$

Here $R$ is the set of admissible states, $\Sigma$ is the set of initial
states, $\sigma$ is the transition function. The transition function in a semiconductor
circuit is implemented as a directed acyclic graph of elementary library
functions or gates. Gates can implement functions such as AND, OR,
NOR, XOR, multiplexer, selector, etc. At a lower level, gates consist
of transistors. System state is stored in registers.

Such a level of detail in describing logic circuits is called Register 
Transfer Level.

\section{CMOS Device Power} \label{sect1_2} For power estimation we will
rely on a basic model described in chapter two ~\cite{Power}.

For one gate, power is defined as
$$
P = P_{a} + P_{s}.
$$

Here $P_a$ is active power, depending on data, $P_s$ is static
power, independent of data.
\begin{eqnarray*}
P_{a} = P_{d}+P_{sc}, \\
P_{s} = P_{leak}
\end{eqnarray*}

Here $P_d$ is dynamic power, $P_{leak}$ is power losses from leakage
currents, $P_{sc}$ is power losses as a result of short circuit during
transistor switching.
\begin{equation}
\label{eq:d_pow} P_{d} \simeq \alpha CV^2f 
\end{equation} 

Here
$C$ is parasitic capacitance, $V$ is supply voltage, $f$ is clock frequency,
$\alpha$ is switching frequency coefficient, showing average probability
of gate opening at each cycle. For strobe signal $\alpha=1$, for others on
average $0.1$. Maximum clock frequency of circuit operation at a given supply
voltage is related to delay on the critical path, i.e., chain of gates
having the largest switching delay.
$$
f \leq 1/d_{crit},\qquad d_{crit}=\sum d_i
$$ 

Delay on one gate is also related to voltage by the following formula:
\begin{equation}\label{eq:delay} 
d(V) \propto \frac{C_L V}{(V-V_T)^2} \simeq \frac{C}{V}, \qquad V >> V_T. 
\end{equation}

Here $C_L$ is load capacitance, which consists of input-charged
gate capacitances of transistors and parasitic wire capacitances.
$$
C_T=C_{ox}WL
$$
Here $C_T$ is transistor gate capacitance, $V_T$ is threshold voltage, $C_{ox}$ is 
specific gate oxide capacitance, $W$, $L$ are transistor gate width and length.

Due to manufacturing defects, digital circuits work unstably at
voltages close to $V_T$; there exists a minimum voltage $V_0 >> V_T$,
ensuring stable circuit operation. At this voltage, the circuit can operate
at maximum clock frequency $f_0=1/d_{crit}(V_0)$. The circuit can also operate
at lower clock frequency, if its operating speed is sufficient for solving
the problem. On frequency interval $f \in (0, f_0]$, power growth with frequency is linear. At
large clock frequencies
$$
V \propto \frac{1}{d_{crit}} \propto f,\qquad f > f_0.
$$

Thus, combining these two estimates, we obtain an estimate
$$
P_d(f) \propto \left\{ \begin{array}{ll} f, & f \leq f_0\\ f^3, & f > f_0
\end{array} \right.
$$

Power losses as a result of short circuit during transistor switching
are determined by the following formula:
$$
P_{sc} \simeq \alpha T_{sc}I_{sc}V, \qquad T_{sc}=\frac{T_r + T_f}{2}
$$

Here $T_r$, $T_f$ are signal rise and fall times, respectively. For logic
circuits $P_{sc}$ is small and may not be accounted for. We will not account for it in
further calculations.

Static power is mainly due to leakage currents flowing through
gates in closed state and is defined for one gate at room
temperature as
\begin{equation}\label{eq:s_pow} 
P_{leak} \simeq VI_{sub}, \qquad I_{sub} \propto C_{ox}W/L 
\end{equation}

Thus, with decreasing geometric norms, leakage currents grow. Best
energy efficiency is achieved at $V=V_0$. Threshold voltage $V_T$
increases with increasing gate length $L$, which reduces leakage current.

In component library for one technology, there can be gates with different
geometries and different threshold voltage $V_T$. Increase of threshold
voltage leads to decrease of performance.
 
\section{Energy Consumption Optimization Methods}

Energy consumption optimization can occur at various levels.
Sherazi~\cite{disser:Sherazi} gives an overview of such
optimizations. Optimizations are listed in table~\ref{table:energy_opt}.

\begin{table} [htbp] \centering \caption{Energy consumption optimization methods.}
\label{table:energy_opt} \begin{tabular}{|l|l|} 
\hline 
Level&Optimization\\
\hline
Transistor&Reducing supply voltage\\ 
&Multiple supply voltages\\
&Transistor scaling\\
&Substrate potential shift\\ 
\hline
Gate&Multiple clock signals\\ 
&Clock gating\\
&Reducing transients\\ 
&Transistor stacking\\ 
&Gates with different threshold voltages\\ 
\hline 
Microarchitectural&Block duplication\\ 
&Pipelining\\ 
&Resource allocation\\ 
&Data representation optimization\\ 
&Arithmetic optimizations\\ 
&Power gating\\
&Time-multiplexing resources\\ 
\hline 
System&Dynamic frequency\\ 
&and supply voltage management\\ 
&Use of parallel algorithms\\ 
\hline 
\end{tabular} \end{table}

Transistor and gate levels are considered in many works and are sufficiently
well studied. Upper levels and their interaction with each other
are rarely considered. In this work we will mainly consider system and
architectural levels of optimization. The main type of system optimization is
algorithmic optimization. It is easy to see that algorithm complexity influences its
energy consumption. However, complexity is defined within the framework of some model
computations, which, in turn, is determined by the architecture
of the semiconductor device and its large functional blocks. The greatest
influence on energy consumption is exerted by implementation of complex computational
algorithms. These often include digital signal processing algorithms.

A brief description of energy consumption optimization methods is given in 
appendix~\ref{Appendix:optim}.

\section{Logical Synthesis Methods}\label{sect1_1} A logical circuit
is synthesized from library elements, which are part of
the technological process. The library usually contains various types of gates and
various implementations of a gate depending on the required delay. Different
implementations of one gate can differ in area, delay, and
energy consumption by several times.

There are several methods for synthesizing logical circuits. The oldest is manual
assembly of logical circuits from basic library gates. Today, manual assembly
is used only for high-performance systems produced in mass
series.

For low-power systems, synthesis from high-level
hardware description languages Verilog, VHDL is mainly used. The logical circuit is
described at the level of connecting library blocks through interfaces,
non-addressable registers storing variables, arithmetic and logical
functions, such as multiplication, addition, Boolean algebra operations. The logical
synthesis program based on high-level description finds a locally
optimal circuit for connecting library gates under given constraints on
clock frequency and circuit geometry on the chip. For arithmetic functions,
the synthesis program can perform equivalent arithmetic transformations,
arithmetization of branch computation and find optimal configurations of addition
and multiplication circuits. Moreover, automatic insertion and
removal of registers and transfer of computations between pipeline stages can be performed. These
transformations allow balancing the length of the critical path and other paths in
the circuit and using the most economical types of gates. Due to the large
search space, the synthesis program usually cannot find a globally
optimal solution, so synthesis results are often unstable to small
changes in the circuit description.

Synthesis from Verilog, VHDL has some disadvantages. The computational model
of the algorithm is usually written in imperative high-level languages, such as C,
C++, Matlab. Languages Verilog, VHDL have a significantly lower level of
abstraction, which requires serious rewriting of the model and subsequent
verification. Verification is performed using extremely slow logical
simulators. These reasons increase development effort, often lead to
errors and prevent making changes to the algorithm after converting the model to
Verilog, VHDL.

A new method of logical synthesis from high-level
languages (High Level Synthesis) is gradually gaining popularity. The general idea is automation
of converting an algorithm into a logical circuit, so that the developer
mainly needs to specify interfaces for interaction between blocks and
memory structure, while insertion of registers for storing intermediate results and
pipelining of computations are performed automatically. There exist industrial
tools for logical synthesis from algorithmic high-level languages Matlab,
C, C++, SystemC, Bluespec Verilog. There also exist experimental tools
for synthesis from languages Hardware Join Java, Haskell and others. Some disadvantage
of high-level synthesis is even less stability of synthesis results
to model changes due to the impossibility of finding a globally optimal hardware
implementation of the algorithm. This often leads to somewhat worse synthesis quality
compared to synthesis of an equivalent implementation from languages Verilog, VHDL. However,
the time gain allows either obtaining an acceptable result significantly
faster, or more fully exploring the space of admissible implementations and obtaining
significantly better synthesis results through algorithmic optimizations.
Synthesis can be performed either in one stage from a high-level language to a graph
of library gates, or in two stages with intermediate translation to Verilog,
VHDL. In the work, two-stage synthesis was used with Synopsys Design Compiler Ultra at the final
stage, which can account for the actual geometry
of components on the chip, unlike current implementations of HLS tools. Accounting for
actual chip geometry for lithographic processes with geometric
norms 45 nm and less is important due to the large area of inter-gate
connections and clocking circuits, whose influence is not accounted for when synthesizing to
a gate graph~\cite{6089686}.

In the work, the language
SystemC~\cite{citeulike:1136497} was used for describing high-level models, which is a language for modeling parallel
systems and hardware based on C++, intended for system-level
modeling and verification. SystemC is an IEEE standard and allows modeling
a system including software and hardware components at various levels from
abstract synchronous and asynchronous transactions to precise modeling
of hardware. An advantage of the SystemC language is the absence of the need
to rewrite computational models if they are written in C++, speed and
simplicity of verification, simplicity of including software components
of the modeled system in the model.

Experience shows that proper hierarchical decomposition of the circuit, use of
idioms, such as separation of logical and arithmetic expressions and
maximization of arithmetic computation graphs can influence the
resulting circuit area and power by up to two times. Even greater influence is exerted by parameters
of synthesis tools: order of module synthesis, permission
to transfer computations between pipeline stages, constraint on synthesis time, and
so on.

\section{Typical Accelerator Architecture} \label{sect1_3} \begin{figure}[ht]
\center \includegraphics[width=0.5\textwidth]{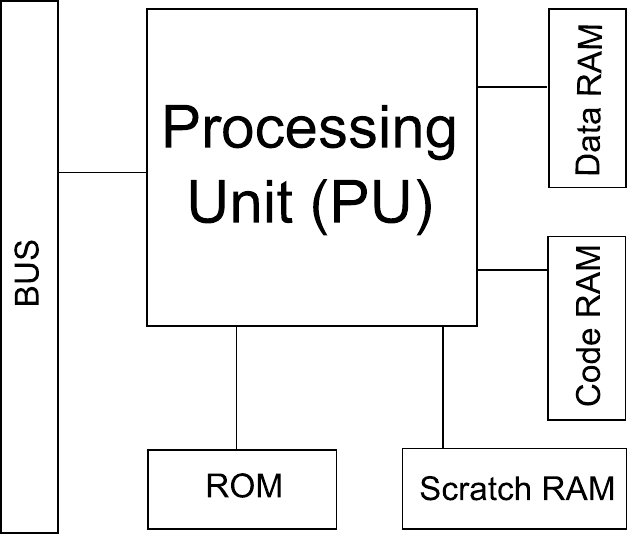} \caption{Basic
block architecture.} \label{fig:base-arch} \end{figure}
        
Typical architecture of modern systems-on-chip includes many
specialized components, united by one or several buses or
a switch. Each specialized block also usually has a typical
structure, shown in fig.~\ref{fig:base-arch}.

Power of such a block consists of static power of components that cannot be disconnected
from power and power of components with controlled power. To
non-disconnectable components belong code memory and state memory, other
components can be disconnected during idle time. To disconnectable components
belong computational devices, intermediate data memory and read-only
memory. Use of non-volatile memory in constantly operating blocks
is currently not justified due to high
overhead for reading and writing.

\subsection{Estimation of Accelerator Dissipated Power at Fixed Problem Size} \label{sect1_3_1} 

Suppose that a fixed algorithm is executed on a computational block, in which 
$p$ processors can be engaged in parallel. It is required to find $p$ at which the consumed power of the block is minimal.

The processed algorithm is characterized by the following constant parameters:
\begin{itemize}
\item $B$ - computational complexity of the algorithm, expressed in the number of elementary operations per
second. An elementary operation is launched for execution each clock cycle. Usually operations are executed in a pipeline,
that is, the execution time of an operation is several clock cycles.
\item $K$ - non-parallelizable fraction of the algorithm, equal to the ratio of the critical path length to the length of the entire algorithm.
\end{itemize}

The computational block is characterized by the following constant parameters:
\begin{itemize}
\item $N_0$ - number of gates in the non-disconnectable part,
\item $N_1$ - number of gates in the disconnectable part without computational components,
\item $\hat{N}_2$ - number of gates in the sequential implementation of computational components of the block,
\item $f$ - clock frequency.
\end{itemize}

The number of gates in computational components of the block taking into account parallelism is defined as $N_2 = p \hat{N}_2$.

In these notations, using formulas~\ref{eq:d_pow} and~\ref{eq:s_pow}, the power of the computational block on the linear growth section from frequency can be estimated as 
$$
P_s=P_{leak}=c_0 (N_0+S(N_1+N_2))
$$
$$
P_a=P_{d}=c_1 f S (N_1 + N_2)
$$
$$
P=P_a+P_s=c_0 (N_0+S(N_1+N_2)) + c_1 f S (N_1 + N_2)
$$
where $c_0$, $c_1$ are constant coefficients, and $S=S(p)\le 1$ is the duty cycle or ratio of computation time to total time, including idle time.

At $p=1$ the duty cycle equals $S=B/f$ by definition. At $p>1$ the function $S(p)$ is replaced by the speedup function from parallelism $l(p)$ in accordance with the equation 
$$
S(p)=\frac{B}{l(p)f}.
$$

The function $l(p)$ can be described using Amdahl's
law~\cite{Amdahl:1967:VSP:1465482.1465560}, which is used in the field of 
parallel programming to estimate speedup at fixed problem
size.
$$
l(p)=\frac{p}{1+K(p-1)}.
$$

The question of optimal choice of parallelism for minimizing energy consumption when
modeling speedup using Amdahl's law was considered by Woo and
Lee~\cite{Woo:2008:EAL:1495784.1495841} for multicore superscalar
processors. Unlike the problem we consider, processors operate at
high frequency and at high supply voltage, which allows not 
accounting for memory energy consumption.

Thus, 
\begin{equation}\label{eq:power} 
P=c_0 N_0 + B (c_0 / f + c_1)\frac{ p \hat{N}_2 + N_1}{l(p)} 
\end{equation}

At the same time, power reaches a minimum at the maximum frequency of linear 
scaling $f_0$:
$$
\arg \min_{f, f\leq f_0}{P}=f_0.
$$

\begin{lemma} \label{lemma:P_min}
Power reaches a minimum in $p$ at the value
\begin{equation}\label{eq:opt-par} 
p_0=\arg \min_p{P}=\max\left(
1,\sqrt\frac{N_1(1 - K)}{\hat{N}_2K} \right).
\end{equation}

At the same time, the minimum power has the value
\begin{equation}\label{eq:opt-power} 
P_{\min}=c_0 N_0 + B (c_0 / f_0 + c_1)
\left((1-K)\hat{N}_2 + KN_1 + 2\sqrt{K N_1 \hat{N}_2(1-K)}
\right). 
\end{equation}
\end{lemma}

\proof
The minimized power function can be represented in the form 
$$
P(p) = \alpha p + \beta + \gamma p^{-1},
$$
with suitable coefficients $\alpha$, $\beta$, $\gamma$. By differentiation we find the minimum condition $\alpha - \gamma p^{-2}=0$. From this
by identical transformations we obtain the conclusion of the lemma. \square

\vspace{3mm}
An important indicator of parallelization quality is the maximum energy savings
$$
\rho = \frac{P(1)}{P_{\min}},
$$
where $P(1)$ is the power of the block in the absence of parallelization. 

\begin{corollary}
At full parallelization (at $K=0$)
\begin{eqnarray*}
\rho & = & 1 + \frac{B (c_0 / f_0 + c_1) N_1}{P_{\min}}, \\
P_{\min} & = & c_0 N_0 + B (c_0 / f_0 + c_1) \hat{N}_2.
\end{eqnarray*}
\end{corollary}

\vspace{3mm}
In engineering practice, work on algorithm parallelization makes sense to conduct
only if energy savings $\rho \geq 1.2$.

Thus, within this architecture, for well-parallelizable
algorithms with dominance of intermediate result memory, increasing
parallelism leads to energy savings. Savings are achieved by reducing
static power of intermediate data memory by disconnecting from
power.

\subsection{Asymptotic Rate of Power Growth with Problem Size Growth}

In addition to the problem of estimating optimal parallelism, consider the problem of estimating power 
growth with increasing problem size. Although usually the problem size in algorithms 
for low-energy devices is known in advance, in some cases it 
can be varied. For example, filter size, number of estimated parameters, 
image resolution, etc.

The power function depends on parallelism $p$ and problem size $n$:
$$
P(p, n)=c_0 N_0(n) + B(n) (c_0 / f + c_1)\frac{p\hat{N}_2 + N_1(n)}{l(p,n)},
$$
where quantities $N_0(n)$, $N_1(n)$ depend on memory size, which 
usually grows with problem size growth. The quantity $\hat{N}_2$ does not depend on $n$, since the computation 
algorithm does not change.

To estimate $l(p,n)$ we use Gustafson-Barsis' law~\cite{Gustafson88}, 
which is used in the field of parallel programming for estimates when 
problem size grows: 
$$
l(p,n) = (1-\alpha)p + \alpha.
$$
Gustafson-Barsis' law assumes that the non-parallelizable part 
of the problem does not grow with dimension growth, which often occurs when
parallel subtasks are independent of each other. 

\begin{lemma}
Let memory size $N_0(n)$, $N_1(n)$ and algorithm complexity $B(n)$ grow linearly with problem size. Then in the absence of parallelization, power $P(1,n)=O(n^2)$, and minimum power $P_{\min}(n)=O(n)$ as $n\to\infty$.  
\end{lemma}

\proof
Substitute linear functions and Gustafson-Barsis' law:
\begin{equation}\label{eq:n-power} 
P(p, n)=c_0 N_0 n + B n (c_0 / f + c_1)\frac{p \hat{N}_2 + N_1 n}{(1-\alpha)p + \alpha}.
\end{equation}

In the absence of parallelization $p=1$. Obviously, in this case the quantity $P(1,n)$ has order $n^2$ as $n\to\infty$. 

If the parallelization coefficient $p=p(n)$ is proportional to $n$, then $P(p(n),n) = O(n)$. \square

\section{Choosing Optimal Memory Type} \label{sect1_4}

Also, minimization of state memory by
recomputing intermediate data, called
rematerialization in compiler theory, and reduction of code memory by specializing the block for
a class of algorithms can be energetically beneficial.

\begin{table} [htbp] \centering 
\caption{Relative cell size, including overhead.} 
\label{table:mem-area} 
\begin{tabular}{|l|r|} 
\hline 
Memory Type&Size\\ 
\hline 
Dynamic eDRAM (1T) 1rw&1\\ 
Static 6T-SRAM 1rw&3\\ 
Register 8T-SRAM 1r1w&6\\ 
\hline 
\end{tabular} 
\end{table}

Another important optimization is choosing an energy-efficient memory type depending
on purpose. For small energy-efficient devices, only
memory types formed on the chip using the same
technological process can be used. Library memory components are organized in
a rectangular array of cells. Other types of memory organization, such as queues,
at large size can be efficiently implemented only based on
addressable memory. External memory is not energy-efficient due to power
losses on the external interface. Memory by type can be read-only,
rewritable, and random access. Read-only memory is suitable only for storing
initial bootloader and constant tables. Storing other data in ROM is associated
with high error cost during development. Today there is no built-in
random access non-volatile memory. Rewritable memory (flash memory) with
block reading and writing can be formed on the chip together with
the logical circuit in some manufacturing processes. Due to block reading
and writing, a buffer of
random access memory must be located between such memory and the computational block.

Random access memory on the chip can be dynamic, static, and
register (static high-speed memory). Register memory can
be addressable or be a set of separate registers. Also, memory can
have several read and write ports for simultaneous access in one clock cycle.
For example, the standard formula for register memory 1r1w - that is, one read port
and one write port. Static memory usually has the formula 1rw - that is, one
read and write port. In some technological processes, memory with
formula 2rw and others is available. Differences are due to the number of transistors in the cell and
overhead for cell switching and regeneration for dynamic memory.
Exact characteristics of various types of memory are trade secrets
of manufacturing firms. Energy consumption is correlated with area,
table~\ref{table:mem-area} gives an idea of the relative area
of various memory types per bit. Thus, the most
energy-efficient is built-in dynamic memory with formula 1rw, which
should be taken into account when developing algorithms.

\section{Choosing Optimal Width and Representation of Numerical Data}
\label{arithmetic_opt}

The main part of algorithms executed on the device performs processing of
numerical data. From the point of view of energy consumption, reduction of data
representation width is important, since this allows reducing energy consumption
of memory and computational components. Most algorithms in the field of digital
signal processing are described by high-level computational models,
using floating-point numbers.

The main influence on energy consumption is exerted by the data storage format in
memory. For storage in memory, fixed-point number format is used,
which can be considered as integers, and floating-point number format in
accordance with IEEE754 standard. Since shared
access to accelerator and programmable processor memory is preferable, the width
of an addressable memory word equals $8\times2^k$, that is, 8, 16, 32, 64 bits, etc.
One addressable memory word can store several data values, the width
of which is usually also determined as a power of 2, for example, a complex number or
quaternion. Data width and format are chosen so as to ensure
stable operation of algorithms in the presence of rounding errors. Width
of intermediate data influences area and, due to carry overflow chains
between digits, on critical path length and clock frequency of the circuit.

Floating-point format has several advantages. These
include: 
\begin{itemize} 
\item Simplicity of transferring an algorithm from model to
platform due to absence of the need for quantization. 
\item Greater
computational stability and predictability compared to integer
implementation. 
\item Greater dynamic range compared to an integer
of the same width, which allows reducing memory. 
\end{itemize}

Thus, use of floating-point numbers allows substantially
reducing development time and, in some cases, memory size. Comparison
of accuracy of various number representations is given in table~\ref{table:fp-vs-int-acc}
Areas of basic adder and multiplier blocks for floating-point and
fixed-point formats are presented in table~\ref{table:fp-vs-int} in
relative form. The circuits were synthesized by the author with geometric
norms 22 nm under a delay constraint of 1.5 ns. Absolute values are
trade secrets of the manufacturer.

\begin{table} [htbp] 
\centering 
\caption{Relative size and delay of basic computational blocks.} 
\label{table:fp-vs-int-acc}
\begin{tabular}{|l|r|r|r|} 
\hline Data Type&Exponent Range (2)&Precision (2)&Precision (10)\\
\hline 
Int16&0&15&~4.5\\ 
Int32&0&31&~9.3\\ 
FP16 (Half)&[-14..15]&11&~3.3\\
FP32 (Single)&[-126..127]&24&~7.2\\ 
FP64 (Double)&[-1022..1023]&53&~15.9\\ 
\hline 
\end{tabular} 
\end{table}

There are also a number of disadvantages of using floating-point numbers, such as:
\begin{itemize} 
\item Additional overhead for performing
computations. Comparison shows that single-precision
floating-point multipliers (32 bit) without support for denormalized numbers are comparable in size
to an integer multiplier of the same width. However, the size of a floating-point
adder increases by 10 times. This is due to the presence of high-speed
controlled shifters at the input and output of the circuit, whose complexity significantly
exceeds the complexity of the adder itself. It should be noted that controlled
shifters have to be inserted into the computational circuit to ensure its
reuse between different algorithms. That is, they are in any case
present in the circuit, but as separate blocks. To amortize overhead
for data shifting, circuits with group exponent are often used, where
the computation unit is not a separate addition or multiplication, but a group
of sequential operations. Shifting is performed only on input and output
data of the circuit, and intermediate data have integer representation. 
\item
Complication of hardware implementation verification due to ambiguity in definition
of floating-point operations between the computational model and implementation.
The requirement of compliance with IEEE754 standard, ensuring equivalence,
leads to an increase in circuit area by 2 times, which is not acceptable for
low-power devices. A similar problem is usually solved when converting
an algorithm from floating-point to fixed-point and remains outside hardware
verification. In the presence of floating-point hardware implementation, checking
is transferred to a later verification stage. Two approaches to solving
this problem are possible: development of tests accounting for allowable rounding error, and
testing the algorithm with an exact software model of hardware blocks
for floating-point computation. 
\end{itemize}

Despite the disadvantages, from the point of view of energy consumption, possibilities
of reuse and simplicity of algorithm implementation, use of single-precision
floating-point numbers for storing data in addressable memory
is justified.

\begin{table} [htbp] 
\centering 
\caption{Relative size and delay of basic computational blocks.} 
\label{table:fp-vs-int} 
\begin{tabular}{|l|r|}
\hline Block Type&Area\\ 
\hline 
ADDSUB 16&1\\ 
ADDSUB 32&2.8\\ 
ADDSUB FP16 w/o denorm&13.11\\ 
ADDSUB FP32 w/o denorm&29.93\\ 
ADDSUB FP32 denorm&43.64\\ 
ADDSUB FP64 denorm&202.8\\ 
\hline 
MUL 16&16.75\\ 
MUL 32&55.14\\ 
MUL FP16 w/o denorm&13.99\\ 
MUL FP32 w/o denorm&46.6\\ 
MUL FP32 denorm&92.2\\ 
MUL FP64 denorm&384.62\\ 
\hline 
\end{tabular} 
\end{table}

When computing intermediate data without saving to addressable memory or if
extended dynamic range of floating-point numbers is not required,
the best choice is use of fixed-point numbers. For
intermediate data, the width of an addressable memory word is not a limiting
factor, and area and energy consumption savings can be achieved
by reducing bits in representation. Reduction is achieved by reducing
the size of non-addressable register memory and arithmetic blocks.

Manual transformation of an algorithm from floating-point to fixed-point
numbers, often called quantization, is a labor-intensive task with high
probability of introducing hard-to-detect errors at boundary points when
overflow or zeroing of values occurs. The transformation can take up to 30\% of
total time for hardware implementation of the algorithm~\cite{Groetker96icspat}.

The problem arises of automatic selection of representation width for intermediate
data under constraints on computation accuracy based on high-level
algorithm specification in a high-level programming language.

The synthesizable subset of SystemC~\cite{SystemC_synth} includes two templates
for representing such numbers $sc\_fixed$ and $sc\_ufixed$ for signed and
unsigned data. Fixed-point data format is determined by word width
and integer part width, which sets the position of the binary point
relative to the most significant bit of the number and the method of handling rounding and overflow.
High-level synthesis tools, such as Calypto Catapult
C~\cite{CatapultC}, Cadence C-to-Silicon Compiler~\cite{C2S}, Forte
Cynthesizer~\cite{ForteCynth} support working with fixed-point
numbers, which reduces labor costs and probability of error when converting
computational models for their hardware implementation.

Integer part width can be determined based on analysis of value ranges
of variables. Finding optimal word length is a more complex task,
since it requires estimating the influence of variable representation accuracy on the final
computation result. Both tasks can be automatically solved by methods
of static analysis and automatic theorem
proving~\cite{conf/date/NayakHCB01, conf/dac/ShiB04, conf/pldi/StephensonBA00,
conf/fccm/BanerjeeBHNKU03}. Another approach consists in dynamic
profiling during program execution~\cite{journals/tsp/SungK95,
conf/date/MallikSBZ06, han2006automating, hainam:inria-00617718}. Static
methods usually give more conservative parameter estimates compared to
dynamic ones, however they usually ensure guaranteed satisfaction of
constraints and work in less time. Dynamic methods are usually
probabilistic, since to ensure guaranteed satisfaction of
constraints requires exhaustive enumeration of input parameters of the optimized
algorithm, which is often unacceptable due to computation time. In practice, both
methods work best for acyclic computation graphs.

Static removal of logical circuits without dependencies is built into software
logical synthesis tools such as Synopsys Design Compiler. It also
can be considered as conservative word width optimization.

\section{Data Rematerialization} Storing results of intermediate
computations for reuse or data tabulation are not always
optimal solutions for energy consumption. Two factors influence this.

\begin{enumerate} 
\item For storing data, random access
memory is used, which consumes energy in idle state. 
\item Energy for
accessing memory can be greater than energy for recomputation.
Such a situation is possible if a specialized
hardware block is used for computations. 
\end{enumerate}

In table~\ref{table:mem-vs-fp32} a comparison of area of a typical
register memory block and floating-point multiplier is given, which allows obtaining
an idea of their energy consumption.

\begin{table} [htbp] 
\centering 
\caption{Relative size of memory and computational blocks.} 
\label{table:mem-vs-fp32} 
\begin{tabular}{|l|r|} 
\hline
Element Type&Size\\ 
\hline MUL FP32 w/o denorm&1\\ 
Register 8T-SRAM 1r1w 256x32&1.5\\ 
\hline 
\end{tabular} 
\end{table}

For comparison with other memory types, one can refer to
table~\ref{table:mem-area}, for comparison with other computational blocks to
table~\ref{table:fp-vs-int}. Thus, data memory energy consumption
can exceed energy for their recomputation. In this case, recomputation
is more beneficial. This technique is called data
rematerialization.

\section{Influence of Prospective Technologies on Energy Efficiency}
\label{sect1_6} Semiconductor circuit manufacturing is a
rapidly developing field. Qualitative estimates of energy efficiency and architecture
choice are tied to the generation of technological processes and to specifics
of technological component libraries. When planning prospective developments,
one should take into account not only the current state, but also prospective technologies,
planned for introduction into production within the next several years. Such
technologies for low-power technological processes today are:

\begin{itemize} 
\item Near-threshold voltage circuits; 
\item Phase Change Memory. 
\item Magnetoresistive memory (MRAM). 
\end{itemize}

\subsection{Near-Threshold Voltage Circuits} \label{sect1_6_1} For
stable operation of digital circuits, supply voltage $V$ must be several times
greater than transistor threshold voltage $V_T$. Reducing supply voltage
leads to reduction of both active and static dissipated power of the circuit.
At the same time, by formula \ref{eq:delay} maximum clock frequency decreases
proportionally to $f_{0}\propto \frac{(V-V_T)^2}{V}$. However, one should remember that
the circuit must perform some fixed computational load. If
the circuit cannot perform this load, then it requires parallelization,
inversely proportional to clock frequency reduction, which leads to potentially
infinite increase in area and power. Let us estimate energy for performing one
operation at frequency $f_0$.
\begin{eqnarray*}
E_{f_0}\propto V^2\left(\alpha C + \beta \frac{I_{sub}}{(V-V_T)^2}\right), \\
\lim_{V \rightarrow V_T}E_{f_0}=\infty.
\end{eqnarray*}

Reducing supply voltage of logical circuits leads to an increase in error
probability, to prevent errors it is required to use end-to-end linear
error-correcting codes and/or multiple computation of results with
voting, which reduces the gain from voltage reduction. Moreover, such
a change in circuit design approach is not supported by development
tools, which potentially leads to an increase in development effort
for specialized circuits for a set of algorithms.

When using the "race-to-sleep" approach, energy consumption is mainly
due to state memory leakage currents.

We can conclude that for these reasons, for computational blocks,
use of near-threshold voltage circuits in the near future will not be
justified in low-power specialized circuits. This corresponds to
current technology positioning for general-purpose microprocessors and microcontrollers
and medical, including implantable equipment.
Since general-purpose microprocessors are produced in large batches,
development costs are amortized in the cost structure. Implantable
equipment is functionally simple and expensive, which also allows
using less technological development methods.
 
Another way of using the technology is development of memory with
reduced supply voltage in data storage mode. Use of such
memory does not present complexity from the point of view of development
tools, since it is a large universal library block and
is connected through a standard interface. Reducing memory supply voltage
during idle time will substantially reduce circuit power with the "race-to-sleep"
approach.

To reflect the influence of memory with reduced supply voltage in idle mode
in formula \ref{eq:power}, the term $c_0N_0$ should be replaced by the sum
$c_0N_0+c_0^*N_0^*$ where the second term accounts for static power
of memory. The formula takes the following form, which does not change its form.
$$
P=c_0 N_0+c_0^*N_0^* + B (c_0 / f + c_1)\frac{(r(p) \hat{N}_2 + N_1)}{l(p)}
$$

\subsection{Phase Change Memory and Magnetoresistive Memory} 
\label{sect1_6_2} Other prospective types of memory
are phase change memory and magnetoresistive memory. These
types of memory change cell resistance when switching state.

Just like flash memory, these memory types are rewritable, that is, they do not
consume energy when storing data. They can be formed on the chip
within the same technological process as logical circuits. Unlike
NAND flash memory, they have random access and a large number of rewrite
cycles, comparable to random access memory. Read energy 
is comparable to this parameter for random access memory, however write energy is several 
times greater than read energy.

These types of memory are well suited for storing tables and program code. To
reflect the influence of new memory types, changes in formula \ref{eq:power} are not
required. It is sufficient to account for the difference in read and write cost in the multiplier
$B$, accounting for algorithm complexity.

\section{Influence of Software Architecture on Energy Efficiency}
\label{sect1_7} Specialized semiconductor circuits for digital
data processing with developed functionality are usually built on the basis of
a parameterized programmable microprocessor or general-purpose
microcontroller, for example, belonging to families ARC, ARM, x86, Tensilica. This
is justified for several reasons:

\begin{itemize}
\item possibility of reusing computational resources
between parts of the algorithm to save chip area and improve
energy efficiency; 
\item reuse of data memory by various parts
of algorithms to reduce its size and reduce the need for copying
data, which leads to improved energy efficiency; 
\item possibility
of correcting errors in the algorithm after chip manufacture, which allows
separating hardware and software development processes to reduce overall
development time; 
\item possibility of reusing the block and architecture
of the chip in other types of devices by changing the algorithm to reduce
development time and production cost; 
\item presence of developed programming tools (compiler,
debugger, OS) to reduce development time. 
\end{itemize}

Inclusion of programmable blocks leads to substantial influence of program
memory on circuit area on the chip. This also leads to an increase in
active power consumed for reading code from memory. Costs for reading program
memory can make a substantial contribution to active power. Influence on
static power may be insignificant in the case when code is executed
from rewritable memory and code cache in random access memory is disconnected from power
during idle time.

For optimizing energy consumption, a reasonable approach is minimization
of program code size on the device. To reduce development costs
of software, programs for low-power devices are written in
high-level languages, such as C, C++, Ada, Fortran and, recently,
Java, C\#. At the same time, programs are executed in a runtime environment, including
basic libraries, type checking and error handling tools. Moreover,
presence of device drivers, such as sensors, Wi-Fi, Bluetooth,
ZigBee and network protocol modules, for example, TCP/IP, is required. All these software
modules increase memory costs and energy consumption of the device.

Another substantial problem is that all sufficiently complex
programs in the process of development and operation contain errors.
There are three different approaches to error detection.

\begin{itemize} 
\item static proof of program correctness based
on specifications; 
\item testing and debugging on software or FPGA
emulator; 
\item error detection and debugging during program execution on
a real device. 
\end{itemize}

The approach based on program correctness analysis is limited due to
combinatorial explosion of states. In practice, it can be applied either for
very simple programs, or for a limited class of errors, or is
heuristic and does not guarantee absence of errors. With complication
of semiconductor devices and increase in their computational performance,
growth of their functionality also occurs, which leads to explosive growth in volume
of software. Most methods of automatic correctness
checking verify the program for compliance with specification. But costs
for creating specification many times exceed costs for development
of program code. Moreover, the specification itself is not guaranteed to be error-free. More
details on static detection of one class of errors is written in
section~\ref{formal_verification}.

Testing and debugging on an emulator provides the possibility of error detection
in a controlled environment using test coverage. However, emulation has
many disadvantages that do not allow detecting a substantial
share of errors by this method:

\begin{itemize} 
\item low emulation speed, not corresponding to the speed
of real device operation, leads to impossibility of testing long
operation scenarios; 
\item equipment and its drivers are represented by abstract
models, which can substantially differ in behavior from real
equipment, which is related to labor intensity of model development; 
\item due to
inaccuracies of equipment models and low emulation speed, temporal
characteristics differ from real ones, which leads to impossibility of detecting
many synchronization errors; 
\item tests are synthetic and do not cover
real usage scenarios of the device; 
\item even 100\% test coverage
by program operators does not guarantee absence of errors, moreover, it
usually does not include equipment drivers. \end{itemize}

Thus, error detection on a running device remains one of
the main practical methods of testing and debugging. The runtime environment
of the program must provide error detection and possibility of their debugging. The situation
is complicated by the limited interface of the device with the external world. This leads
to the need for substantial redundancy in program code and, possibly, in
microprocessor functionality. To perform debugging functions, the runtime
environment must provide early detection and logging of errors with
possibility of determining program state at the moment of error with accuracy to
the operator that caused the error, and state of local variables at the moment of error.
Moreover, protection of error log from destruction as a result of
errors in the program is desirable. All these requirements are usually associated with functionality
of an operating system. Inclusion of an operating system on the device increases
memory costs and energy consumption of the device.

There are several types of operating systems for low-power devices and
runtime environments, partially or fully implementing these requirements:
 
\begin{itemize} 
\item operating systems with paged memory support;
\item operating systems without paged memory support; 
\item managed runtime environments. 
\end{itemize}

\subsection{Operating Systems with Paged Memory Support} To this class
belong all general-purpose operating systems with preemptive
multitasking, such as Linux, FreeBSD and many real-time systems,
such as QNX, VxWorks, WinCE.

Memory controller with paged memory support (Memory Management Unit, MMU)
provides dynamic page swapping and data loading from external
storage devices. In the case of low-power devices, this is rewritable memory.
For some processors, for example, ARM9 architecture, paged
memory management is related to data caching. Moreover, MMU provides memory protection and
separation of address spaces of different processes. Page swapping usually
is not used for real-time systems, to which most
low-power devices belong, due to increase in delays and deterioration
of system operation predictability.

Memory protection and presence of an operating system with preemptive multitasking in
most cases ensure survival of error logging system and
system debugger even at critical failures. This often allows detecting
the cause of the error and conducting debugging on the device. Presence of memory protection
mechanisms does not guarantee immediate error detection. An error inside one
process is often detected after a prolonged period of program operation
on corrupted data. This leads to high labor intensity of determining
the root cause of the error after its detection.

Typical memory size for such operating systems starts from several
megabytes.

Presence of paged memory mechanisms substantially slows down access to
the processor bus due to address translation and can lead to translation buffer
misses, which leads to large delays. These mechanisms are mainly
supported by processors with 32-bit architecture, which have large
area and peak energy consumption, which may be unacceptable for
low-power devices.

\subsection{Operating Systems without Paged Memory Support} These
operating systems can be full-featured multitasking operating
systems or single-process operating systems, connected to
a user program as a library. To this class belong uCLinux,
Free RTOS, eCos RTOS, Nucleus RTOS, Integrity RTOS, LynxOS, ThreadX RTOS and
others.

Presence of error protection is well recognized by developers of processors and
microcontrollers. Memory Protection Unit (MPU)
is embedded in processor architectures without paged memory support.
Careful use of these mechanisms in the operating system allows
obtaining close functionality for error detection on a small amount of
memory compared to paged memory. The main limitation is
the number of memory regions supported by MPU, for example, for ARM-R5, no more than 16
blocks of 8 sub-blocks of fixed size \cite{arm-r5-mpu}. Memory protection
has insignificant influence on processor energy consumption.

Typical memory size for such operating systems starts from several
hundred kilobytes.
 
Comparison of ARM family processors similar in performance by presence of MMU
and energy consumption is given in table~\ref{table:arm}.

\begin{table} [htbp] 
\centering 
\caption{Comparison of ARM processors differing in presence of MMU.} 
\label{table:arm} 
\begin{tabular}{|l|r|r|} 
\hline
Parameter&ARM-A5\cite{arm-a5}&ARM-R5\cite{arm-r5}\\
\hline 
Manufacturing Process&TSMC 40LP&TSMC 40LP\\ 
Frequency & 480 & 480\\ Bus& AXI & AXI\\
Supply Voltage (V)& 1.1 & 1.1\\ 
Performance (DMIPS/MHz)&1.57&1.66\\ 
Memory Controller Type&MMU&MPU\\ 
\hline
Energy Efficiency (DMIPS/mW)&13&24\\ 
\hline 
\end{tabular} 
\end{table}

\subsection{Managed Runtime Environments} A managed runtime environment
(Managed Runtime Environment, MRE) is a runtime environment for a high-level
language, providing potential software recovery from any runtime
errors in managed code of the program. This requirement does not
extend to hardware failures.

Such a definition is idealized, since errors related to
exhaustion of physical resources cannot always be recovered without
using the same type of resources. Implementations can provide approximation to
ideal behavior, guaranteeing recovery from primary errors, that is,
errors that did not arise in the error handler.

Errors in the implementation of the managed runtime environment and in external unmanaged
modules are not guaranteed for recovery. Not all errors can be
recovered according to program logic. Thus, in the program there can
be fatal errors. Fatal errors by definition lead to
premature program termination.

Presence of a managed runtime environment guarantees detection of the root cause
of an error, if it occurred in managed code, and, in most cases,
allows excluding data loss, even with a fatal error.

To managed runtime environments, execution environments of Java and
CLI(.NET) are traditionally referred. But according to the general definition, execution environments
of such languages as Python, JavaScript can be referred to the same class.

Possibility and advantages of systems programming with their use are considered in many
works~\cite{Java-Drivers, Singularity, Jikes, JNode} based on practical
development of drivers, virtual machines and operating systems, including in
works of the author~\cite{salishev2009mre, salishev2008mre,
Frampton:2009:DMH:1508293.1508305} based on practical development
of the Moxie virtual machine.

To main features of managed runtime environments,
according to~\cite{Cierniak03theopen}, belong: 

\begin{itemize} 
\item code verification at load time, guaranteeing strict compliance of high-level language
semantics and executable representation of the compiled program; 
\item strict typing, guaranteeing safe type casting and type checking at
runtime; 
\item automatic memory management based on garbage
collection, providing possibility of working with memory without using
unsafe pointers; 
\item exception handling. \end{itemize}

These properties provide a greater level of safety and stability
of software components to errors compared to binary compilation from
traditional languages for embedded software development, such
as C, C++, without the need for using hardware memory protection means and
multitasking operating system. Absence of these components leads to
reduction in energy consumption.

Implementation of a managed runtime environment on the device requires additional
resources, which leads to concern about inefficiency of such a solution compared
to traditional ones. Indeed, there are reasons in the architecture of managed
runtime environments, leading to some inefficiency:

\begin{itemize} 
\item Strict typing requires additional checks at
runtime, which in languages C, C++ remain on the programmer's conscience.
Modern compilers allow performing these checks once, moving them
out of loops, which makes their influence on program execution speed unnoticeable.
Presence of strict typing and absence of unsafe pointers in many
cases allows compilers to perform more aggressive operations, for example,
related to loop vectorization. 
\item For efficient garbage collection, heap
size must exceed the size of dynamically allocated data in memory by 2 times
\cite{oracle-GC}. For embedded applications related to digital signal
processing, this is not a serious problem, since most data
is allocated in memory statically. 
\item Garbage collection can lead to
delays. Use of incremental algorithms, such as
Metronome~\cite{Metronome-GC} or incremental concurrent marking and
sweep, allows achieving controlled deterministic delay, not
exceeding units of milliseconds. Delay-critical parts of code in any
case traditionally do not use dynamic memory allocation. 
\item
Internal data structures for providing typing, reflection and exception
handling occupy additional memory and computational resources. Reflection and
unlimited exception handling is an integral part of languages Java
and C\#. Since these capabilities are not used for real-time systems,
they may not be supported in specialized implementations. 
\end{itemize}
 
Thus, use of managed runtime environments for systems
programming does not present insurmountable obstacles. In exceptional
cases, to increase efficiency, a managed program can call
external unmanaged libraries through an external function interface. Even in this
case, the managed runtime environment provides greater protection from errors,
since errors are often localized in newly written integration code, and
not in libraries.

For use in embedded devices, technologies
Sunspot~\cite{sunspot} and Micro.NET~\cite{Micro.NET} are now available. The latter technology
is available under open license APL~\cite{APL}, which allows porting it
to any processor architectures. It can be used on devices with
minimum random access memory 300KB, which is comparable to typical requirements
of operating systems without paged memory support.

\section{Formal Verification as a Means of Improving Energy Efficiency}
\label{formal_verification} Presence of errors in programs, including in
programs for hardware synthesis, leads to the need for mechanisms for their
handling at software and hardware levels. Guarantees of absence of errors
provide possibility of removing these mechanisms and saving energy. Most
difficult to detect are synchronization errors and related errors,
associated with queue overflow during asynchronous interaction. Possibility
of such errors leads to the need for transferring synchronization mechanisms
to software components of the device, where they can be corrected after
hardware production or introducing into hardware mechanisms
for controlling queue length and stopping computations to unload the queue. This
leads to substantial overhead and increase in energy consumption.

It should be noted that errors can also arise during device
manufacture, so project verification does not exclude the need for mechanisms
for detecting manufacturing errors, such as Built-In Self-Test
circuits~\cite{BLTJ:BLTJ576}.

To guarantee absence of synchronization errors, in some cases it is possible
to bring the program to a cyclo-static form, constructing for its execution
a fixed schedule. A similar approach is used for automatic
pipelining of logical circuits by means of automatic logical
synthesis~\cite{Retiming}. For development of more complex devices, a
reactive programming model and a family of languages for synchronous
data flow programming such as Lustre~\cite{Lustre},
Estrel~\cite{Estrel}, Streamit~\cite{Streamit} have been proposed. This approach works for
relatively simple programs and is poorly applicable to programs with complex
control logic, not reducible efficiently to data flow. Moreover,
rewriting of the program in a different model compared to ordinary
imperative languages C, C++ is required. This complicates splitting the task into hardware and
software components. These reasons led to the approach not finding
wide industrial application.

Another, more universal approach to synchronization based on Kahn
networks~\cite{KPN} and asynchronous message exchange is often used in
industry, for example, in circuit synthesis in CatapultC. In the general case, the approach requires
infinite memory for guarantees of absence of synchronization errors, which is not
realizable in hardware. In some cases, a static
schedule can be constructed, but memory size can be very large. Moreover, the Kahn networks
model substantially limits use of synchronous event mechanisms, such
as interrupts.

Thus, tools for formal checking of programs for presence
of synchronization errors within industrially used models of hardware
development and programming, such as C++ and SystemC, are in demand. Use of such
tools may allow substantially reducing energy consumption
of low-power specialized semiconductor devices.

The task of testing and debugging synchronization errors is one of the main
problems in software development. For sequential programs,
for which synchronization is not required, the necessary quality level can be
achieved by means of full code coverage by unit tests. For parallel
programs this is not so, since small changes in the execution environment
unpredictably influence execution schedule and lead to reproduction of previously
non-reproducible errors~\cite{LEE}.

For software-hardware systems, the problem acquires even greater importance.
Such systems have greater complexity compared to purely hardware
systems and greater number of connections between software and hardware
components compared to systems based on general-purpose processors.
To save resources, software can operate without an operating
system and with limited debugging and error logging capabilities. In
the case when a synchronization error affects hardware components, its
correction requires manufacturing a batch of chips of a new version. Use
of high-level synthesis tools reduces development difficulties
of software-hardware systems, which allows developing systems with more
complex hardware parts in short time. This, in turn, increases
the risk of synchronization errors in hardware components.

Usually three types of synchronization errors are distinguished: 
\begin{enumerate} 
\item Deadlock,
when the system or its part comes to a state of eternal waiting for an impossible
event without using computational resources; 
\item Livelock, when the system
or its part is in a state of eternal loop without performing useful work;
\item Concurrent data modification, when several system components as a
result of violation of data write order violate their integrity.
\end{enumerate}

In the general case, with infinite memory, the task of detecting synchronization
errors is algorithmically undecidable~\cite{Netzer}. On finite
memory, the task has exponential complexity in the worst case.

There are 4 main approaches to automatic program correctness checking.
\begin{enumerate} \item Approaches based on automatic theorem proving;
\item Model checking; \item Approaches based on static analysis; \item
Approaches based on dynamic checking (instrumentation and execution).
\end{enumerate}

These approaches have different balance of accuracy, completeness and computational
complexity. Approaches based on model checking and correctness proving
in the general case are absolutely accurate and complete, but computational complexity limits
their application to large projects or requires manual extraction of a reduced
model or manual imposition of constraints and contracts, which leads to
additional labor costs and risk of error.

The dynamic approach has in the general case the least computational
complexity at the cost of reducing completeness and accuracy, since it completely depends on
completeness of external test coverage.

With the goal of automatic detection of synchronization errors, with participation of the author,
the system ``Aegis for SystemC'' was developed based on static analysis and
heuristic patterns of correct SystemC synchronization models. This approach
allows, by applying heuristics for merging system states, to achieve
practically realizable complexity even for large systems. Complexity reduction
is achieved at the cost of reducing analysis accuracy, that is, false
positives can occur.

In the first version, an approach usually applied in software
development was implemented, based on analysis of order of calls of synchronization primitives
\textit{wait, notify}\cite{6089227}. This approach allows detecting
only deadlocks and is not applicable to synchronous-asynchronous systems, including
clocked hardware models and asynchronous software controlled
by events due to exponential growth of complexity.

Subsequently, a universal approach was developed based on analysis of reachability
of operators in a model of a correct program on
SystemC\cite{Glukhikh:2013:SAA:2547165.2547276}. This approach allows
detecting all 3 types of synchronization errors and is applicable to synchronous and
asynchronous models and their combination.

To reduce computational complexity, problem conditions were limited.
A synchronous-asynchronous model with a set of stochastic tests was considered.
The goal was detection of only those errors that can be reproduced
by the test system. Thus, the system ``Aegis for SystemC'' can be
considered as an extension of system test coverage. A fundamental difference
of the system from ordinary testing is guaranteed detection of errors,
even if the probability of their reproduction by a test is negligibly small, but different
from zero.

Conditions of absence of deadlocks and concurrent modification are automatically extracted
from the model. Since the concept of useful work depends on the algorithm, for
detecting livelocks, user annotations in source code are used.
User annotations have the form of conditional expressions in C++ language, which
substantially simplifies their application by developers compared to contract
languages based on temporal logics.

A problem of the approach based on static analysis is a large number
of false positives~\cite{Ball}. To reduce the influence of this problem, the system
supports user-defined contracts for marking
unrealizable states in source code, thus, the system will not
detect false positives repeatedly.

The tool ``Aegis for SystemC'' supports a large
subset of C++, SystemC and STL. In the testing process, it was applied to 21
models of software-hardware systems with more than 30K lines of code (not counting
comments) and demonstrated possibility of detecting real synchronization
errors.

The applied approach uses the technique of abstract interpretation of source
code~\cite{C96} for determining the set of possible states of the model. This
technique is similar to code execution, but variable values are described not as
scalar values, but as abstract value domains, which allows reducing
analysis complexity by merging execution paths and excluding
exponential growth of state space.

\clearpage			
\chapter{Hardware Acceleration of Elementary Function Computation Using Specialized Computational Blocks}
\label{chapt2}

Computation of elementary functions from data arrays is an important component
of many digital signal processing algorithms. 
Computation can be implemented in software on a processor or  
as a specialized logic circuit. 
Algorithmically, expansions in
series on the entire interval, piecewise-polynomial interpolation, splines can be used. 
Software implementation
is not energy-efficient for large volumes of computations, since
overhead for instruction fetch and decode and memory access
to tables is comparable to computation costs. In hardware implementation, often
a CORDIC scheme is used with computation speed of one cycle per bit of precision.
The CORDIC scheme is highly efficient but very slow. With large volumes
of computations, low speed can take the circuit beyond the bounds of the
linear power scaling region with frequency. In addition, other components in
idle time will consume energy due to leakage currents, 
since power disconnection is possible only for large blocks as a whole. 
In the case of large volumes
of computations that cannot be tabulated, other methods are applied. For
precisions less than 18 bits, piecewise-linear approximation algorithms are usually applied, 
since they can be implemented without using multipliers, only with
the help of addition and shift.
For larger precisions, more efficient are hardware-implemented algorithms 
of piecewise-polynomial approximation, providing computation of one value with 
startup delay of
one cycle and arbitrary precision. Such computational components as
multipliers and adders can also be reused in other algorithms,
for example, FFT. 

When developing a specialized circuit, precisions of tables and intermediate data
are not limited by machine word width and can be arbitrary. For such circuits, 
important from the point of view of energy efficiency is minimization of table size and
computation precision while maintaining the required result precision. 

The problem of piecewise-polynomial approximation is generally considered solved for
floating-point computations on programmable devices.
Using the Remez method, one can find an interpolation polynomial of best
uniform approximation and its interpolation nodes. An estimate of approximation error
by the residual at interpolation nodes, which arises due to rounding
of polynomial coefficients, is also known.

In the practical formulation of the problem when developing a specialized
device, it is required to find minimal table widths under a given constraint 
on approximation precision. The Remez method does not allow solving
this problem. Besides rounding, another way to reduce table size is
reusing coefficients between segments. For approximation
of functions, smooth splines are often used. The approximation problem described above
does not impose smoothness requirements, however, additional constraints
allow reducing table size. Smoothness requirements are excessively
strong and can lead to an increase in the number of approximation segments.
Therefore, we will consider an almost smooth quasyspline with minimal residuals.

To find piecewise-polynomial approximation with minimal
tables and approximation by an almost smooth quasyspline with minimal residuals, we reduce
the problem to a mixed integer programming problem. 

Using the proposed methods leads to a substantial reduction in table
size. Prototyping shows the practical influence of this reduction on
area and energy consumption of the computational block.

\section{Problem Statement for Approximation with Given Precision}\label{sect3_1}

Define precision requirements when computing at exactly representable points
$$
\|f-g\|=\sup_{x\in[a,b]}|(f-g)(x)|<\alpha=2^{-k}.
$$

Here $g$ is the approximating function, $k$ is the number of bits in the fractional part
of fixed-point representation.

\begin{definition}
We call a rounded (quantized) computation method conservative when
the following inequality is satisfied:
$$
\|f-g\| \leq \|f-p\| + \|p-g\| = \varepsilon_m + \varepsilon_q <\alpha,
$$
where $p$ is the best approximation with hardware-representable coefficients, 
$g$ is hardware-computable values of $p$ with rounding error, $\varepsilon_m$ is
the error of approximation and coefficient rounding, $\varepsilon_q$ is the error
of computation rounding. 
\end{definition}

For computation rounding error, the following lower bound holds:
$$
\varepsilon_q \geq \frac{\alpha}{2}\left(1+2^{-l_q}\right),
$$
where $l_q$ is the number of additional guard bits in the computational block.
Computation rounding error consists of rounding error when computing with 
additional bits and final rounding to $k$ bits. 

One can choose computation rounding parameters such that the
inequality
$$
\varepsilon_q \leq \frac{\alpha}{2}\left(1+2^{-l_m}\right),
$$
holds, where $l_m$ is the number of additional bits of table precision. Thus,
one can establish the following a priori requirement on the precision of the approximation method:
$$
\varepsilon_m \leq \frac{\alpha}{2}\left(1-2^{-l_m}\right).
$$

The requirement of conservatism can be abandoned, since points with maximum
approximation error and computation rounding error usually do not coincide. Such
optimization allows additionally reducing the area
of computational components of the block by 10\% or more. But this requires a complete
enumeration of all representable points, which at precisions greater than 25 bits leads to
large computation time. For large precisions, the conservatism requirement allows
solving each problem separately by analytical methods, which reduces
search space and makes them practically solvable. 

\section{Problem of Reducing Table Size}

Let function $f$ be given on the interval $[a, b]$. Let an increasing set of points $(x_i)_{i=0}^{k}$ partition this interval 
into $k$ equal segments. Denote the restriction of function $f$ as $f_i = f|_{[x_i, x_{i+1}]}$ for $0\le i\le k-1$. 

\begin{definition}
Let $\varepsilon > 0$. We call a set of polynomials $\{p_i\}_{i=0}^{k-1}$ a quasyspline of degree $N$ approximating $f$ with precision $\varepsilon$, if in the uniform metric
$$
\|f_i-p_i\|_{[x_i, x_{i+1}]}\leq \varepsilon, \qquad \deg p_i \le N, \qquad 0\le i\le k-1.
$$
\end{definition}

The problem of computing elementary functions reduces to selecting approximating quasysplines of given precision and minimal complexity, which in this case is table length. If the required approximation does not exist, then the domain of $f$ is partitioned into smaller segments, and the same problem is considered for each of them.

Consider the smoothness defects of the quasyspline
$$d_{i,\nu} = p^{(\nu)}_{i}(x_i)-p^{(\nu)}_{i-1}(x_i), \qquad 0 < i < k.$$

If for any $i, \nu < N$, $d_{i,\nu} = 0$, then the quasyspline $\{p_i\}$ is a smooth spline 
of minimal defect. From the existence of a quasyspline for a given set $(f, N, k, \varepsilon)$, in general it does not follow
the existence of a spline of minimal defect for the same set $(f, N, k, \varepsilon)$.

The quasyspline is completely determined by coefficients $(p_s, \{d_{i,\nu}\})$. 
Storing $p_s$ requires a table on average $k$ times smaller than storing all $\{p_i\}$.
Under the assumption of smoothness of $f$, one can also expect that all $d_{i,\nu}$ are small. 
If $k=2^m$, then hardware implementation of computing the quasyspline from the set of coefficients 
$(p_s, \{d_{i,\nu}\})$ does not significantly complicate compared to $\{p_i\}$,
since it requires only additional integer adders and shifts, 
which are substantially simpler than the already used multipliers.
Thus, storing coefficients $(p_s, \{d_{i,\nu}\})$ instead of $\{p_i\}$ leads to 
substantial reduction in complexity of hardware implementation of approximation.

Suppose that a table in hardware
is represented as a logical function in disjunctive normal form (DNF). 
Then for table size one can take the number of nonzero data bits in 
the table. The number of nonzero bits for defect coefficients at intermediate nodes 
of the quasyspline $L$ does not exceed
$$
L \leq \sum_{i=0}^{2^{k_0}-1} \sum_{\nu=0}^{N-1} {\max(0, 2+\lceil l_{m}+\log_2{|d_{i,\nu}|} \rceil)},
$$
where $l_{m}$ is the minimal allowable length of the fractional part of a coefficient of polynomial approximation in one segment.
An additional unit is required to store the sign of $d_{i,\nu}$.

Set the problem of minimizing tables for $d_{i,\nu}$. The nonlinear functional $L$ is inconvenient for minimization. 
Consider a simpler linear problem
$$\arg \min_{\{d_{i,\nu}\}} \sum_{\nu=0}^{N}\sum_{i=1}^{k-1}\alpha_{i,\nu}|d_{i,\nu}|.$$
Parameters $\alpha_{i,\nu}$ are chosen such that to zero the largest 
number of coefficients.

\section{Estimate of Approximation Precision on One Segment}\label{sect3_2}
To estimate the error of the approximation method $\varepsilon_m$, we
will need information from interpolation theory, see,
for example,~(\cite{computational1995}, p. 122-141). It is presented in 
appendix~\ref{Appendix:poly}.

Let us use these results to estimate the precision of approximation
by a polynomial on one segment in the presence of additional constraints.

\subsection{Approximation Error of Interpolation
Polynomial with Inequality-Type Constraints}\label{sect3_1_4}

Introduction of additional constraints violates the formulation of the problem of finding
polynomials of best uniform approximation. Let us pose an analogous in meaning
problem that preserves correctness in the presence of constraints. Replace the requirement
of uniform approximation on a segment with the requirement of uniform approximation on
a grid of points. The uniform approximation problem is a limiting case of the new
problem as the grid step tends to 0.

Let us find an estimate of polynomial approximation error under a constraint on
deviation of the polynomial from the approximated function on a grid with sufficiently fine
step.

Introduce several definitions
\begin{definition}
Denote $W^n(M, [a,b])$ the class of all functions on segment $[a,b]$, whose 
derivative of order $n$ is bounded by a given number:
$$
|f^{(n)}(x)| \leq M.
$$
\end{definition}

\begin{definition}\label{definition:lebesgue}
Let $(x_{1},\ldots, x_{n})$ be a set of interpolation nodes on interval $[a,b]$. 
We call the Lebesgue numbers for this set 
$$
\lambda_{n,\nu}=\sup_{x\in[-1,1]}\sum_{k=1}^{n} |l_k^{(\nu)}(x) |, \qquad
0\leq\nu<r,\quad r \geq 1,
$$
where $l_k(x)$ are the fundamental polynomials from definition~\ref{definition:fundamental} in Appendix B.
\end{definition}

Let us formulate a theorem on the estimate of approximation precision by an interpolation polynomial.

\begin{theorem}\label{theorem:approx_main}
Let $f\in W^r(M;I)$, $r \geq 2$ and a system of nodes $x=\{x_0,\ldots,x_n\}, x_i \in
[a, b]$ is given, containing interpolation nodes $\hat{x}=\{\hat{x}_1,\ldots,\hat{x}_r\}$,
$\hat{x} \subset x$ and satisfying the condition
$$
x_0=a, x_n=b, x_i<x_{i+1}, x_{i+1} - x_{i} < \delta, \qquad 0 \leq i < n.
$$

Consider an interpolation polynomial $p$ constructed from interpolation data $(\hat{x}, \xi)$,
such that $\sup |\xi_i-f(x_i)|=\varepsilon_0$.

Then the inequality
$$
\varepsilon_0\leq\|f-p\| 
\leq \varepsilon_0  + \delta ^ 2 
\left(
{\frac{(b-a)^{r-2}}{8(r-2)!}}M + \frac{\lambda_{r, 2}(I_0)}{2(b-a)^{2}} \varepsilon_0
\right). 
$$
\end{theorem}

{\sf Proof.}
Denote $\varepsilon=\|f-p\|=|f(x_*)-p(x_*)|$. Point $x_*$ exists, 
since a continuous function on a compact set attains its extremum.

Point $x_*$ may coincide with node $x_k$ for some $k$, then $\varepsilon=\varepsilon_0$.

Otherwise, $x_*$ will be an extremal point of a smooth function, since the boundaries of the segment are nodes and  
$$
f'(x_*)-p'(x_*) = 0.
$$

Choose the node $x_k$ closest to $x_*$,
$$
\sigma=|x_*-x_k|\leq {\frac{\delta}{2}}.
$$ 

Consider the first term and remainder of Taylor series expansion
$$
f(x_k)-p(x_k)=f(x_*)-p(x_*) + {\frac{\sigma^2}{2}} (f''(\psi)-p''(\psi)).
$$

Using the triangle inequality, write the following estimate: 
$$
|f(x_*)-p(x_*)| \leq |f(x_k)-p(x_k)| + {\frac{\delta ^ 2}{8}}
|f''(\psi)-p''(\psi)|.
$$

Replacing pointwise estimates with norms, we obtain
$$
\varepsilon \leq \varepsilon_0 + {\frac{\delta ^ 2}{8}} \|f''-p''\|.
$$

Using the estimate of approximation precision for the second derivative from theorem~\ref{theorem:approx_acc}, rewrite the inequality as 
$$
\varepsilon \leq \varepsilon_0 + \delta ^ 2  
\left(
{\frac{(b-a) ^{r-2}}{8(r-2)!}}M + \frac{\lambda_{r, 2}(I_0)}{2(b-a)^{2}} \varepsilon_0
\right). 
$$
\square

The theorem in this form allows application without change of variables and transition to
the interval $[-1,1]$. For practical application, it is necessary to compute in advance
$\lambda_{r, 2}(I_0)$ for a known system of nodes on the interval $[-1,1]$.

\subsection{Error of Quadratic and Cubic Approximation by Interpolation
Polynomial}\label{sect3_1_5}

From a practical point of view, we are interested in quadratic and cubic interpolation
with symmetric arrangement of nodes on the interval $[-1,1]$.

For quadratic interpolation, the only such system is $(-1,0,1)$. 
The system $(-1,0,1)$ coincides with the system of extended Chebyshev nodes for $k=3$.

\begin{lemma}
For the system of nodes $(-1, 0, 1)$, $\lambda_{3,2} = 4$.
\end{lemma}

{\sf Proof.}
In accordance with definition~\ref{definition:lebesgue},
\begin{eqnarray*}
\lambda_{3,2} & = & \sup_{x\in[-1,1]}\sum_{k=1}^{3} |l''_k(x)| \\
\omega_{I_0}(x) & = & (x-1)x(x+1) = x^3-x, \qquad \omega'_{I_0}(x)=3x^2-1 \\
l(x) & = & \left({\frac{x(x-1)}{2}}, -(x-1)(x+1), {\frac{x(x+1)}{2}}\right)\\
l''(x) & = & (1, -2, 1), \qquad\lambda_{3,2}=\sum_{k=1}^{3} |l''_k(x)|=4.
\end{eqnarray*}\square

These nodes are contained in a grid of nodes with step $\delta=2^{-k}$.

For cubic interpolation, symmetric systems on $[-1,1]$ are described as
$(-1,-d, d,1)$ with parameter $d \in (0,1)$. For $d=\sqrt 2 - 1$, the system coincides
with the system of extended Chebyshev nodes for $k=4$.

\begin{lemma}
For $I=[-1, 1]$ and system of nodes $(-1, -d, d, 1)$, $\min_{d\in
(0,1)}{\lambda_{4,2}} = 24$, and is attained at $d=0.5$.
\end{lemma}

{\sf Proof.}
By definition~\ref{definition:lebesgue}
\begin{eqnarray*}
\lambda_{4,2} & = & \sup_{x\in[-1,1]}\sum_{k=1}^{4} |l''_k(x)|, \\
\omega_{I_0}(x) & = & -d^2 x^2+d^2+x^4-x^2,  \qquad 
\omega'_{I_0}(x)=-2x-2d^2 x+4x^3, \\
l(x) & = & \Bigg({\frac{(x+d)(x-d)(x-1)}{2d^2-2}}, {\frac{(x+1)(x-d)(x-1)}{2d-2d^3}}, \\
&&
{\frac{(x+1)(x+d)(x-1)}{2d^3-2d}}, {\frac{(x+d)(x-d)(x+1)}{2-2d^2}} \Bigg), \\
&=&
\Bigg({\frac{-d^2 x+d^2+x^3-x^2}{2 d^2-2}}, {\frac{d x^2-d-x^3+x}{2 d^3-2 d}}, \\
&&
{\frac{d x^2-d+x^3-x}{2 d^3-2 d}}, {\frac{d^2 x+d^2-x^3-x^2}{2 d^2-2}}\Bigg)\\
l''(x) & = &\left({\frac{6x-2}{2 d^2-2}}, {\frac{-6x+2d}{2 d^3-2 d}}, 
{\frac{6x+2d}{2d^3-2 d}}, {\frac{-6x-2}{2 d^2-2}}\right), \\
\lambda_{4,2} &=&{\frac{1}{1-d^2}} \sup_{x\in[-1,1]} \left(
|3x-1|+|{\frac{3}{d}}x-1|+|{\frac{3}{d}}x+1|+|3x+1|
\right).
\end{eqnarray*}

The function is piecewise-linear and, consequently, attains
extremes at the endpoints of the interval and at break points $x \in \{1, -1, 1/3,-1/3,
d/3,-d/3\}$.
From symmetry with respect to 0, we can consider only points 
$x \in {1, 1/3, d/3}$
$$
\lambda_{4,2}={\frac{1}{1-d^2}} \max \{
6+{\frac{6}{d}}, 2+{\frac{2}{d}}, 4
\} = {\frac{6}{d-d^2}}.
$$

The minimum is attained at an extremal point 
$$
{\frac{\partial \lambda_{4,2}}{\partial d}} = 
{\frac{6 (-1 + 2 d)}{(-1 + d)^2 d^2}} = 0.
$$

$\min_{d\in
(0,1)}{\lambda_{4,2}} = 24$, and is attained at $d=0.5$
\square

\vspace{3mm}
An additional advantage is
that these nodes are always contained in a uniform grid with step $\delta=2^{-k}$.
Such a system of nodes is not an extended Chebyshev system.

Using an exact error estimate, we can find the allowable
polynomial approximation of degree $N$ by solving a linear
programming problem. At the same time, a constraint of
representability using fixed-point numbers is imposed on polynomial coefficients.

\section{Calculation of Tables Using Integer Linear Programming}

Suppose the following method of calculating approximations of a given function $f$ on a given domain interval $\Delta$ is fixed. 
Interval $\Delta$ is partitioned into segments, and within each segment the function is approximated by a polynomial. 
Memory stores parameters necessary for calculating values of polynomials on all segments.  It is required to minimize the size of this memory. 

In accordance with the statement of theorem \ref{theorem:approx_main}, uniform precision 
of approximation by an interpolation polynomial is guaranteed on the entire domain segment 
of the function if approximation precision on a given grid of samples $(x_i)$ is ensured. 
Approximation on a grid of samples is determined by a finite set of linear inequalities, 
and linear programming problems can be associated with it. 
To obtain rounded values of coefficients, we will apply the method of integer linear programming. 
Necessary information is presented in appendix~\ref{Appendix:lp}.

\subsection{Integer Approximation on One Segment}\label{sect3_4}

In this section, notation for one of the integer linear programming problems, which will be needed later for main formulations, is presented. 

Suppose it is required to approximate function $f$ by a polynomial $p(x)$ of degree $N$ on a given grid of samples: 
$$
\left\{ \begin{array}{l}
		\varepsilon=\max_{i \in [0..P]}|f(x_i)-p(x_i)|, x_0=0, x_i > x_{i-1}, x_P=1,\\
        {\varepsilon}^* = \min_p{\varepsilon}, p=\sum_{k=0}^{N}p_kx^k.
\end{array} \right.
$$

The problem of optimal approximation on a grid of samples can be formulated in canonical form as follows:
$$
\left\{ \begin{array}{l}
		\left(\begin{array}{rr} V&-1\\-V&-1 \end{array}
		\right)\hat{x}\leq\left(\begin{array}{r} f\\-f \end{array} \right),\\
        {\varepsilon}^* = \min_{\hat{x}}{l\hat{x}}.
\end{array} \right.
$$

Here $V=\{x_i^k\}_{i=0,k=0}^{P,N}$ is the Vandermonde matrix,
$f=(f(x_0),\ldots,f(x_P))'$ is the vector of values at nodes, 
$\hat{x}=(p_0,\ldots p_N,\varepsilon)'$ is the vector of variables, $l\hat{x}$ is
the objective function, where $l=(0,\ldots,0,1)$.

To find rounded coefficients of the polynomial, the method of mixed
integer linear programming is used. Additional constraints
on variables are introduced: 
$\hat{x}\in (Q_{l_m})^{N+1}\times R_+$, where $Q_{l_m}$ is the set of binary fractions with $l_{m}$ digits after the decimal point.
Using integer programming provides substantially better precision compared
to rounding coefficients to the nearest representable point. This
allows saving 1-2 bits on each coefficient.

Unlike the Remez algorithm, the method based on linear
programming allows not only approximating optimal in
precision interpolation polynomials, but also solving the problem under constraints
on sufficient approximation precision and on values of polynomials at nodes. This
is achieved at the cost of additional computational expenses for solving the problem on
a large set of nodes.

\subsection{Case of Quasyspline}\label{sect3_5}

A function composed of approximating polynomials on adjacent segments forms a quasyspline. 
The method of reducing table size studied in this section is based on grouping several adjacent interpolation segments and then jointly encoding coefficients of polynomials on these segments. 

The idea of reducing table size of polynomial interpolation is related to the fact that approximating polynomials on adjacent segments can have redundant information that can be taken into account. 
Thus, each section of the data table contains parameters of some quasyspline.

At the end of this section, it will be shown that using the solution of an auxiliary linear programming problem and 
analytical estimates by theorem \ref{theorem:approx_main}, table lengths can be substantially reduced in this way for a number of elementary functions.

For approximation
of functions, smooth splines of minimal defect are often used. 
The approximation problem does not impose
requirements on smoothness, however, additional constraints allow reducing
table size by reusing coefficients between segments.

Consider an approximating quasyspline of degree $N-1$ with $2^{k_0}$ links. We will consider two- and four-link
quasysplines. Powers of two are used, since in this
case, the higher bits of the argument are used as an address in the coefficient table, and
lower bits as the interpolation argument, without the need for modulo computations.
A two-link quasyspline allows reusing all coefficients, 
without adding additional computations. Quasysplines with more than 2 segments add
additional adders to the computational block, thus,
energy efficiency of data reuse decreases due to growth
in the number of computations.

The smoothness requirement is excessively strict, and its fulfillment can
lead to an increase in the number of segments of piecewise-polynomial interpolation
and doubling of table size, which is not compensated by their subsequent reduction. To
preserve the number of segments, residuals are allowed at quasyspline nodes, which in
practice are equal to $0$ on most segments. Since
constraints are not strict, such optimization does not lead to a decrease
in grid step compared to the problem without constraints. 

Since part of the coefficients is zeroed, the proposed method can be considered a method 
of non-uniform tabulation, since on segments where the function has a smaller modulus 
of the derivative, less data is used for its tabulation.
For one quasyspline, it is necessary to store $N$ polynomial coefficients and $N(2^{k_0}-1)$
additional values of residuals at internal nodes of the quasyspline. The appearance of residuals
leads to the need for additional tables and additional adders for
introducing corrections to coefficients.

Such an approximation algorithm is a generalization of a smooth two-link spline, 
considered by Strollo et al.~\cite{10.1109/TC.2010.127}. Using more 
than two links of the quasyspline and relaxing smoothness requirements allows substantially 
reducing table size. For example, transitioning to a four-link quasyspline compared 
to a two-link one gives table reduction of more than 30\%.

Let us formulate the optimization problem of finding residuals for an almost smooth
quasyspline.

\subsection{Optimization Problem}

Consider the problem of approximating function $f$ by a multi-link quasyspline of given degree with a given number of links and with given precision. 
In this section, a linear programming problem is formulated, aimed both at achieving approximation precision and at reducing table length.

For brevity of notation, we will assume that lengths of all segments of polynomial interpolation are equal to $1$. Thus, boundaries of segments of polynomial interpolation are integers:
$$
x_i = i, \qquad 0\le i< 2^{k_0},
$$
where $2^{k_0}$ is the number of links of an almost smooth quasyspline. The quasyspline will be denoted $\{p_i(x)\}$. 
It is determined by coefficients of its polynomials, where $p_i(x)$ is the polynomial of the $i$-th link. 

Within each link, choose a grid of samples $\bar{x}_j=2^{-k_2}j$ for $0\le j<2^{-k_2}$ with step $\delta = 2^{-k_2}$. 

Additional constraints on precision of representation of polynomials
$p_{ij}\in Q_{l_m}$ are introduced, where $Q_{l_m}$ is the set of binary fractions with fractional part of $l_{m}$ digits. 

\begin{theorem} \label{theorem:approx_optim_cont}
Let $N>2$, $f\in W^N(M, [0, 2^{k_0}])$ and $\varepsilon_m>0$ be the given approximation precision. Define
$$
\bar{\varepsilon}=
\left(
\varepsilon_m  - {\frac{\delta ^ 2 M}{8(N-2)!}} 
\right)
\left(
1 + \frac{\delta ^ 2 \lambda_{N, 2}(I_0)}{2}
\right) ^{-1}.
$$

Suppose there exists a solution to the following mixed integer linear programming problem with respect to 
coefficients of an almost smooth quasyspline $\{p_i(x)\}$ on segment $[0, 2^{k_0}]$ under constraint on representation $p_{ij}\in Q_{l_m}$:
$$
\left\{ \begin{array}{l}
\varepsilon=
\sum_{\nu=0}^{N}\sum_{i=1}^{2^{k_0}-1}\alpha_{i,\nu}|d_{i,\nu}|\\
\varepsilon_i \leq \bar{\varepsilon}\\
\varepsilon^*=\min_{p_i, i \in [0..2^{k_0}-1]}\varepsilon 
\end{array} \right.
$$

where
\begin{eqnarray*}
\varepsilon_i & = & \sup_{0\le j < 2^{k_2}}|f_i(x_i+\bar{x}_j)-p_i(x_i+\bar{x}_j)|, \qquad 0\le i< 2^{k_0}, \\
d_{i,\nu} &= & p^{(\nu)}_{i}(x_i)-p^{(\nu)}_{i-1}(x_i), \qquad 0 < i <
2^{k_0}.
\end{eqnarray*}

Then the quasyspline $\{p_i(x)\}$ provides the given uniform approximation precision of function $f$:
$$
\sup_{x\in[0, 2^{k_0}]} |f(x) - p(x)| \le \varepsilon _m,
$$
and the number of nonzero bits for coefficients at intermediate nodes of the quasyspline $L$ does not exceed
$$
L \leq \sum_{i=0}^{2^{k_0}-1} \sum_{\nu=0}^{N-1} {\max(0, 2+\lceil l_{m}+\log_2{|d_{i,\nu}|} \rceil)},
$$
where $l_{m}$ is the minimal allowable length of the fractional part of a coefficient of polynomial approximation in one segment.
\end{theorem}

{\sf Proof.}
By theorem~\ref{theorem:approx_main}  
$$
\|f_i-p_i\| \leq \bar{\varepsilon}  + \delta ^ 2 
\left(
{\frac{(b-a) ^{r-2}}{8(r-2)!}}M + \frac{\lambda_{r, 2}(I_0)}{2(b-a)^{2}} \bar{\varepsilon}
\right).
$$
Substitute values $\bar{\varepsilon},\quad b-a =1,\quad r=N$, then  $\|f_i-p_i\| \leq \varepsilon_m$.

Consider the case $d_{i,\nu} \neq 0$, then $d_{i,\nu}$ has at most $L_{i,\nu}$ nonzero bits. 
$$L_{i,\nu} = 2 + \lceil l_{m}+\log_2{|d_{i,\nu}|} \rceil$$
An additional unit in the estimate accounts for storing the sign.

If $d_{i,\nu} = 0$, then 0 bits are required for its storage, and $L_{i,\nu} = -\infty$.

\square

\vspace{3mm}
First, by enumeration, find the minimal $l_{m}$ for which a basic plan exists. 
Then the problem is solved by the method of integer mixed linear programming similarly to the previous one.

To solve the problem by standard methods, it is required to bring the problem to canonical form.
\begin{theorem}
The optimization problem from theorem~\ref{theorem:approx_optim_cont} can be reduced to an equivalent problem
 of linear programming in canonical form.
$$
\min_x l^T\hat{x}, \qquad A\hat{x} \leq b, 
\qquad
A=\left(\begin{array}{rr} 
       \bar{V}&0\\
       -\bar{V}&0\\
       \bar{D}&I\\
       -\bar{D}&I\\       
   \end{array}\right), 
\qquad
b=\left(\begin{array}{r} 
      \bar{f}+\bar{\varepsilon}\\
      -\bar{f}+\bar{\varepsilon}\\
      0\\
      0      
    \end{array}\right)
$$
$$
\bar{V}=
\left(\begin{array}{rrr} 
       V_0&0&\ldots\\
       0&\ddots&0\\
       0&\ldots&V_{2^{k_0}-1}       
\end{array}\right),
\qquad
V_i=\{(x_i+\bar{x}_j)^k\}_{j=0,k=0}^{2^{k_0}-1,N}
$$
$$
\bar{D}=
\left(\begin{array}{rrrr} 
       D_1^0&-D_1^0&0&\ldots\\
       \vdots&\vdots&\vdots&\vdots\\
       D_1^{P-1}&-D_1^{P-1}&0&\ldots\\
       0&\ddots&\ddots&0\\
       0&\ldots&D_{2^{k_0}-1}^0&-D_{2^{k_0}-1}^0\\       
       \vdots&\vdots&\vdots&\vdots\\
       0&\ldots&D_{2^{k_0}-1}^{P-1}&-D_{2^{k_0}-1}^{P-1}       
\end{array}\right)
$$
$$
D_i^{\nu} p_m=p_m^{(\nu)}(x_i)
$$
$$
\bar{f}=(f(x_0), f(x_0+\bar{x}_1), \ldots, f(x_{2^{k_0}}))^T
$$
$$
\hat{x}=(p_0^0,\ldots,p_0^P,p_{1}^0,\ldots,p_{2^{k_0}-1}^P,d_{1,0},
\ldots,d_{1,P-1},d_{2,0},\ldots,d_{2^{k_0}-1,P-1})^T
$$
$$
l=(0,\ldots,\alpha_{1,0},\ldots,\alpha_{1,P-1},\alpha_{2,0},\ldots,\alpha_{2^{k_0}-1,P-1})
$$
Here $V_i$ are Vandermonde matrices on segments, $\bar{f}$ is the vector of values at 
nodes on all segments. 
\end{theorem}
{\sf Proof.}
Perform substitution 
$$
\left\{ \begin{array}{l}
   \left(\begin{array}{rr} 
       \bar{V}&0\\
       -\bar{V}&0\\
       \bar{D}&I\\
       -\bar{D}&I\\       
   \end{array}\right)
   \hat{x}\leq
   \left(\begin{array}{r} 
      \bar{f}+\bar{\varepsilon}\\
      -\bar{f}+\bar{\varepsilon}\\
      0\\
      0      
    \end{array}\right)\\
    {\varepsilon}^* = \min_{\hat{x}}{l\hat{x}}
\end{array} \right.
$$
After simplification, we obtain the inequalities from theorem~\ref{theorem:approx_optim_cont}.
\square

Coefficient tables for a function include coefficients for each quasyspline.

Tables can be synthesized as fixed logic, then area is proportional to 
the number of nonzero bits in the representation of coefficients. 
ROM can also be used, then for a coefficient the maximum width across all quasysplines is chosen.

For functions with monotonic derivatives on part of segments at the beginning or end of the table, correction 
coefficients are equal to zero, since the function grows slower than on the remaining segments, error does not 
have time to accumulate and does not require correction. 
When synthesizing tables from fixed logic, this leads to a decrease 
in table area on a semiconductor chip, therefore zeroing part of coefficients at the beginning or 
end of the table affects its size. 

Thus, when calculating table size, 
the product of coefficient width in bits by the number of nonzero coefficients was used.
Also, if a coefficient for all quasysplines has the same sign, then the sign bit was not counted in the width 
of the table for this coefficient.

Such an approximation algorithm is a generalization of a smooth two-link spline, 
considered by Strollo et al.~\cite{10.1109/TC.2010.127}. Using more 
than two links of the quasyspline and relaxing smoothness requirements allows substantially 
reducing table size. For example, transitioning to a four-link quasyspline compared 
to a two-link one gives table reduction of more than 30\%.

\section{Prototyping Results}\label{sect3_6}
Tables were constructed for quadratic approximation of functions $\sin, \ln, 1/x,
\sqrt{x}$ in the precision range 24-32 bits. Reduction of table size for
two-link quasyspline is more than 40\%, for four-link --- more than 60\%. 
Data for two-link quasyspline correspond to results of Strollo et al.
~\cite{10.1109/TC.2010.127}

\begin{figure}[htbp]
	\center
    \includegraphics[width=0.4\textwidth]{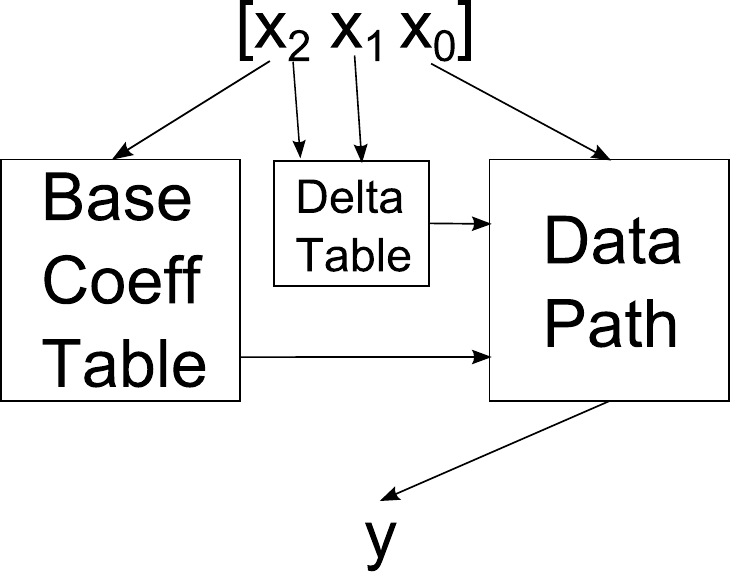}
    \caption{Architecture of the function interpolation block}
    \label{fig:interp}
\end{figure}

Prototyping of an interpolation block with one-cycle delay
based on the proposed algorithm and synthesis of virtual topology for
low-power semiconductor fabrication process with
geometric norms 22 nm was performed. The block architecture is shown in 
figure~\ref{fig:interp}. Tables due to small size were synthesized 
as combinational circuits. Comparison of various implementation variants by area 
is shown in table~\ref{table:interp-area} and by delay - in 
table~\ref{table:interp-speed}. Data is presented in relative form, 
since absolute data is company trade secret. Data for
cubic interpolation is given for comparison with commercial library
DesignWare~\cite{DesignWare} of Synopsys company. Since direct comparison contradicts
the library license agreement, our own implementation
of the block was performed based on documentation and semiconductor circuit synthesis. This implementation
was used for comparison. 

Comparison shows that reduction of table
size in practice leads to reduction in size of the semiconductor block
of function approximation. Computational components can be reused
for computing several functions, in this case for storing tables one should
use a ROM block, which will lead to additional area reduction.
In table~\ref{table:interp-comp}, comparison of block areas and
floating-point multiplier FP32 is given. The practical gain in area 
of the logic circuit is substantially smaller than the estimate of table size reduction. 
This is an expected result, since the logic circuit besides tables contains 
computational elements as well, whose area does not decrease. 
At the same time, a gain of 35-45\% is sufficiently significant to justify replacing 
piecewise-polynomial approximation with the proposed scheme.

\begin{table} [htbp]
  	\centering
	\caption{Relative block size at delay 5 ns} 
	\label{table:interp-area}
	\begin{tabular}{|r|r|r|r|r|}
	\hline
	Precision&$N=2$, q-spline-4&$N=2$, q-spline-2&$N=2$&$N=3$\\
	\hline
	24&55\%&72\%&110\%&100\%\\
	32&68\%&84\%&112\%&100\%\\
	\hline
	\end{tabular}
\end{table}

\begin{table} [htbp]
  	\centering
	\caption{Relative period of maximum clock frequency} 
	\label{table:interp-speed}
	\begin{tabular}{|r|r|r|r|r|}
	\hline
	Precision&$N=2$, q-spline-4&$N=2$, q-spline-2&$N=2$&$N=3$\\
	\hline
	24&70\%&74\%&78\%&100\%\\
	32&78\%&74\%&78\%&100\%\\
	\hline
	\end{tabular}
\end{table}

\begin{table} [htbp]
  	\centering
	\caption{Relative block size compared to floating-point multiplier} 
	\label{table:interp-comp}
	\begin{tabular}{|l|r|}
	\hline
	Block type&Size\\
	\hline
	MUL FP32 w/o denorm&1\\
	SINCOS24, $N=2$, q-spline-4&1.5\\
	SINCOS32, $N=2$, q-spline-4&3.2\\
	\hline
	\end{tabular}
\end{table}
\chapter{Streaming FFT Algorithm on Multi-Bank Memory} \label{chapt3}
This chapter studies algorithms for implementing FFT with arbitrary radices on Hartenstein's anti-machine. 

According to Hartenstein's classification~\cite{hartenstein1991novel}, an anti-machine is a coarse-grained reconfigurable computing machine.
The fundamental difference of an anti-machine is the absence of a program in the form of a sequence of instructions. 
The program is specified by choosing the configuration of coarse-grained components over a prolonged time interval. 
This provides simplification of the computation block by reducing program memory and instruction decoder, which can take up to half of the area and energy of the computational block.
An anti-machine is not universal in the general case; simplification leads to limitation in algorithm execution.
Only specially designed algorithms that take into account the anti-machine architecture can be executed.

A characteristic property of the FFT algorithm and other frequency transforms is the connection of all points in input and output data.
Parallelizing such algorithms is associated with complicating the memory architecture to ensure simultaneous reading and writing of several points according to a complex
addressing law.

\begin{definition}
A butterfly is an algorithm for calculating DFT without splitting by definition 
$$
X_n = \sum_{k=0}^{R-1} x_k\,e^{-\frac{2\pi i}{R} kn}, \qquad 0\le n\le R-1,
$$
where $R$ is the size or radix of the butterfly, $(x_k)_{k=0}^{R-1}$ is the input array, $(X_n)_{n=0}^{R-1}$ is the output array.
\end{definition}

Streaming algorithms provide computation of one FFT butterfly with a startup interval of one cycle. They are well suited for implementation on an anti-machine.
The advantages of streaming algorithms include:
\begin{itemize}
\item high speed;
\item ability to vary FFT length during operation;
\item use of energy-efficient library memory components;
\item possibility of reusing computational components and memory in other
algorithms.
\end{itemize}

\begin{figure}[htbp]
	\center
    \includegraphics[width=0.8\textwidth]{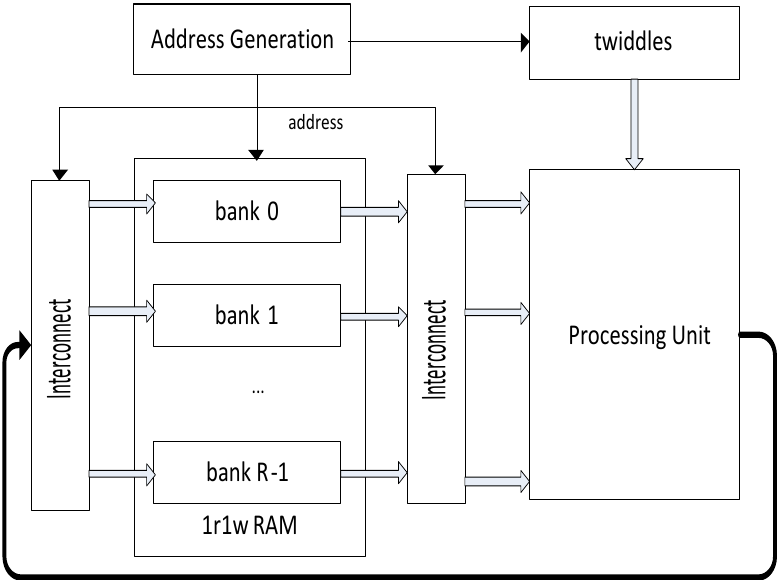}\\
    \caption{Architecture of streaming FFT computation block with 1r1w memory.}
    \label{fig:fft-base}
\end{figure}

The block diagram of a streaming FFT computation block with a butterfly of size $R$ is shown in fig. \ref{fig:fft-base}. This block is a variant of anti-machine implementation.
The processing unit (Processing Unit) executes a butterfly operation with a startup interval of one cycle. The input data for this operation are $R$ numbers, one from each data bank. 
Through block reconfiguration, computational resources involved in computing FFT butterflies can also be used to perform vector operations
of complex and real addition and multiplication, computation of elementary and logical functions, filtering and data convolution.

Suppose it is required to perform FFT on $N=R^q$ samples with some natural $q$. 
If permutations in Interconnect blocks do not lead to additional delays and 
if reading and writing in memory blocks is performed in the same cycle, then the total DFT computation time equals
$$
T(N) = \frac{N}{R} \log_R(N) + C_p,
$$
where $C_p$ is the pipeline length inside the processing unit. 

This section studies ways of implementing FFT on Hartenstein's anti-machine, requiring exactly this minimal computation time $T(N)$. 

When developing algorithms, the following additional constraints were set:
\begin{itemize}
\item absence of additional memory for intermediate values;
\item use of RAM with 1rw formula;
\item full utilization of memory access speed; 
\item absence of memory caching;
\item absence of arbitration of simultaneous accesses;
\item change of parallelism at the circuit design stage;
\item efficient change of FFT length during circuit operation; 
\item absence of a separate data reordering step.
\end{itemize}

In the first section, formulas representing FFT in terms of Kronecker product are proved. 
In the second section, algorithms for implementing FFT with a fixed set of memory banks are described. 
The third section is devoted to self-sorting FFT algorithms with read and write constraints.

\section{General Approach to Developing Streaming FFT Accelerators} 

Consider a streaming architecture of an FFT accelerator based on
multi-port memory with random access. Multi-port memory is used
to ensure simultaneous writing and reading of several data at different
addresses. Suppose that memory banks in fig. 1 contain 1r1w elements, in which one read and one write can be performed in one cycle. 

A memory access conflict is a situation of several simultaneous accesses at different addresses to one port of a memory bank. 
Since a memory bank port can provide access to only one value, memory conflict leads to the need to implement logic 
for arbitration (ordering) of simultaneous accesses and computation delays.
Let us pose the problem of finding such a way of reading and writing after each butterfly, 
in which conflicts will not occur. 
Thus, the processor and memory will be fully loaded, which minimizes both computation time and the amount of memory used.

\subsection{Synchronous Data Flow Graphs}
We formalize architecture requirements using the theory of synchronous data flow graphs~\cite{73588}. Introduce several definitions.
\begin{definition}
We call a synchronous data flow graph (SDF) a directed graph $G=(V,E, prod, cons, d)$,
where $V$ is a set of nodes representing actors (computational blocks),
$E$ is a set of arcs representing directed data transmission channels.

Actors transmit discrete messages over data channels.
Actor $v_i \in V$ is activated when there are $cons(e_{ki}), e_{ki} \in E$ 
messages on its input arcs. At the moment of activation, actor $v_i$ erases $cons(e_{ij})$ input messages
and writes $prod(e_{ij}), e_{ij} \in E$ output messages. 
Arcs can contain an initial set of messages $d(e_{ij})$.
\end{definition}
\begin{definition}
We call a schedule of synchronous data flow graph $G$ a sequence of node activations $S=v_{i_1}v_{i_2}..v_{i_n}$, $n > 0$, 
returning the graph state to its initial state. 
\end{definition}
\begin{definition}
We call an iteration of synchronous data flow graph $G$ the execution of schedule $S$.
\end{definition}
Schedule execution for a synchronous data flow graph can be modeled in discrete time.

\begin{definition}
We call a homogeneous synchronous data flow graph (HSDF) a data flow graph $G=(V,E, prod, cons, d)$,
if $prod(e_{ij}) = cons(e_{ij}) = 1, e_{ki} \in E$. 
\end{definition}
Consider a set of data operations implementable by this data flow graph $O$.
Each node can be configured to perform one operation from a subset implementable
by this node.
\begin{definition}
We call an architectural template a homogeneous synchronous data flow graph $G=(V,E, d, p, op)$,
where $op(v_i) \subset O$, $v_i \in V$ is the set of operations supported by the node,
$p \geq 0$ is the internal pipeline length (execution delay in cycles).
\end{definition}
Delay means the node firing time. 
Zero delay means immediate execution in the same time quantum.
For a schedule to exist, the total delay in any cycle must be greater than 0.

\begin{definition}
We call $conf$ a configuration of architectural template $G$, 
if $conf(v_i) \in op(v_i)$, $v_i \in V$.
\end{definition}
\begin{definition}\label{defn:instruction}
We call $I=(conf, k)$ an instruction of architectural template $G$, 
where $conf$ is a configuration and $k$ is a constant number of iterations of the synchronous data flow graph,
determined by the configuration.
\end{definition}
\begin{definition}
We call $P=I_1..I_n, n > 0$ a program of architectural template $G$, 
where $I_k$ are instructions.
\end{definition}
We will say that an algorithm is implementable by an architectural template on a data set if there exists a program of the architectural template, 
implementing execution of this algorithm on this data set.

\begin{figure}[htbp]
	\center
    \includegraphics[width=0.8\textwidth]{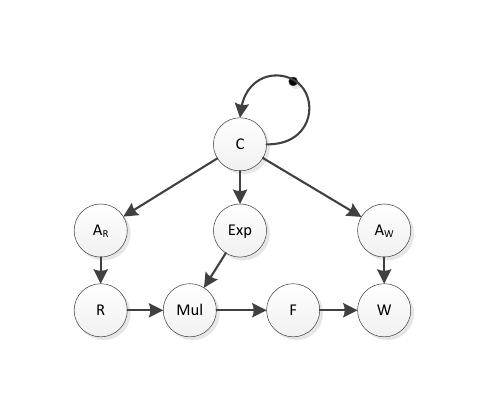}\\
    \caption{Architectural template of streaming FFT computation block.}
    \label{fig:fft-sdf}
\end{figure}

Fig. \ref{fig:fft-sdf} shows an architectural template corresponding to the architecture in fig. \ref{fig:fft-base} without accounting for multi-bank memory.

Here $C$ is a counter, $Exp$ is a complex exponential computation block, $A_R$, $A_W$ are read and write address generators, 
$R$, $W$ are read and write ports, $Mul$ is a vector complex multiplier, $F$ is a butterfly computation block.

\subsection{FFT Splitting Rule}

Let $n$ be a natural number and $\omega_n=e^{-\frac{2\pi i }{n}}$. The discrete Fourier transform (DFT) of order $n$ is a linear operator with matrix
$$
{\mathcal F}_n=[\omega^{kl}_n]_{0\leq k,l<n}.
$$
For arbitrary integers $N>0$ and $0\le j<N$, introduce notation $e_{j,N}$ for the $j$-th standard unit vector in ${\sf R}^N$. 

Further, $\otimes$ denotes the Kronecker product: if $C=A\otimes B$, dimensions of matrices $A$ and $B$ are $m\times n$ and $k\times \ell$, 
respectively, then $C_{ik+j, p\ell+q} = A_{i,j} B_{p,q}$ for all $0\le i< m$, $0\le j< n$, $0\le p< k$, $0\le q< \ell$.

Further, we will use the following properties of the Kronecker product:
\begin{eqnarray*}
(A \otimes B)^T & = & A^T \otimes B^T, \\
(A \otimes B) (C \otimes D) & = & (AC) \otimes (BD), \\
(A \otimes B) \otimes C & = & A \otimes (B \otimes C).
\end{eqnarray*}
The second equality is valid only under the condition of correctness of products $AC$ and $BD$. 
From the last equality, in particular, it follows that $(A \otimes I_k) \otimes I_{\ell} = A \otimes I_{k\ell}$ and $I_k \otimes (I_{\ell} \otimes A) = I_{k\ell} \otimes A$.

Let $m, k$ be natural numbers and $n=km$. The set of vectors $(e_{i,k}\otimes e_{j,m})$ for $0\le i<k$, $0\le j<m$ forms a standard basis in ${\sf R}^n$. 

Introduce notation for permutation matrix $L^n_k$ and for diagonal matrix $W^n_m$ of order $n$, defined by formulas
\begin{eqnarray*}
L^n_k (e_{i,k} \otimes e_{j,m}) & = & e_{j,m} \otimes e_{i,k}, \qquad 0\le i<k, \quad 0\le j<m, \\
W^n_m (e_{i,k} \otimes e_{j,m}) & = & \omega_n^{ij} (e_{i,k} \otimes e_{j,m}), \qquad 0\le i<k, \quad 0\le j<m.
\end{eqnarray*}

Permutation $L^n_k$ performs the following index transformation:
$$
l^n_k : im+j \to jk+i, \qquad 0 \leq i < k, \quad 0 \leq j < m.
$$
From the definition of $L^n_k$ it follows that for any vectors $x\in {\sf R}^k$ and $y\in {\sf R}^m$: $L^n_k (x\otimes y) = y \otimes x$.

Matrix $W^n_m$ can also be explicitly represented as 
$$
W^n_m = \left(
\begin{array}{lll}
	(V^m_n)^0& & \\
	&\ddots& \\
	& &(V^m_n)^{k-1} \\
\end{array}
\right), \qquad
V^m_n=\left(
\begin{array}{lll}
	(\omega_n)^0& & \\
	&\ddots& \\
	& &(\omega_n)^{m-1} \\
\end{array}
\right).
$$

These matrices have the following properties.

\begin{lemma} \label{aux_LW}
Let $n=km$. Then
\begin{eqnarray*}
L^n_k & = & (L^n_m)^{-1} = (L^n_m)^T, \\
W^n_m & = & L^n_m W^n_k L^n_k.
\end{eqnarray*}
\end{lemma}

\proof The statements follow directly from the definitions. \square

\vspace{3mm}
The Kronecker product is non-commutative. Permutation of $I_k$ and ${\mathcal F}_m$ in the Kronecker product is performed using permutation matrices $L^n_k$.

\begin{lemma} \label{commutat}
Let $n=km$. Then
$$
L^n_k (I_k \otimes {\mathcal F}_m) L^n_m = {\mathcal F}_m \otimes I_k.
$$
\end{lemma}

\proof
Let $0\le i<k$ and $0\le j < m$. Then
\begin{eqnarray*}
L^n_k (I_k \otimes {\mathcal F}_m) L^n_m (e_{j,m}\otimes e_{i,k}) & = & L^n_k (e_{i,k} \otimes ({\mathcal F}_m e_{j,m})) = ({\mathcal F}_m e_{j,m}) \otimes e_{i,k} \\
& = &
({\mathcal F}_m \otimes I_k)(e_{j,m}\otimes e_{i,k}),
\end{eqnarray*}
which proves the lemma statement. \square

\vspace{3mm}
Fast Fourier transform (FFT) algorithms are based on factorization of the DFT matrix using the splitting
rule formulated in the following statement.

\begin{lemma} \label{FFT_split}
Let $n=km$, where $k$, $m$ are natural numbers. Then
$$
{\mathcal F}_n = L^n_k (I_k \otimes {\mathcal F}_m) W^n_m ({\mathcal F}_k \otimes I_m).
$$
\end{lemma}

\proof
Let $0\le i<k$, $0\le j < m$. Then
$$
W^n_m({\mathcal F}_k \otimes I_m) (e_{i,k} \otimes e_{j,m}) = W^n_m(({\mathcal F}_k e_{i,k}) \otimes e_{j,m}) = ((\omega_k^{ip} \omega_n^{pj})_{p=0}^{k-1} \otimes e_{j,m}).
$$
Since $(L^n_k)^T=L^n_m$, for $0\le p<k$, $0\le q < m$ we similarly obtain that
$$
(e_{q,m}^T \otimes e_{p,k}^T) L^n_k (I_k \otimes {\mathcal F}_m) = \bigg((I_k \otimes {\mathcal F}_m^T)L^n_m (e_{q,m} \otimes e_{p,k})\bigg)^T = \bigg((e_{p,k} \otimes {\mathcal F}_m^T e_{q,m})\bigg)^T.
$$
Hence
$$
(e_{q,m}^T \otimes e_{p,k}^T) L^n_k (I_k \otimes {\mathcal F}_m)W^n_m({\mathcal F}_k \otimes I_m) (e_{i,k} \otimes e_{j,m}) = \omega_k^{ip} \omega_n^{pj} (e_{q,m}^T{\mathcal F}_m e_{j,m}) = \omega_n^{ipm + pj + kjq}.
$$
From the definition of $\omega_n$ it follows that $\omega_n^n=1$. Therefore
$$
\omega_n^{ipm + pj + kjq} = \omega_n^{(p+qk)(im+j) - iqn} = (e_{q,m}^T\otimes e_{p,k}^T) {\mathcal F}_n (e_{i,k} \otimes e_{j,m}),
$$
which proves the lemma statement. \square

\begin{corollary} \label{FFT_split_IF}
Let $n=km$, where $k$, $m$ are natural numbers. Then
$$
{\mathcal F}_n = L^n_k(I_k \otimes {\mathcal F}_m) W^n_m L^n_m (I_m \otimes {\mathcal F}_k) L^n_k.
$$
\end{corollary}

\proof The formula is obtained from lemma \ref{FFT_split} by substituting the statement of lemma \ref{commutat}. \square 

\subsection{Index Inversion}

The general formula of the FFT algorithm is obtained by iterating the splitting rule from corollary \ref{FFT_split_IF}. This formula includes a mapping generated by inversion of multi-indices, which is defined below.

Let $\alpha = (n_K, n_{K-1}, \ldots, n_0)$ be some ordered set of natural numbers, called a multi-index below, and $N=\prod_{k=0}^K n_k$. Each integer from $0$ to $N-1$ is uniquely represented in the numeral system generated by multi-index $\alpha$, as a multi-index $p=(p_K, \ldots, p_0)$ from the condition
$$
n = p_0 + n_0(p_1 + \cdots + n_{K-2}(p_{K-1} + n_{K-1}p_K) \cdots ), \qquad 0\le p_j < n_j, \quad 0\le j\le K.
$$
In this case, the relationship between multi-index $p$ and number $n$ is denoted as follows: $p = n_{\alpha}$, $n = p^{\alpha}$. We will call multi-index $\alpha$ generating the numeral system, and multi-index $p$ consistent with this system. The number of encoded numbers will be denoted $N=|\alpha|$.

From the definition it immediately follows that for any generating multi-indices $\alpha$, $\beta$ and for any multi-indices $p$, $q$ consistent with them, respectively,
$$
e_{p^{\alpha}, |\alpha|} \otimes e_{q^{\beta}, |\beta|} = e_{(p, q)^{(\alpha, \beta)}, |\alpha|\, |\beta|}.
$$

By inversion of multi-index $p=(p_K, \ldots, p_0)$ we call the same set of components, but in reverse order: $p^{\star} = (p_0, \ldots, p_K)$. Multi-index $\alpha^{\star}=(n_0, \ldots, n_K)$ also generates numbering of numbers from $0$ to $N-1$. By permutation of inversion of multi-index $\alpha$ we call a permutation $P_{\alpha}$ of numbers from $0$ to $N-1$, defined by the rule
$$
P\bigg(p_K + n_K(p_{K-1} + \cdots + n_2(p_1 + n_1 p_0) \cdots )\bigg) = p_0 + n_0(p_1 + \cdots + n_{K-2}(p_{K-1} + n_{K-1}p_K) \cdots )
$$
for $0\le p_j < n_j$, $0\le j\le K$. This rule can also be written as $P_{\alpha}n = ((n_{\alpha^{\star}})^\star)^{\alpha}$. Thus, permutation $P_{\alpha}$ maps numbers from $0$ to $|\alpha|-1$, written in the numeral system generated by multi-index $\alpha^{\star}$, to numbers with inverted digit representation in the numeral system generated by multi-index $\alpha$.

Permutation $P_{\alpha}$ generates a linear mapping in ${\sf R}^N$, which will be denoted $S_{\alpha}$. It is fully characterized by values $S_{\alpha} e_{n,N} = e_{P_{\alpha}n, N}$ for $0\le n <N$. 

From the definition of permutation matrix $L^{mn}_n$ it follows that if $\alpha=(m,n)$ has only two components, then $S_{\alpha} = L^{mn}_n$. 

\begin{lemma} \label{inverse_permut}
Let $\alpha$ be a multi-index and $N = |\alpha|$. Let $M>0$ and $\beta = (M, \alpha)$ be an extended multi-index. Then
$$
S_{(M, \alpha)} = (I_M \otimes S_{\alpha}) L^{NM}_N, \qquad S_{(\alpha, M)} = L^{NM}_M (I_M \otimes S_{\alpha}).
$$
\end{lemma}

\proof  Let $0\le m<M$ and $0\le n<N$. To prove the first equality, introduce notation for multi-index $\beta=(M, \alpha)$. Perform auxiliary transformations:
$$
mN + P_{\alpha}n = (m, (n_{\alpha^{\star}})^{\star})^{(M, \alpha)} = [(n_{\alpha^{\star}},m)^{\star}]^{\beta} = [((nM+m)_{\beta^{\star}})^{\star}]^\beta = P_{\beta} (m+nM).
$$
Substitute this equality in the following transformation:
\begin{eqnarray*}
&&
(I_M \otimes S_{\alpha}) L^{NM}_N (e_{n, N} \otimes e_{m,M}) = (I_M \otimes S_{\alpha}) (e_{m, M} \otimes e_{n, N}) = e_{m, M} \otimes (S_{\alpha} e_{n, N}) \\
& = &
e_{m, M} \otimes e_{P_{\alpha} n, N} = e_{mN+P_{\alpha}n, MN} = e_{P_{\beta}(m+nM)} = S_{\beta} e_{m+nM, MN} = S_{\beta} (e_{n, N} \otimes e_{m,M}),
\end{eqnarray*}
which is equivalent to the first statement of lemma \ref{inverse_permut}. 

The second statement is proved similarly. Denote $\gamma = (\alpha, M)$. Then
$$
m + M P_{\alpha}n = ((n_{\alpha^{\star}})^{\star}, m)^{(\alpha, M)} = [(m, n_{\alpha^{\star}})^{\star}]^{\gamma} = [((mN+n)_{\gamma^{\star}})^{\star}]^{\gamma} = P_{\gamma} (mN+n).
$$
Substitute this equality in the following transformation:
\begin{eqnarray*}
&&
L^{NM}_M (I_M \otimes S_{\alpha}) (e_{m, M} \otimes e_{n, N}) = L^{NM}_M (e_{m, M} \otimes S_{\alpha} e_{n, N}) = L^{NM}_M (e_{m, M} \otimes e_{P_{\alpha} n, N}) \\
& = &
e_{P_{\alpha} n, N} \otimes e_{m, M} = e_{NP_{\alpha}n+m, MN} = e_{P_{\beta}(mN+n)} = S_{\beta} e_{mN+n, MN} = S_{\beta} (e_{m, M} \otimes e_{n, N}),
\end{eqnarray*}
which is equivalent to the second statement of lemma \ref{inverse_permut}. \square

\subsection{FFT Formula of Arbitrary Dimension in Time Domain}
We use the splitting rule to bring the FFT algorithm to a form suitable for implementation on the architectural template of fig. \ref{fig:fft-sdf}.

Suppose the dimension $N$ of DFT can be factored: $N = \prod_{k=0}^K n_k$. 
Then the DFT computation algorithm can contain only butterflies of length $n_0, \ldots n_K$. 
The FFT algorithm, which is usually applied for dimensions $N$ equal to a power of $2$, is based on this property. 
In the following statement, this property is proved for arbitrary dimensions $n_0, \ldots, n_K$. 

\begin{theorem} \label{FFT_iter}
Let $\alpha = (n_K, \ldots, n_0)$ be a set of natural numbers and $N = \prod_{k=0}^K n_k$. For each $k=0, \ldots, K$, define $N_k=\prod_{j=0}^k n_j$. Then
$$
{\mathcal F}_N = \left(\prod_{k=0}^K D_{k,\alpha}\right) S_{\alpha},
$$
where in the product of matrices $D_{k,\alpha}$, matrices with smaller indices are on the right, 
$$
D_{k,\alpha} = A_k^{-1} (I_{N/n_k} \otimes {\mathcal F}_{n_k}) \widehat{W}_k A_k, \qquad 0\le k\le K,
$$
permutation matrices $A_k$ and diagonal matrices $\widehat{W}_k$ are defined by equalities
$$
A_k = I_{N/N_k} \otimes L^{N_k}_{n_k}, \qquad \widehat{W}_k = I_{N/N_k} \otimes W^{N_k}_{n_k}.
$$
\end{theorem}

\proof
We prove the theorem by induction. Let $K=0$. Then $N=n_0$. By definition, $L^{n_0}_{n_0}=I_{n_0}$ and therefore $A_0 = I_N$. Moreover, $W^{n_0}_{n_0} = I_{n_0}$, and therefore $\widehat{W}_0 = I_N$. The conclusion of the theorem becomes an identity: ${\mathcal F}_N={\mathcal F}_{n_0}$.

We prove the induction step. Suppose the statement is proved for $K-1\ge 0$. Let us prove for $K$. Apply corollary \ref{FFT_split_IF} with $m=n_K$, $k=N_{K-1}$, $n=N$:
$$
{\mathcal F}_N = L^N_{N_{K-1}}(I_{N_{K-1}} \otimes {\mathcal F}_{n_K}) W^N_{n_K} L^N_{n_K} (I_{n_K} \otimes {\mathcal F}_{N_{K-1}}) L^N_{N_{K-1}}.
$$
By definition, $A_K = L^N_{n_K}$ and by lemma \ref{aux_LW}: $A_K^{-1} = L^N_{N_{K-1}}$. Therefore
$$
{\mathcal F}_N = D_{k, \alpha} (I_{n_K} \otimes {\mathcal F}_{N_{K-1}}) L^N_{N_{K-1}}.
$$
Let $\alpha_{K-1} = (n_{K-1}, \ldots, n_0)$. From the induction hypothesis and from the property of Kronecker products it follows that
$$
{\mathcal F}_{N_{K-1}} = \left(\prod_{k=0}^{K-1} D_{k,\alpha_{K-1}}\right) S_{\alpha_{K-1}}.
$$
Hence
$$
(I_{n_K} \otimes {\mathcal F}_{N_{K-1}}) L^N_{N_{K-1}} = \left(\prod_{k=1}^{K-1} (I_{n_K} \otimes D_{k,\alpha_{K-1}}) \right) (I_{n_K} \otimes S_{\alpha_{K-1}}) L^N_{N_{K-1}}.
$$
By lemma \ref{inverse_permut} 
$$
(I_{n_K} \otimes S_{\alpha_{K-1}}) L^N_{N_{K-1}} = S_{\alpha}.
$$
By properties of the Kronecker product
\begin{eqnarray*}
I_{n_K} \otimes (I_{N_{K-1}} \otimes L^{N_k}_{n_k}) & = & I_{N_K} \otimes L^{N_k}_{n_k} = A_k, \\
I_{n_K} \otimes (I_{N_{K-1}} \otimes L^{N_k}_{N_{k-1}}) & = & I_{N_K} \otimes L^{N_k}_{N_{k-1}} = A_k^{-1}, \\
I_{n_K} \otimes (I_{N_{K-1}} \otimes W^{N_k}_{n_k}) & = & I_{N_K} \otimes W^{N_k}_{n_k} = \widehat{W}_k, \qquad 1\le k\le K-1.
\end{eqnarray*}
Therefore
$$
I_{n_K} \otimes D_{k, \alpha_{K-1}} = D_{k, \alpha}, \qquad 1\le k\le K-1.
$$
Substitution of these expressions leads to the formula
$$
(I_{n_K} \otimes {\mathcal F}_{N_{K-1}}) L^N_{N_{K-1}} = \left(\prod_{k=1}^{K-1} D_{k,\alpha} \right) S_{\alpha},
$$
which proves the induction step. \square

All matrices in the theorem condition have structure $I_{N/n_k} \otimes X$, which allows placing the entire chain of operations for one butterfly ${\mathcal F}_{n_k}$ 
on one iteration of the architectural template. In this case, one FFT stage $D_{k,\alpha}$ will correspond to one instruction of the architectural template.

\vspace{3mm}
The digital representation of FFT input signal indices, consistent with the algorithm from theorem \ref{FFT_iter}, is determined by multi-index $\alpha^{\star}=(n_0, \ldots, n_K)$ with higher digits corresponding to radix $n_0$. The FFT result is obtained with digital representation of indices written in the numeral system generated by multi-index $\alpha$. FFT stages $D_{k,\alpha}$ in the algorithm of theorem \ref{FFT_iter} start with processing lower digits.

Permutation matrices $A_k$ from theorem \ref{FFT_iter} can be interpreted in terms of the numeral system. Matrix $L^{N_k}_{n_k}=S_{(N_{k-1}, n_k)}$ performs permutation of index $n_k$ to the end of the multi-index, which is formulated in the following lemma \ref{L_permut}.

For an arbitrary multi-index $\alpha = (n_K, \ldots, n_0)$ and an arbitrary integer $k$ from $0$ to $K$, define multi-index $\bar{\alpha}_k$ by equality
$$
\bar{\alpha}_k = (n_K, \ldots, n_{k+1}, n_{k-1}, \ldots, n_0, n_k).
$$
Since $|\alpha|=|\bar{\alpha}_k|$, both multi-indices generate numeral systems on the same set of integers from $0$ to $N=|\alpha|-1$. Denote the transformation of multi-indices
$$
p = (p_K, \ldots, p_0) \quad \rightarrow \quad \bar{p}^k = (p_K, \ldots, p_{k+1}, p_{k-1}, \ldots, p_0, p_k).
$$ 

\begin{lemma} \label{L_permut}
Let $A_k$ for $1\le k\le K$ be matrices from the statement of theorem \ref{FFT_iter}. For each $k=0, \ldots, K$, introduce multi-index 
$$
\beta_k = (n_K, n_{K-1}, \ldots, n_{k+1}, n_0, n_1, \ldots, n_k).
$$

Then for $k=1, \ldots, K$: $\beta_k = \overline{(\beta_{k-1})}_k$ and matrix $A_k$ performs the following permutation $P$:
$$
n = p^{\beta_{k-1}} \quad \rightarrow \quad P n = (\bar{p}^k)^{\beta_k}, \qquad 0\le n\le |\alpha|-1.
$$
\end{lemma}

\proof Let $1\le k\le K$. The equality $\beta_k = \overline{(\beta_{k-1})}_k$ follows from the definition of operation on multi-indices $\bar{\alpha}_k$ and from the definition of multi-indices $\beta_j$ for $0\le j\le K$.
By definition,
$$
A_k = I_{N/N_k} \otimes L^{N_k}_{n_k} = I_{N/N_k} \otimes S_{(N_{k-1}, n_k)}.
$$
Let $p = (p_K, \ldots, p_k, p_0, p_1 \ldots, p_{k-1})$ be a multi-index in the numeral system generated by $\beta_{k-1}$ and $n = p^{\beta_{k-1}}$ be the index value. Then
$$
A_k e_{n,N} = e_{m, N/N_k} \otimes S_{(n_k, N_{k-1})} e_{\ell, N_k}, 
$$
where
\begin{eqnarray*}
m & = & (p_K, \ldots p_{K+1})^{(n_K, \ldots, n_{k+1})}, \\
\ell & = & (p_k, p_0, \ldots p_{k-1})^{(n_k, n_0, \ldots, n_{k-1})} = (p_k, q)^{(n_k, N_{k-1})}, \qquad q = (p_0, \ldots p_{k-1})^{(n_0, \ldots, n_{k-1})}.
\end{eqnarray*}
Therefore 
$$
S_{(N_{k-1}, n_k)} e_{\ell, N_k} = S_{(N_{k-1}, n_k)} (e_{p_k, n_k} \otimes e_{q, N_{k-1}}) = e_{q, N_{k-1}} \otimes e_{p_k, n_k} = e_{(q, p_k)^{(N_{k-1}, n_k)},N_k}
$$
and
$$
(q, p_k)^{(N_{k-1}, n_k)} = (p_0, \ldots p_k)^{(n_0, \ldots, n_k)}.
$$
Thus,
$$
A_k e_{n,N} = e_{m, N/N_k} \otimes e_{(p_0, \ldots p_k)^{(n_0, \ldots, n_k)},N_k} = e_{(\bar{p}^k)^{\beta_k}, N},
$$
which proves the statement of lemma \ref{L_permut}. \square

\subsection{FFT Formula of Arbitrary Dimension in Frequency Domain}

The DFT computation algorithm described in the formula of theorem \ref{FFT_iter} is commonly referred to as FFT implementation in the time domain. 
A feature of this algorithm is permutation $S_{\alpha}$ of the input vector and multiplication by complex roots of unity before the butterfly. 

The dual statement is commonly referred to as FFT implementation in the frequency domain. 
In it, the index inversion permutation is performed after all butterflies. 
The formula is obtained by transposing the expression from theorem \ref{FFT_iter}.

\begin{corollary} \label{FFT_iter_transpos}
Let $\alpha = (n_K, \ldots, n_0)$ be a set of natural numbers and $N = \prod_{k=0}^K n_k$. For each $k=0, \ldots, K$, define $N_k=\prod_{j=0}^k n_j$. Then
$$
{\mathcal F}_N = S_{\alpha}^{-1} \left(\prod_{k=0}^K \widehat{D}_{k,\alpha}\right) ,
$$
where in the product of matrices $\widehat{D}_{k,\alpha}$, matrices with smaller indices are on the left,
$$
\widehat{D}_{k,\alpha} = A_k^{-1} \widehat{W}_k (I_{N/n_k} \otimes {\mathcal F}_{n_k}) A_k, \qquad 0\le k\le K,
$$
permutation matrices $A_k$ and diagonal matrices $\widehat{W}_k$ are defined by equalities
$$
A_k = I_{N/N_k} \otimes L^{N_k}_{n_k}, \qquad \widehat{W}_k = I_{N/N_k} \otimes W^{N_k}_{n_k}.
$$
\end{corollary}

\proof
From the definition of matrix ${\mathcal F}_N$ it follows that it is symmetric: ${\mathcal F}_N=({\mathcal F}_N)^T$. 
The conclusion of the corollary is obtained by transposing the formula in the statement of theorem \ref{FFT_iter} and substituting obvious equalities $A_k^T = A_k^{-1}$, $(S_{\alpha})^T=S_{\alpha}^{-1}$ and $(W^m_n)^T=W^m_n$. \square

\vspace{3mm}
The following theorem presents a formula of the FFT algorithm in the frequency domain, in which the order of stage execution is determined by radices $m_0, \ldots, m_K$. 

\begin{theorem} \label{FFT_iter_freq}
Let $\beta = (m_K, \ldots, m_0)$ be a set of natural numbers and $N = \prod_{k=0}^K m_k$. For each $k=0, \ldots, K$, define $M_k=\prod_{j=k}^K m_j$. Then
$$
{\mathcal F}_N = S_{\beta} \prod_{k=0}^K \widetilde{D}_{k,\beta},
$$
where in the product of matrices $\widetilde{D}_{k,\beta}$, matrices with smaller indices are on the right, 
$$
\widetilde{D}_{k,\beta} = B_k^{-1} \widetilde{W}_k (I_{N/m_k} \otimes {\mathcal F}_{m_k}) B_k, \qquad 0\le k\le K,
$$
permutation matrices $B_k$ and diagonal matrices $\widehat{W}_k$ are defined by equalities
$$
B_k = I_{N/M_k} \otimes L^{M_k}_{m_k}, \qquad \widetilde{W}_k = I_{N/M_k} \otimes W^{M_k}_{m_k}.
$$
\end{theorem}

\proof
Define $\alpha=\beta^{\star} = (m_0, \ldots, m_K)$, so that $m_k=n_{K-k}$ for $0\le k\le K$. Then $S_{\alpha}^{-1} = S_{\alpha^{\star}}=S_{\beta}$ and in the notation of corollary \ref{FFT_iter_transpos}: 
$$
N_k = \prod_{j=0}^k m_{K-j} = \prod_{i=K-k}^K m_i = M_{K-k}, \qquad 0\le k\le K.
$$
Hence $\widehat{W_k} = \widetilde{W}_{K-k}$, $A_k = B_{K-k}$ and, consequently, $\widehat{D}_{k, \alpha} = \widetilde{D}_{K-k, \beta}$. \square

\vspace{3mm}
The digital representation of FFT input signal indices, consistent with the algorithm from theorem \ref{FFT_iter_freq}, is determined by multi-index $\beta^{\star}$ with higher digits corresponding to radix $m_0$. The FFT result is obtained with digital representation of indices written in the numeral system generated by multi-index $\beta$. FFT stages $\widetilde{D}_{k,\beta}$ in the algorithm of theorem \ref{FFT_iter_freq} start with processing higher digits.

In FFT implementation in the frequency domain by theorem \ref{FFT_iter_freq}, multiplication by complex exponentials in the diagonal matrix $W^n_k$ is performed after the butterfly at each stage $\widetilde{D}_{k,\beta}$. 
However, to improve computation precision, it is desirable to compute the butterfly after multiplication, since the processor's internal memory has a longer mantissa. 
This order of operations substantially improves computation precision~\cite{4626107}. 
Moreover, this computation order allows using the architectural template of fig. \ref{fig:fft-sdf} for computing FFT in the frequency domain.
The following statement presents a modification of theorem \ref{FFT_iter_freq}, in which at each FFT stage, multiplication is performed first, then the butterfly.

Let $k, m, n$ be natural numbers. Define diagonal matrix $V_{k,m,n}$ of size $N=kmn$ by equations
$$
V_{k,m,n} (e_{i,k} \otimes e_{j,m} \otimes e_{\ell, n}) = \omega_{kmn}^{i(\ell m+j)} (e_{i,k} \otimes e_{j,m} \otimes e_{\ell, n})
$$
for $0\le i<k$, $0\le j<m$, $0\le \ell<n$.

\begin{theorem} \label{FFT_iter_freqW}
Let $\beta = (m_K, \ldots, m_0)$ be a set of natural numbers and $N = \prod_{k=0}^K m_k$. For each $k=0, \ldots, K$, define $M_k=\prod_{j=k}^K m_j$. Then
$$
{\mathcal F}_N = S_{\beta} \prod_{k=0}^K E_{k,\beta},
$$
where in the product of matrices $E_{k,\beta}$, matrices with smaller indices are on the right, 
$$
E_{k,\beta} = B_k^{-1} (I_{N/m_k} \otimes {\mathcal F}_{m_k}) \widetilde{X}_k B_k, \qquad 0\le k\le K,
$$
permutation matrices $B_k$ and diagonal matrices $\widetilde{X}_k$ are defined by equalities
$$
B_k = I_{N/M_k} \otimes L^{M_k}_{m_k}, \qquad \widetilde{X}_k = I_{N/M_k} \otimes V_{m_k, m_{k+1}, M_{k+2}}, \qquad 0\le k\le K,
$$
with extension $m_{K+1}=M_{K+1}=M_{K+2}=1$.
\end{theorem}

\proof
In the FFT formula from theorem \ref{FFT_iter_freq}, between adjacent butterflies $(I_{N/m_{k+1}} \otimes {\mathcal F}_{m_{k+1}})$ and $(I_{N/m_k} \otimes {\mathcal F}_{m_k})$ for $0\le k\le K{-}1$ stands the product
$$
H_k = B_{k+1} B_k^{-1} \widetilde{W}_k = (I_{N/M_{k+1}} \otimes L^{M_{k+1}}_{m_{k+1}}) (I_{N/M_k} \otimes L^{M_k}_{M_{k+1}}) (I_{N/M_k} \otimes W^{M_k}_{m_k}).
$$
We prove that $H_k = \widetilde{X}_k B_{k+1} B_k^{-1}$, after which the theorem statement is obtained by redistributing factors in the formula for ${\mathcal F}_N$. For $k=K$, stage $\widetilde{D}_{K, \beta}$ does not contain multiplication by complex exponentials, since $M_K=m_K$ and $W^{M_K}_{m_K} = I$ is the identity matrix. 

Let $0\le k\le K-1$. By properties of the Kronecker product and by lemma \ref{aux_LW}
\begin{eqnarray*}
B_k^{-1} \widetilde{W}_k & = & I_{N/M_k} \otimes (L^{M_k}_{M_{k+1}}W^{M_k}_{m_k}) = I_{N/M_k} \otimes (W^{M_k}_{M_{k+1}} L^{M_k}_{M_{k+1}}) \\
& = &
(I_{N/M_k} \otimes W^{M_k}_{M_{k+1}}) (I_{N/M_k} \otimes L^{M_k}_{M_{k+1}}) = (I_{N/M_k} \otimes W^{M_k}_{M_{k+1}}) B_k^{-1}.
\end{eqnarray*}
Since the common factor $I_{N/M_k}$ in Kronecker products can be factored out, it remains to prove that
$$
(I_{m_k} \otimes L^{M_{k+1}}_{m_{k+1}}) W^{M_k}_{M_{k+1}} = V_{m_{k+1}, m_k, M_{k-1}} (I_{m_k} \otimes L^{M_{k+1}}_{m_{k+1}}).
$$
Let $0\le N_{k+2} < M_{k+2}$, $0\le n_k < m_k$, $0\le n_{k+1} < m_{k+1}$. Then $0\le n_{k+1} M_{k+2} + N_{k+2} <M_{k+1}$, and therefore components of the left-hand side of this equation are
\begin{eqnarray*}
&&
(I_{m_k} \otimes L^{M_{k+1}}_{m_{k+1}}) W^{M_k}_{M_{k+1}} (e_{n_k,m_k} \otimes (e_{n_{k+1},m_{k+1}} \otimes e_{N_{k+2},M_{k+2}})) \\
& = &
\omega_{M_k}^{n_k(n_{k+1} M_{k+2} + N_{k+2})} (e_{n_k,m_k} \otimes L^{M_{k+1}}_{m_{k+1}}(e_{n_{k+1},m_{k+1}} \otimes e_{N_{k+2},M_{k+2}})) \\
& = &
\omega_{M_k}^{n_k(n_{k+1} M_{k+2} + N_{k+2})} (e_{n_k,m_k} \otimes e_{N_{k+2},M_{k+2}} \otimes e_{n_{k+1},m_{k+1}}) \\
& = &
V_{m_k, m_{k+1}, M_{k+2}} (e_{n_k,m_k} \otimes e_{N_{k+2},M_{k+2}} \otimes e_{n_{k+1},m_{k+1}}) \\
& = &
V_{m_k, m_{k+1}, M_{k+2}} (I_{m_k} \otimes L^{M_{k+1}}_{m_{k+1}}) (e_{n_k,m_k} \otimes (e_{n_{k+1},m_{k+1}} \otimes e_{N_{k+2},M_{k+2}})),
\end{eqnarray*}
which completes the proof. \square

\subsection{Circular Convolution Implementation}

When implementing the circular convolution algorithm, first direct DFT is performed, then component-wise multiplication, and then inverse DFT. In this case, direct DFT is performed in the frequency domain, and inverse DFT in the time domain, and inversion permutations $S_{\alpha}$ cancel each other and can be removed from the computation algorithm. 

As is known, DFT is a unitary transform up to a factor: 
$$
{\mathcal F}_N^{-1} = \frac{1}{N}{\mathcal F}_N^* = \frac{1}{N}\overline{{\mathcal F}_N^T} = \frac{1}{N}\overline{{\mathcal F}_N},
$$
where the bar above denotes complex conjugation. Introduce the component-wise vector multiplication operation: $X.*Y=Z=(Z_k)_{k=0}^{N-1}$, if $Z_k=X_k Y_k$, where $X = (X_k)_{k=0}^{N-1}$, $Y = (Y_k)_{k=0}^{N-1}$.

\begin{theorem} \label{FFT_conv}
Let $\alpha = (n_K, \ldots, n_0)$ be a set of natural numbers and $N = \prod_{k=0}^K n_k$. Let $z=(z_i)_{i=0}^{N-1}$ be the circular convolution of vectors $x=(x_i)_{i=0}^{N-1}$ and  $y=(y_i)_{i=0}^{N-1}$. 

Define matrices $D_{k, \alpha}$ and $E_{k, \alpha^{\star}}$, as in theorems \ref{FFT_iter} and \ref{FFT_iter_freqW}. Then
$$
z = \frac{1}{N} \overline{\left(\prod_{k=0}^K D_{k, \alpha}\right) (\overline{X}.*\overline{Y})},
$$
where
$$
X = \left(\prod_{k=0}^K E_{k, \alpha^{\star}}\right) x, \qquad Y = \left(\prod_{k=0}^K E_{k, \alpha^{\star}}\right) y
$$
and in all products, factors with smaller indices are on the right.
\end{theorem}

\proof
Define $\beta=\alpha^{\star}=(m_K, \ldots, m_0)=(n_0, \ldots, n_K)$. Then by theorem \ref{FFT_iter_freqW}, the DFT result of vector $x$ is
$$
{\mathcal F}_N x = S_{\beta} (E_{K, \beta} \cdots E_{0, \beta}) x = S_{\beta} X,
$$
and similarly, ${\mathcal F}_N y = S_{\beta} Y$. Permutation operator $S_{\beta}$ obviously has the property of distributivity with respect to component-wise vector product: 
$$
(S_{\beta} X) .* (S_{\beta} Y) = S_{\beta} (X.*Y).
$$
Since circular convolution corresponds to component-wise product of Fourier transforms, then
$$
z = {\mathcal F}_N^{-1} [({\mathcal F}_N x).*({\mathcal F}_N y)] = {\mathcal F}_N^{-1} S_{\beta} (X.*Y) = \frac{1}{N} \overline{{\mathcal F}_N S_{\beta} (\overline{X}.*\overline{Y})}.
$$
Substituting the FFT formula from theorem \ref{FFT_iter}, we obtain
$$
z = \frac{1}{N} \overline{ D_{K, \alpha} \cdots D_{0, \alpha} S_{\alpha} S_{\beta} (\overline{X}.*\overline{Y})},
$$
which coincides with the conclusion of the theorem, since $S_{\alpha^{\star}}=S_{\alpha}^{-1}$. \square

\vspace{3mm}
In the formula from theorem \ref{FFT_conv}, vectors entering as input to both direct FFT and inverse FFT are written in digital representation generated by multi-index $\alpha$. 
However, the order of stage execution is different: starting with higher digits in inner FFTs and starting with lower digits in outer FFT.

\section{Multi-Bank Memory Organization}

Suppose FFT is required to be implemented with memory whose number of cells coincides with the number of input data. 
Moreover, cells are divided into memory banks, which impose additional constraints on reading and writing.

\begin{figure}[htbp]
	\center
    \includegraphics[width=0.8\textwidth]{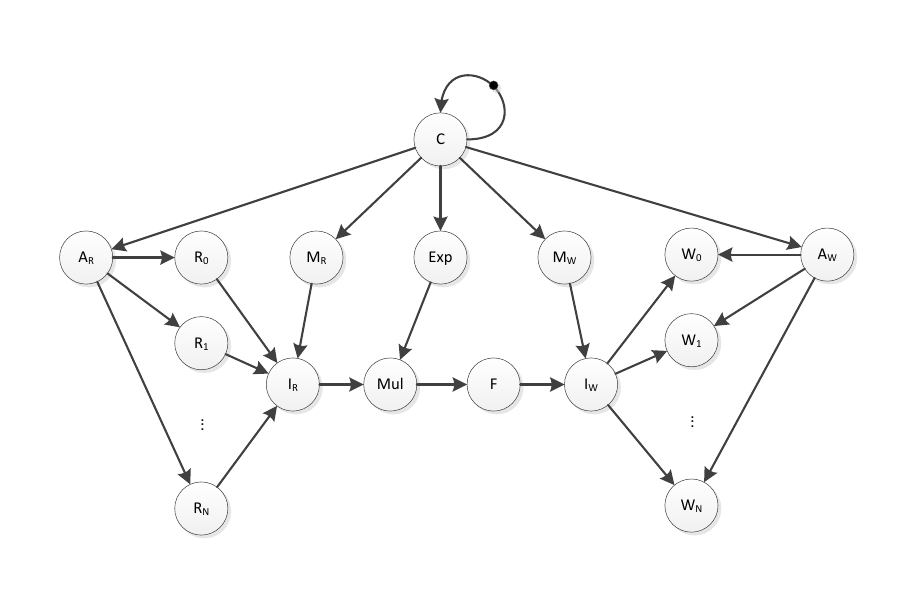}\\
    \caption{Architectural template of streaming FFT computation block with 1r1w memory.}
    \label{fig:fft-1r1w-sdf}
\end{figure}

Consider a more detailed architectural template of fig. \ref{fig:fft-1r1w-sdf}, reflecting the presence of multi-bank dual-port memory.

Here $C$ is a counter, $Exp$ is a complex exponential computation block, $A_R$, $A_W$ are read and write address generators, 
$R$, $W$ are read and write ports, $Mul$ is a vector complex multiplier, $F$ is a butterfly computation block,
$M_R$, $M_W$ are bank number generators, $I_R$, $I_W$ are controlled switches,
$P$ is a pipeline. 

We modify the FFT algorithm for implementation on the architectural template of fig. \ref{fig:fft-1r1w-sdf}.

\subsection{Problem Statement}

Consider an FFT algorithm on $N=\prod_{i=0}^{n-1} n_i$ points, where $n_i$ is the order of the butterfly at stage with ordinal number $i$. Let $R$ be the least common multiple of the set of numbers $(n_i)_{i=0}^{n-1}$, so that $q_i=R/n_i$ are integers.

Suppose that in one cycle the processor executes butterflies with a total number of wings $R$. This means that at the $i$-th stage, $q_i$ butterflies are simultaneously executed, each of order $n_i$. Moreover, all data is stored in $N$ cells distributed across $R$ memory banks. Each memory bank in one cycle can perform only one read operation and one write operation. 

It is required to distribute input data across memory banks in such a way that at each cycle the processor is fully loaded and processes $R$ samples. 
For this, obviously, at each cycle, input samples of all butterflies must be in different banks, and results must be written to the same cells from which samples were read. 
The requirement of absence of conflicts is equivalent to homogeneity of the synchronous data flow graph in the definition of the architectural template.

Main special cases are identical butterfly radices, $n_i=R$ for all $i$, as well as the case $N=R^{n-1} r$, where number $r$ divides $R$. In the latter case, the algorithm is commonly called FFT with radix $r/R$. The Radix-2 algorithm is widely used, in which $R=2$, as well as the Radix-4 algorithm with radix $2/4$.

The result of each butterfly is written to the same addresses from which $R$ numbers were taken, and in the standard order determined by exponents of complex exponentials in multipliers $W$. The input array of numbers of length $N$ must be distributed across banks in such a way that at each stage, exactly one sample from each bank participates in simultaneously executed butterflies. A situation in which this is not the case is called a conflict. It is required to find a distribution of the input array of numbers in which there are no conflicts at all stages.

A solution to this problem is known for $R=r$~\cite{johnson1992conflict}. In~\cite{jo2005new} an algorithm for conflict-free distribution for FFT with radix $2/4$ is formulated. 

In~\cite{hsiao2010generalized} an algorithm for arbitrary mixed radices with launching one butterfly per cycle is formulated, which does not fully utilize memory bandwidth.

In this section, a conflict-free distribution algorithm for an arbitrary set $(n_i)_{i=0}^{n-1}$ of divisors of number $R$ with maximum utilization of memory bandwidth is formulated and proved. 

Let $x$ be the sample number in the input array of numbers. Distribution across memory banks is determined by bank number $m(x)$ and cell number $a(x)$ within the bank.

At each stage with number $i$, several butterflies are executed. Let us order these butterflies by wing number with unit multiplier. The sequence of butterfly execution will be denoted by a set of integers $T_i(k)$, where $k$ is the ordinal number of the operation, and $T_i(k)$ is the number of the butterfly processed at this operation. 

To ensure the necessary rate of parallel computations, the computational block
has an internal pipeline of length $p$. The first operation will be reading, the last --
writing. Pipeline length is the number of cycles of butterfly execution, not including
writing. Writing is not included, since it overlaps with reading in time
to ensure maximum performance. In this case, simultaneous reading and
writing in one cycle of one memory cell yields a correct result.

If read and write operations do not overlap, the maximum computation rate
drops by half to one butterfly in two cycles. 

\subsection{FFT Accelerator with Mixed Radix and 1r1w Memory}\label{sect:fft_1r1w}

Suppose that the FFT algorithm sequentially executes butterflies of orders $n_0, n_1, \ldots, n_K$. The input array length is $N=\prod_{i=0}^K n_i$. Memory is divided into $R$ memory banks and number $R$ is the least common multiple of radices $(n_k)_{k=0}^K$. At each cycle, one or several butterflies are computed with the result written to the same memory cells from which source data for executing butterflies were read. 

It is required to find a distribution of input data across banks $m(x)$, in which at each cycle exactly one sample from each bank is selected, as well as all sets of simultaneously executed butterflies at stage $k$, if $n_k<R$. 

Further, the greatest common divisor of an arbitrary set of natural numbers $(m_j)_{j=0}^J$ will be denoted $\mu((m_j)_{j=0}^J)$, and the least common multiple of the set --- $\nu((m_j)_{j=0}^J)$. By definition, $R=\nu((n_k)_{k=0}^K)$. 

Suppose that in the FFT algorithm, stages are sequentially executed by radices $n_0, n_1, \ldots, n_K$. If indices of the input array have digital representation $n=p^{\alpha}$, where
$$
p=(p_K, \ldots, p_0), \qquad \alpha = (n_K, \ldots, n_0),
$$
and the butterfly of the $k$-th stage of the FFT contains samples whose numbers differ only in component $p_k$, then the FFT algorithm will be called beginning with lower digits of multi-index $\alpha$. 

If indices of the input array have digital representation $n=p^{\alpha}$ with
$$
p=(p_0, \ldots, p_K), \qquad \alpha = (n_0, \ldots, n_K)
$$
and the butterfly of the $k$-th stage of the FFT contains samples whose numbers differ only in component $p_k$, then the FFT algorithm will be called beginning with higher digits of multi-index $\alpha$. 

The main result of this section is presented in the following theorem.

\begin{theorem} \label{bank_stages}
Suppose that in the FFT algorithm, stages are sequentially executed by radices $n_0, n_1, \ldots, n_K$. If the FFT algorithm begins with lower digits, then define $\alpha=(n_K, \ldots, n_0)$ and digital representation of number $n$ of input array component $p=n_{\alpha}=(p_K, \ldots, p_0)$. If the FFT algorithm begins with higher digits, then define $\alpha=(n_0, \ldots, n_K)$ and digital representation of number $n$ of input array component $p=n_{\alpha}=(p_0, \ldots, p_K)$. 

The input array of dimension $N=\prod_{k=0}^K n_k$ is written to $R$ data banks, where $R=\nu(n_0, \ldots, n_K)$ is the least common multiple of butterfly radices. 

For each $k=0,1,\ldots,K$, introduce notation
$$
M_k=\nu((n_i)_{i=0}^k), \quad d_k = \frac{M_k}{M_{k-1}}, \qquad s_{i,k} = \frac{M_i}{\mu(M_i, n_k)}, \quad v_{i,k}=\frac{s_{i,k}}{s_{i-1,k}}, \quad 0\le i < k,
$$
with extension $M_{-1}=s_{-1,k}=1$. 

Define bank number $m(n)$, into which an element of the input array with number $n$ for $n=0,1,\ldots,N-1$ is placed, by formula
$$
m(n) = \sum_{k=0}^K p_k q_k \mod R, 
$$
where $p=n_{\alpha}$ and $q_k=R/n_k$.

Define the set of butterflies executed simultaneously. At stage $k$ for $0\le k\le K$, all butterflies with numbers $\ell$ are simultaneously executed, in whose digital representation the following quantities coincide: integer parts $\lfloor p_j/v_{j,k}\rfloor$ for $0\le j <k$ and integer parts $\lfloor p_j/d_j\rfloor$ for $k<j\le K$.

Then when executing the FFT algorithm, no memory conflicts occur and all $R$ memory banks are used at each cycle.
\end{theorem} 

The proof of the theorem is divided into several auxiliary statements, which are formulated and proved below. It is important to note that for a fixed digital representation of component numbers of the input array in the numeral system generated by an arbitrary multi-index, the memory bank distribution function $m(n)$ is the same for FFT algorithms beginning with higher digits and for those beginning with lower digits. Therefore, implementation of theorem \ref{bank_stages} in the FFT algorithm for circular convolution by theorem \ref{FFT_conv} does not require permutations in memory banks. Only the order of butterfly calls depends on the order of digits of FFT stages.

The most significant special case is DFT with mixed radix, including main radix $R$ and its divisors. Suppose it is required to perform DFT of length $N=rR^{n-1}$, where $r, R$ are natural numbers and $R$ is divisible by $r$, $R=rq$. Consider FFT containing butterflies of types ${\mathcal F}_r$ and ${\mathcal F}_R$. Such an algorithm is commonly called FFT with radix $r/R$. A corollary of theorem \ref{bank_stages} is the following statement.

At each stage, the input array of each butterfly is a set of samples whose digital representations differ only in one component. Let this be component $p_k$. Then the set of remaining components $p^k$ determines the butterfly number $\ell$ within this stage as follows:
$$
\ell=(p^k)^{\alpha^k}, \qquad p^k = (p_{n-1}, \ldots, p_{k+1}, p_{k-1}, \ldots, p_0),
$$
where $\alpha^k=(R, \ldots, R, r)$ for $1\le k \le n{-}1$ and $\alpha^k=(R, \ldots, R)$ for $k=0$.

\begin{theorem}\label{theorem:fft_1r1w}
Suppose that in the FFT algorithm with radix $r/R$, component numbers of the input vector are written in the numeral system generated by multi-index $\alpha=(R, \ldots, R, r)$ of length $n$.  Memory bank number $m(n)$ for input sample with number $n$ and cycle number $T_k(\ell)$ of execution of butterfly $\ell$ at stage $k$ are defined as
\begin{eqnarray*}
m(n) & = & \left(\sum_{i=1}^{n-1}p_i+qp_0\right) \mod R, \\
T_k(\ell) & = & \left\{ \begin{array} {ll} \ell, & 1\le k\le n-1, \\
\left\lfloor \frac{\ell}{q} \right\rfloor, & k=0.
\end{array} \right.
\end{eqnarray*}

Such a choice of traversal order and bank distribution function ensures
absence of conflicts for the architecture of a streaming accelerator for FFT with radix $r/R$ with $R$ banks
of 1r1w memory, regardless of whether it begins with higher or lower digits.
\end{theorem}

\proof 
Apply theorem \ref{bank_stages}. Let FFT begin with lower digits: $n_0=r$ and $n_k=R$ for $1\le k\le n-1$. Then $q_0=R/r=q$ and $q_k=1$ for $1\le k\le n-1$. Therefore, distribution across memory banks $m(n)$ is the same as in theorem \ref{bank_stages}.

To calculate the cycle number of butterfly execution, note that if the butterfly length at some stage is $R$, only one butterfly is executed, so no memory conflict occurs. 

Let $k=0$. From the definition it follows that $d_0=r$, $d_1=R/r=q$ and $d_k=1$ for $k>1$. In accordance with the statement of theorem \ref{bank_stages}, butterflies are simultaneously executed in which integer parts $\lfloor p_1/q \rfloor$ and all numbers $p_i$ for $i\ge 2$ coincide. Since component $p_1$ is the lower digit in the representation of butterfly number $\ell$, the cycle number of execution of this butterfly is $\lfloor \ell/q \rfloor$, as stated in the conclusion of theorem \ref{theorem:fft_1r1w}.

Let FFT begin with higher digits. Then in the notation of theorem \ref{bank_stages}, $n_k=R$ for $0\le k\le n-2$ and $n_{n-1}=r$. At processing stages with numbers $k\le n-2$, only one butterfly per cycle is executed, so no memory conflict occurs. 

Let $k=n{-}1$. From definitions it follows that $s_{i,n-1}=q$ for all $i\ge 0$. However, $s_{-1,n-1}=1$, therefore $v_{i,n-1}=1$ for $1\le i\le n{-}1$ and $v_{0,n-1}=q$. In accordance with the statement of theorem \ref{bank_stages}, butterflies are simultaneously executed in which integer parts $\lfloor p_{n-1}/q \rfloor$ and all numbers $p_i$ for $1\le i\le n-2$ coincide. Quantities $p_1$ and $p_{n-1}$ enter symmetrically into the formula for memory distribution $m(n)$ across data banks. Therefore, when replacing $p_{n-1}$ with $p_1$ in the formula for $T_k(\ell)$, no memory conflicts arise. \square

\vspace{3mm}
To prove theorem \ref{bank_stages}, we introduce a number of notations. 

Fix stage number $k$, $0\le k\le K$. Taking into account notations introduced in theorem \ref{bank_stages}, define
$$
f_{i,k} = \mu(M_i, n_k),  \qquad 0\le i\le k.
$$
Quantities $M_i$, $f_i$, $s_{i,k}$ and $f_{i,k}$ have the following properties, following from their definitions:
$$
M_i = f_{i,k} s_{i,k}, \qquad s_{i,k} = \prod_{j=0}^i v_{j,k}, \qquad 0\le i\le k, \qquad s_{k,k} = \frac{M_k}{n_k}, \qquad v_{k,k}=1.
$$

\begin{lemma} \label{resid_div}
For each $i=0,1, \ldots,k$ and for arbitrary non-negative integers $r_i < f_{i,k}$ and $p_j\le v_{j,k}-1$, define integer function
$$
h_i(r_i,(p_j)_{j=0}^i) = r_i s_{i,k} + \sum_{j=0}^i p_j \frac{M_i}{n_j}.
$$
Then remainders from division of $h_i(r_i,(p_j)_{j=0}^i)$ by $M_i$ are all different and constitute all integers from $0$ to $M_i-1$.
\end{lemma}

\proof
The number of these remainders equals
$$
f_{i,k} \prod_{j=0}^i v_{j,k} = f_{i,k} s_{i,k} = M_i.
$$
Therefore, from the fact that they are different, it follows that they fill the entire set of remainders from division of integers by $M_i$: from $0$ to $M_i-1$. 

Since the remainder operator is linear in variables $r_i$ and $(p_j)_{j=0}^i$, uniqueness of the remainder is equivalent to triviality of the kernel of this operator: from the condition 
$$
h_i(r_i,(p_j)_{j=0}^i) = 0 \mod M_i
$$
it follows that $r_i=0$ and $p_j=0$ for $0\le j\le i$.

We prove triviality of the kernel by induction on $i$. Let $i=0$. Since $v_{0,k}=s_{0,k}$ and $M_0=n_0$, then 
$$
h_0(r_0, p_0) = r_0 s_{0,k} + p_0, \qquad 0\le r_0 < s_{0,k}, \quad 0\le p_k < f_{0,k}.
$$
If $h_0(r_0, p_0)=0$, then $p_0=0$, as remainder from division by $s_{0,k}$. Consequently, $r_0 s_{0,k}=0$ and $r_0=0$, since $s_{0,k}\ge 1$.

We prove the induction step. Let $1\le i\le k$ and the statement is proved for $i-1$. Let us prove it for $i$. Number $f_i=\mu(M_i,n_k)$ divides $f_{i-1}=\mu(M_{i-1},n_k)$, since number $M_k$ divides $M_{k-1}$. Denote $\delta_{i,k} = f_{i,k}/f_{i-1,k}$. Then
$$
d_i = \frac{M_i}{M_{i-1}} = \frac{f_{i,k} s_{i,k}}{f_{i-1,k} s_{i-1,k}} = \delta_{i,k} v_{i,k}.
$$
Number $r_i$ is less than $f_{i,k}=\delta_{i,k}f_{i-1,k}$. Therefore it can be represented in a unique way as
$$
r_i = \delta_{i,k} p_{i-1,k} + \gamma_i, \qquad 0\le p_{i-1,k} < f_{i-1,k}, \qquad 0\le \gamma_i < \delta_{i,k}.
$$

Perform transformations:
\begin{eqnarray*}
h_i(r_i,(p_j)_{j=0}^i) & = & r_i s_{i,k} + \sum_{j=0}^i p_j \frac{M_i}{n_j} = r_i s_{i,k} + p_i \frac{M_i}{n_i} + d_i \sum_{j=0}^{i-1} p_j \frac{M_{i-1}}{n_j} \\
& = & 
(\delta_{i,k} r_{i-1} + \gamma_i) s_{i-1,k} v_{i,k} + p_i \frac{M_i}{n_i} + d_i \sum_{j=0}^{i-1} p_j \frac{M_{i-1}}{n_j} \\
& = &
\gamma_i s_{i-1,k} v_{i,k} + p_i \frac{M_i}{n_i} + d_i \left(r_{i-1} s_{i-1,k} + \sum_{j=0}^{i-1} p_j \frac{M_{i-1}}{n_j} \right) \\
& = & 
\gamma_i s_{i-1,k} v_{i,k} + p_i \frac{M_i}{n_i} + d_i h_{i-1}(r_{i-1}, (p_j)_{j=0}^{i-1}).
\end{eqnarray*}
By the induction hypothesis, equality $h_{i-1}(r_{i-1}, (p_j)_{j=0}^{i-1})=0$ is possible only with zero arguments. Therefore it suffices to prove that from divisibility of number
$$
w = \gamma_i s_{i-1,k} v_{i,k} + p_i \frac{M_i}{n_i} 
$$
by $d_i$ for $0\le \gamma_i < \delta_{i,k}$, $0\le p_i < v_{i,k}$ it follows that $\gamma_i=p_i=0$. 

Let $w$ divide by $d_i=v_{i,k} \delta_{i,k}$. First we prove that numbers $M_i/n_i$ and $v_{i,k}$ are coprime. Indeed, by general properties of the least common multiple, for any natural numbers $a$ and $b$, the pair of numbers $\nu(a,b)/a$ and $\nu(a,b)/b$ is coprime. Since $M_i=\nu(M_{i-1}, n_i)$, number $M_i/n_i$ is coprime with number $M_i/M_{i-1} = d_i = v_{i,k} \delta_{i,k}$ and, consequently, with number $v_{i,k}$.

If $w$ divides by $v_{i,k}$, then from the definition of $w$ it follows that number $p_i M_i/n_i$ also divides by $v_{i,k}$. But $p_i<v_{i,k}$ by definition, and number $M_i/n_i$ is coprime with $v_{i,k}$. Consequently, $v_{i,k}=1$.

From inequality $p_i<v_{i,k}=1$ it follows that $p_i=0$ and $\delta_{i,k}=d_i$. Hence $w=\gamma_i s_{i-1,k}$ and $\gamma_i<\delta_{i,k}$. Since $w$ divides by $d_i=\delta_{i,k}$, to complete the proof of the lemma it remains to establish that numbers $s_{i-1}$ and $\delta_{i,k}$ are coprime. 

By general properties of the greatest common divisor, for any natural numbers $a$ and $b$, the pair of numbers $a/\mu(a,b)$ and $b/\mu(a,b)$ is coprime. Apply this property to numbers $a=M_{i-1}$ and $b=f_{i,k}=\mu(M_i, n_k)$. Compute
$$
\mu(M_{i-1}, f_{i,k}) = \mu(M_{i-1}, \mu(M_i, n_k)) = \mu(M_{i-1}, M_i, n_k) = \mu(M_{i-1}, n_k) = f_{i-1,k}.
$$
Consequently, the pair of numbers
$$
s_{i-1} = \frac{M_{i-1}}{f_{i-1,k}} = \frac{a}{\mu(a,b)}, \qquad \delta_{i,k} = \frac{f_{i,k}}{f_{i-1,k}} = \frac{b}{\mu(a,b)}
$$
is coprime. \square

\begin{corollary} \label{list_resid}
For arbitrary non-negative integers $p_k < n_k$ and $p_j\le v_{j,k}-1$ for $0\le j\le k-1$, define
$$
h((p_j)_{j=0}^k) = \sum_{j=0}^k p_j \frac{M_k}{n_j}.
$$
Then remainders from division of $h((p_j)_{j=0}^k)$ by $M_k$ are all different and constitute all integers from $0$ to $M_k-1$.
\end{corollary}

\proof
The statement is obtained from lemma \ref{resid_div} by substituting $i=k$. Indeed, in the statement of lemma \ref{resid_div}, $v_{k,k}=1$ and therefore $p_k=0$. It remains to replace $r_i$ with a new variable $p_k$ from this corollary, as well as substitute $s_{k,k}=M_k/n_k$ and $f_{k,k}=n_k$. \square

\begin{lemma} \label{order_bank}
For $0\le i\le K$, define set
$$
P_i = \left\{ \begin{array} {ll} \{0, 1, \ldots, v_{i,k}-1\}, & 0\le i < k, \\ \{0, 1, \ldots, n_k-1\}, & i=k, \\ \{0, 1, \ldots, d_i-1\}, & k < i \le K. \end{array} \right.
$$
On the direct product of sets $P=\prod_{i=0}^K P_i$, define function
$$
h(p) = \sum_{i=0}^K p_i q_i \mod R, \qquad p = (p_i)_{i=0}^K.
$$
Then $h$ is a bijection from $P$ onto set $\{0, 1, \ldots, N{-}1\}$.
\end{lemma}

\proof
We prove a more general statement by induction: for $k\le j\le K$, mapping
$$
h_j(p) = \sum_{i=0}^j p_i \frac{M_i}{n_j}, \qquad p = (p_i)_{i=0}^j\in \prod_{i=0}^j P_i \mod M_j,
$$
is a bijection from set $\prod_{i=0}^j P_i$ onto $\{0, 1, \ldots, M_j{-}1\}$. Then for $j=K$ we obtain the lemma statement, since $M_K=R$. Since the number of elements in $\prod_{i=0}^j P_i$ equals the number of residues modulo $M_i$, it suffices to prove triviality of the kernel of mapping $h_j$. 

The induction base for $j=k$ is proved in corollary \ref{list_resid}. We prove the induction step. Suppose the statement is proved for $j-1\ge k$. Let $p=(p_i)_{i=0}^j\in \prod_{i=0}^j P_i$. Select all components except the last: $\widetilde{p} = (p_i)_{i=0}^{j-1}\in \prod_{i=0}^{j-1} P_i$. Then
$$
h_j(p) = d_j h_{j-1}(\widetilde{p}) + p_j \frac{M_j}{n_j}, \qquad 0\le p_j < d_j.
$$
Let $h_j(p)=0$. Then $p_j M_j/n_j$ divides by $d_j$. For any natural numbers $a$ and $b$, numbers $\nu(a,b)/a$ and $\nu(a,b)/b$ are coprime. For $a=M_{j-1}$ and $b=n_j$, we obtain that $\nu(a,b)=M_j$ by definition. Consequently, numbers $M_j/M_{j-1}=d_j$ and $M_j/n_j$ are coprime. Since $p_j M_j/n_j$ divides by $d_j$ and $0\le p_j<d_j$, then $p_j=0$. Hence $d_j h_{j-1}(\widetilde{p})=0$ and by the induction hypothesis $\widetilde{p}=0$. \square

\vspace{3mm}
{\sf Proof of theorem \ref{bank_stages}}.

Consider stage with number $k$, $0\le k\le K$, by radix $n_k$. We will use digital representation $(p_K, \ldots, p_0)$ of sample numbers of the input array in the numeral system generated by multi-index $\alpha=(n_K, \ldots, n_0)$. Each butterfly processes all samples whose numbers have the same values of components $(p_K, \ldots, p_{k+1}, p_{k-1}, \ldots, p_0)$, constituting the digital representation of number $\ell$ of this butterfly, and component $p_k$ runs through all admissible values from $0$ to $n_k-1$. 

Consider all samples participating in butterflies executed simultaneously. In accordance with the rule from theorem \ref{bank_stages}, components of the digital representation of these samples can be written in the following form:
$$
p_i = \left\{ \begin{array} {lll} \bar{p}_i v_{i,k} + \delta_i, & 0\le \delta_i < v_{i,k}, & i < k, \\
\delta_k, & 0\le \delta_k < n_k, & i=k, \\
\bar{p}_i d_i + \delta_i, & 0\le \delta_i < d_i, & k < i \le K, \end{array} \right.
$$
where $(\bar{p}_i)_{i=0}^K$ are fixed numbers and $\bar{p}_k=0$. 

These samples are written in memory banks, whose numbers in accordance with the definition from theorem \ref{bank_stages} equal
$$
m(n) = \left(\sum_{i=0}^K \delta_i q_i + \bar{m}\right) \mod R, \qquad \bar{m} = \sum_{i=0}^{k-1} \bar{p}_i v_{i,k} q_i + \sum_{i=0}^{k-1} \bar{p}_i d_i q_i.
$$
By lemma \ref{order_bank}, remainders from division of the first sum by $R$ are different and run through the entire set of remainders from $0$ to $R{-}1$. Adding number $\bar{m}$ does not change this set of remainders. Thus, all samples are taken from different memory banks and the number of these samples equals $R$. \square

\subsection{FFT Accelerator with Mixed Radix and 1rw Memory}

\begin{figure}[htbp]
	\center
        \includegraphics[width=0.8\textwidth]{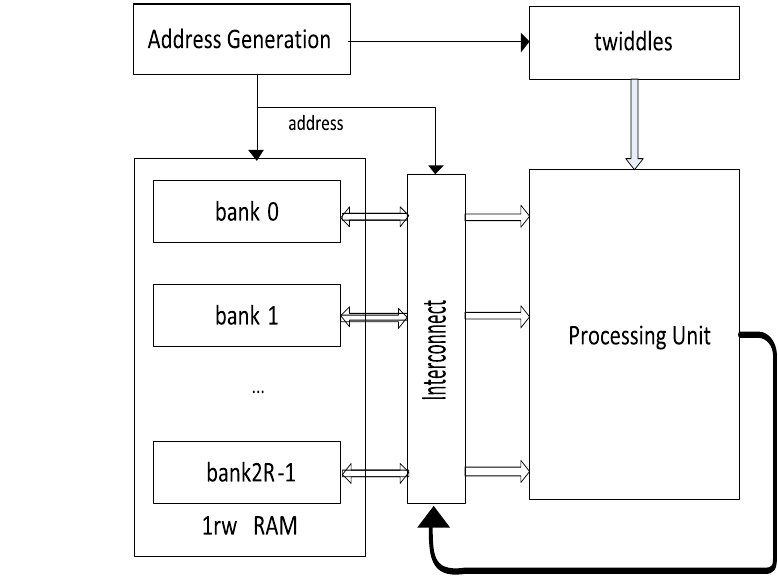}\\
    \caption{Architecture of streaming FFT computation block with 1rw memory.}
    \label{fig:fft-base2}
\end{figure}

We modify the architecture from section~\ref{sect:fft_1r1w} to use 1rw memory, in which reading and writing to one memory bank cannot be performed in one cycle. 
For this, we use $2R$ memory banks and require that writing and reading be performed in non-intersecting sets of banks. 

\begin{figure}[htbp]
	\center
    \includegraphics[width=0.8\textwidth]{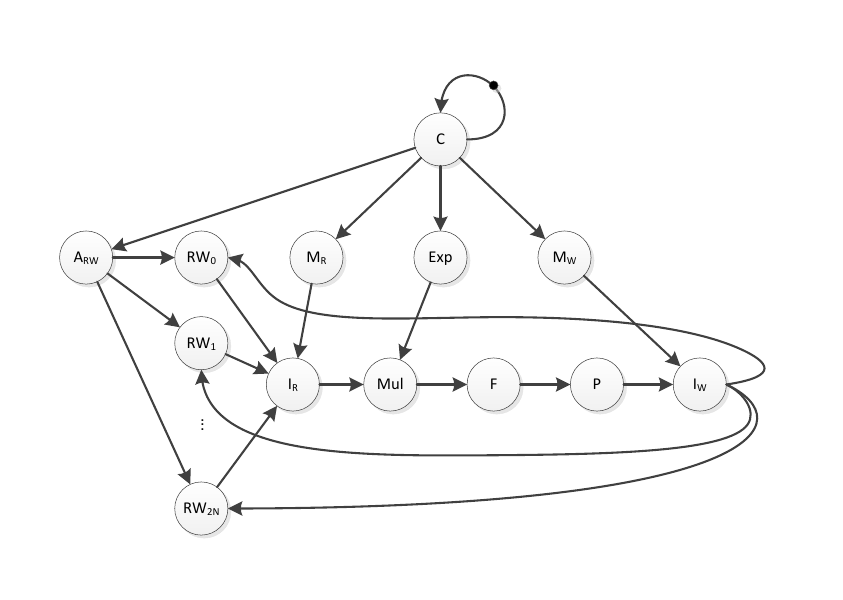}\\
    \caption{Architectural template of streaming FFT computation block with 1rw memory.}
    \label{fig:fft-1rw-sdf}
\end{figure}

Consider the architectural template of fig. \ref{fig:fft-1rw-sdf}, reflecting the presence of multi-bank single-port memory.

Here $C$ is a counter, $Exp$ is a complex exponential computation block, $A_{RW}$ is a read and write address generator, 
$RW$ are read/write ports, $Mul$ is a vector complex multiplier, $F$ is a butterfly computation block,
$M_R$, $M_W$ are bank number generators, $I_R$, $I_W$ are controlled switches.

We modify the FFT algorithm for implementation on the architectural template of fig. \ref{fig:fft-1rw-sdf}.

\begin{theorem}\label{theorem:fft_1rw}
Let the pipeline length be odd, $p \mod~2 =1$. Consider the following traversal order and bank index distribution function:
\begin{eqnarray*}
T^2_i(k) & = & \left\{ \begin{array}{ll}
		t(k), & i = 0, \\
        k, & i > 0,
\end{array} \right. \\
t(k^0) & = & \left[\ldots, \left(\sum_{i=2}^{n-1}k_i + \bar{k}_1 +
r\bar\bar{k}_1\right) \mod~R \right], \\
m_2(x) & = & \left(2\left(\sum_{i=1}^{n-1}x_i+qx_0\right) - (x_0 \mod 2) \right) \mod~2R.
\end{eqnarray*}

Such a choice of traversal order and bank distribution function ensures
absence of conflicts for the architecture of a streaming FFT accelerator with $2R$ banks
of 1rw memory when executing one butterfly per cycle.
\end{theorem}

\proof
Let $i$ be the stage number by radix $R$. In one cycle, exactly one butterfly is processed. The higher digit of numbers of all samples of one butterfly is the same. The parity of this digit determines the group of $R$ memory banks from which reading is performed at this cycle. The higher digit together with the group of $R$ memory banks change at each cycle in accordance with the definition of $T_i(k)$ and the decomposition of number $k$ in the numeral system generated by multi-index $\alpha^{\star}$. Since the pipeline length is odd, at each cycle writing results "in place" is performed not to the group of banks from which reading is performed. Therefore, no memory conflict between read and write operations occurs.

For bank numbers of two samples to coincide, $m_2(k)=m_2(\ell)$, it is required that numbers $m(k)=m(\ell)$ from theorem \ref{theorem:fft_1r1w} coincide and, moreover, that the parity of the lower component in the numeral system generated by multi-index $\alpha^{\star}$ coincide. 

At the stage by radix $R$, the traversal order of butterflies is the same as in theorem \ref{theorem:fft_1r1w}. Consequently, no memory conflict during reading arises by theorem \ref{theorem:fft_1r1w}. 

At each stage, one memory element participates in only one butterfly.

The lower digit of butterfly number $p^i$ is number $k_0$, determining the parity of the memory bank number. The parity of the cycle number coincides with the parity of the memory bank number in the read operation. If the write delay is odd, then the write operation is performed to memory banks with opposite parity. Hence, no conflict of read and write operations arises.
\square

\section{Self-Sorting FFT}

Index inversion $S_{\alpha}$ in the FFT algorithm requires an additional memory pass. Since
points with inverted indices may lie in the same memory bank, due to
memory access conflicts the time of this pass equals the time of executing two FFT stages. 
At the same time, in many algorithms using FFT, it is impossible to use
computational blocks on this pass. Such idle time is not energy-efficient, and
it is desirable to avoid it.

To solve this problem, there exists a class of self-sorting FFT algorithms that
do not require an explicit index inversion step.
For this, a modification of the self-sorting Johnson-Burrus algorithm~\cite{Hegland:1994:SIF:201897.201901} was developed. 
In the Johnson-Burrus FFT algorithm, the stage with small radix is performed in the middle. 
From the point of view of streaming architecture, this algorithm can be represented as follows:
$$
{\mathcal F}_N = \prod_{i=0}^{n-1} D_i,
$$
where $D_i$ are stage matrices, defined as 
$$
D_i = \left\{ \begin{array}{ll}
		A_0^{-1} (I_{N/r} \otimes {\mathcal F}_r)A_0, & i = 0, \\		
        A_i^{-1}(I_{N/R} \otimes {\mathcal F}_R)W^N_iA_i, & 0 < i < \lfloor\frac{n+1}{2}\rfloor, \\
        Q_{i+1}A_i^{-1}(I_{N/R} \otimes {\mathcal F}_R)W^N_iA_iQ_i^{-1}, & \lfloor\frac{n+1}{2}\rfloor \leq i < n - 1, \\            
        \hat{A}_{i}^{-1}(I_{N/R} \otimes {\mathcal F}_R)W^N_i\hat{A}_{i}, & i=n-1,                    	    	        	        
\end{array} \right.
$$
and $Q_i$ is a permutation of two digits and
$$
Q_{\lfloor\frac{n+1}{2}\rfloor} = I, \qquad Q_{n-1}=(S_N^R)^{-1}.
$$

Half of the stages are not computed in
place. However, if we consider groups of butterflies of $R$ elements, then they
are executed in place. For absence of precedence conflicts between butterflies
in one group, it is sufficient that the internal pipeline of the computational block
satisfies the condition
$$
p \geq R-1.
$$

This section presents a generalization of the Johnson-Burrus algorithm to FFT with execution of the small stage at the beginning.

\subsection{Self-Sorting FFT Accelerator with 1r1w Memory}

To implement this algorithm on the architectural template of fig. \ref{fig:fft-sdf}, it is sufficient to change addressing functions and increase the internal pipeline length. 

\begin{lemma} \label{commut_kron}
Let $a$ and $b$ be square matrices, and matrices $A = I_n \otimes a$, $B = b\otimes I_m$ have the same size $N$. 

Then if $m n$ divides $N$, then $A B = B A$.
\end{lemma}

\proof 
Denote the size of matrix $a$ as $k$, and the size of matrix $B$ as $\ell$. From the condition it follows that $N=kn=\ell m$. By condition, $q=mn/N$ is an integer. Hence 
$$
m = \frac{qN}{n} = kq, \qquad n = \frac{qN}{m} = q\ell.
$$
Therefore $I_n=I_{\ell}\otimes I_q$ and $I_m = I_q \otimes I_k$ and
\begin{eqnarray*}
A B & = & (I_{\ell} \otimes (I_q \otimes a)) (b \otimes (I_q \otimes I_k)) = b \otimes I_q \otimes a, \\
B A & = & (b \otimes (I_q \otimes I_k)) (I_{\ell} \otimes (I_q \otimes a)) = b \otimes I_q \otimes a.
\end{eqnarray*}
\square

\vspace{3mm}
Let $N=R^{n-1}r$ and the sequence of FFT stages be determined by multi-index
$$
\alpha = (\alpha_{n-1}, \ldots, \alpha_0), \qquad \alpha_0 = r, \qquad \alpha_k = R, \quad 1\le k\le n{-}1,
$$
so that the small stage with radix $r=R/q$ is executed at the beginning. In FFT algorithms both in time and frequency domains, indices of input array components are written in the numeral system generated by multi-index $\alpha^{\star}=(r, R, \ldots, R)$. 

By theorem \ref{FFT_iter_freqW}, the FFT algorithm in the frequency domain is described by formula
$$
{\mathcal F}_N = S_{\alpha} E_{n-1,\alpha} E_{n-2,\alpha} \cdots E_{0,\alpha},
$$
where $S_{\alpha}$ is a permutation of vector components in accordance with inversion of multi-index $\alpha$. The goal of the self-sorting algorithm is to replace the initial permutation with gradual partial permutations at each stage, performed with minimal pipeline length. 

Introduce an operator for permutation of initial and final digits in the index of an arbitrary vector. Let the vector have length $N$ and number $N$ divide by $k^2$, where $k\ge 1$. Then permutation matrix $Y^N_k$ is defined by equations
$$
Y^N_k (e_{i,k} \otimes e_{\ell,N/k^2} \otimes e_{j, k}) = e_{j,k} \otimes e_{\ell,N/k^2} \otimes e_{i, k}, \qquad 0\le i,j < k, \quad 0\le \ell < N/k^2.
$$

\begin{theorem} \label{self_sort}
Let $N=rR^{n-1}$, where $R=rq$. Define permutation matrices
$$
Q_k = \left\{ \begin{array} {ll} I, & k=0, \\
I_{r R^{k-1}} \otimes L^R_r \otimes I_{R^{n-k-1}}, & 1\le k\le \lfloor \frac{n-1}{2} \rfloor, \\
I_{R^{n/2-1}} \otimes L^{r R}_r \otimes I_{R^{n/2-1}}, & k = \frac{n}{2}, \\
I_{R^{n-k-1}} \otimes [ (I_{r R^{2k-n}} \otimes L^R_r) Y^{r R^{2k+1-n}}_R ] \otimes I_{R^{n-k-1}}, & \lceil \frac{n+1}{2} \rceil \le k \le n-1. \end{array} \right.
$$

Then
$$
{\mathcal F}_N = G_{n-1} \cdot \ldots \cdot G_0,
$$
where $G_k = Q_k E_{k,\alpha}$.
\end{theorem}

\proof
For $k\ge 1$, from the definition of $N_k$ it follows that $N_k=R^{n-k}$ and $N/N_k=rR^{k-1}$. Matrix $E_{k,\alpha}$ can be transformed to the form
$$
E_{k,\alpha} = I_{rR^{k-1}} \otimes a_k, \qquad a_k = [L^{N_k}_{N_{k+1}} (I_{N_{k-1}} \otimes {\mathcal F}_{n_k}) V_{n_k, n_{k+1}, N_{k+2}} L^{N_k}_{n_k}].
$$
We prove that $E_{j,\alpha} Q_k = Q_k E_{j,\alpha}$ for $0\le k < j\le n-1$. The pair of matrices $I_{rR^{j-1}} \otimes a_j$ and $b_k \otimes I_{R^{n-k-1}}$ for any matrix $b_k$ of size $rR^k$ satisfies the conditions of lemma \ref{commut_kron}, since number 
$$
r R^{j-1} \cdot R^{n-k-1} = r R^{n-1 + (j-k-1)}
$$
divides evenly by $N=r R^{n-1}$. Therefore, the product of the pair of matrices $I_{rR^{j-1}} \otimes a_j$ and $b_k \otimes I_{R^{n-k-1}}$ is commutative. Matrices $Q_k$ have the form $b_k \otimes I_{R^{n-k-1}}$ for any $k$, therefore $E_{j,\alpha} Q_k = Q_k E_{j,\alpha}$ for $0\le k < j\le n-1$. Hence
\begin{eqnarray*}
G_{n-1} \cdot \ldots \cdot G_0 & = & Q_{n-1} E_{n-1, \alpha} Q_{n-2} E_{n-2, \alpha} \cdot \ldots \cdot Q_1 E_{1, \alpha} Q_0 E_{0, \alpha} \\
& = &
(Q_{n-1} Q_{n-2} \cdot \ldots \cdot Q_0) \cdot (E_{n-1,\alpha} \cdot E_{0, \alpha}) = (Q_{n-1} Q_{n-2} \cdot Q_0) S_{\alpha}^{-1} {\mathcal F}_N
\end{eqnarray*}
by theorem \ref{FFT_iter_freqW}. It remains to prove that 
$$
Q_{n-1} Q_{n-2} \cdot \ldots \cdot Q_0 =  S_{\alpha}.
$$
All matrices entering this formula perform permutations of vector elements, and the product of matrices corresponds to composition of permutations. Therefore it suffices to prove that the composition of index permutations on the left-hand side leads to index inversion permutation $P_{\alpha}$.

The input FFT vector has length $N=rR^{n-1}$. Further, numbers from $0$ to $N-1$ will be represented in different numeral systems, generating multi-indices of which contain only numbers $r$ or $q=R/r$. In this case, notation of component $\bar{k}$ with one upper bar will mean that the corresponding component of the generating multi-index equals $q$, and notation $\bar\bar{k}$ with two upper bars --- that it equals $r$. 

Let the index of element $k$ of some vector of length $N$ have digital representation
$$
s_0 = (\bar\bar{k}_0, \bar\bar{k}_1, \bar{k}_1, \ldots, \bar\bar{k}_{n-1}, \bar{k}_{n-1}).
$$
From the notation it follows that the generating multi-index is $(r, r, q, \ldots, r, q)$. In the numeral system generated by multi-index $\alpha^{\star}=(r, R, \ldots, R)$, the same index has representation $k_{\alpha^{\star}}=(k_0, k_1, \ldots, k_{n-1})$, where $k_0=\bar\bar{k}_0$ and $k_i=q \bar\bar{k}_i + \bar{k}_i$ for $1\le i\le n-1$. 

Let $m=\lfloor (n-1)/2 \rfloor$. Transformations $(Q_m \cdot \ldots \cdot Q_0)$ reduce to replacing the index, whose digital form is $s_0$, with an index, whose digital form is
$$
s_1 = (\bar\bar{k}_0, \bar{k}_1, \bar\bar{k}_1, \ldots, \bar{k}_m, \bar\bar{k}_m, \bar\bar{k}_{m+1}, \bar{k}_{m+1}, \ldots, \bar\bar{k}_{n-1}, \bar{k}_{n-1}).
$$

{\bf 1.} Let number $n$ be odd. Then in vector $s_1$, the first $m$ pairs have the form $(\bar\bar{k}_{m-j}, \bar{k}_{m-j+1})$ for $j = m, m{-}1, \ldots, 1$. Then in vector $s_1$ follows number $\bar\bar{k}_m$, then another $m$ pairs of the form $(\bar\bar{k}_{m+j}, \bar{k}_{m+j})$ for $j=1, \ldots, m$. 

Transformation $Q_{m+j}$ for $j=1, \ldots, m$ reduces to the following permutation in the digital form of indices:
$$
(\ldots, \bar\bar{k}_{m-j}, \bar{k}_{m-j+1}, \ldots, \bar\bar{k}_{m+j}, \bar{k}_{m+j}, \ldots) \rightarrow (\ldots, \bar\bar{k}_{m+j}, \bar{k}_{m+j}, \ldots, \bar{k}_{m-j+1}, \bar\bar{k}_{m-j}, \ldots),
$$
where ellipsis marks digits that do not change under permutation. As a result, transformation $(Q_{n-1} \cdot \ldots \cdot Q_{m+1})$ permutes the digital record $s_1$ to the form
$$
s_2 = (\bar\bar{k}_{n-1}, \bar{k}_{n-1}, \ldots, \bar\bar{k}_{m+1}, \bar{k}_{m+1}, \bar\bar{k}_m, \bar{k}_m, \bar\bar{k}_{m-1}, \ldots, \bar\bar{k}_1, \bar{k}_1, \bar\bar{k}_0),
$$
which in the numeral system generated by multi-index $\alpha$ has digital representation $(k_{n-1}, \ldots, k_1, k_0)$, which is the inversion of the input digital representation $k_{\alpha^{\star}}$. Therefore $Q_{n-1}\cdots Q_0 = S_{\alpha}$.

{\bf 2.} Let number $n$ be even. Then $m{+}1=n/2$ and operator $Q_{n/2}$ performs the following transformation of the digital form of indices:
$$
(\ldots, \bar\bar{k}_m, \bar\bar{k}_{m+1}, \bar{k}_{m+1}, \ldots) \rightarrow (\ldots, \bar\bar{k}_{m+1}, \bar{k}_{m+1}, \bar\bar{k}_m, \ldots).
$$
Ellipsis marks digits that do not change under permutation. On the left side these are $m$ pairs of digits of the form $(\bar\bar{k}_{m-j}, \bar{k}_{m-j+1})$ for $j = m, m{-}1, \ldots, 1$. On the right side these are $m$ pairs of digits of the form $(\bar\bar{k}_{m+j}, \bar{k}_{m+j})$ for $j=1, \ldots, m$. As in the case of odd $n$, operation $Q_{m+j}$ permutes these pairs of digits and inverts the order in the right pair. As a result, vector $s_2$ is again obtained. \square

\vspace{3mm}
The sequence of butterfly execution at each stage determines the necessary volume of internal memory and pipeline length. The sequence of butterfly execution in the self-sorting algorithm must be changed to avoid conflicts when accessing memory.

Let $\alpha=(\alpha_{n-1}, \ldots, \alpha_1, \alpha_0)=(R, \ldots, R, r)$ be a multi-index indicating the sequence of FFT stage execution, starting from the stage with butterflies by radix $\alpha_0=r$ and up to the stage with butterflies by radix $\alpha_{n-1}=R$. 
In accordance with theorem \ref{FFT_iter_freqW}, numbers of input vector elements are written in the numeral system generated by the inverse multi-index $\alpha^{\star} = (r, R, \ldots, R)$. The digital form of number $k$ is
$$
s = (k_0, k_1, \ldots, k_{n-1}), \qquad k=s^{\alpha^{\star}} = k_{n-1} + R(k_{n-2} + \ldots + R(k_1 + Rk_0) \ldots ),
$$
where $0\le k_0<r$ and $0\le k_i<R$ for $1\le i\le n-1$. As in the proof of theorem \ref{self_sort}, split each component: $k_i=\bar\bar{k}_i q + \bar{k}_i$, where $0\le \bar\bar{k}_i < r$, $0\le \bar{k}_i < q$.

Let $i\ge (n{+}1)/2$ be the FFT stage number. From the proof of theorem \ref{self_sort} it follows that at the $i$-th stage, numbers of array components have digital representation
$$
p = (\bar\bar{k}_0, \bar{k}_1, \bar\bar{k}_1, \ldots, \bar{k}_{n-1}, \bar\bar{k}_{n-1})
$$
in the numeral system generated by multi-index $\gamma=(r,q,r,\ldots,q,r)$. To the input of each butterfly is supplied a set of samples, whose numbers have all components $\bar{k_j}$ and $\bar\bar{k}_j$ the same, except components $(\bar{k}_i, \bar\bar{k}_i)$, which run through all possible values. Therefore, the natural number of the butterfly is $k = (p^i)^{\gamma_0}$, where $\gamma_0=(r,q,r,\ldots,q,r)$ is shorter than vector $\gamma$ by two components and
$$
p^i = (\bar\bar{k}_0, \bar{k}_1, \bar\bar{k}_1, \ldots, \bar{k}_{i-1}, \bar\bar{k}_{i-1}, \bar{k}_{i+1}, \bar\bar{k}_{i+1}, \ldots,  \bar{k}_{n-1}, \bar\bar{k}_{n-1}).
$$
The cycle number at which the butterfly at stage $i$ with number $k=(p^i)^{\gamma_0}$ is executed will be denoted $T_i(k)$. 

Operation $Q_i$ consists in inverting pair $(\bar\bar{k}_{n-1-i}, \bar{k}_{n-i})$ and permuting the obtained components with pair $(\bar{k}_i, \bar\bar{k}_i)$. Define operation of permuting this pair to the end of multi-index $p^i$ of butterfly number $k=(p^i)^{\gamma_0}$:
$$
U_i(p^i) = (\bar\bar{k}_0, \bar{k}_1, \ldots, \bar{k}_{n-1-i}, \bar\bar{k}_{n-i}, \ldots,\bar{k}_{i-1}, \bar\bar{k}_{i-1}, \bar{k}_{i+1}, \bar\bar{k}_{i+1}, \ldots, \bar{k}_{n-1}, \bar\bar{k}_{n-1}, \bar{k}_{n-i}, \bar\bar{k}_{n-1-i})
$$
for $i\ge (n{+}1)/2$.

If number $n$ is even and $n>2$, then operation $Q_m$ for $m=n/2$ consists in permutation: 
$$
(\bar\bar{k}_{m-1}, \bar\bar{k}_m, \bar{k}_m) \rightarrow (\bar\bar{k}_m, \bar{k}_m, \bar\bar{k}_{m-1}).
$$
Therefore, define permutation in the digital representation of butterfly number:
$$
U_m(p^m) = (\bar\bar{k}_0, \bar{k}_1, \ldots, \bar{k}_{m-2}, \bar\bar{k}_{m-2}, \bar{k}_{m+1}, \bar\bar{k}_{m+1}, \ldots, \bar{k}_{n-1}, \bar\bar{k}_{n-1}, \bar{k}_{m-1}, \bar\bar{k}_{m-1}).
$$

\begin{theorem}
Let $n>2$ and the distribution function of the input vector across data banks $m(n)$ be determined by theorem \ref{theorem:fft_1r1w}. Order of butterfly traversal at the $i$-th FFT stage is given by cycle number $T_i(k)$ of execution of butterfly with number $k=(p^i)^{\gamma_0}$:
$$
T_i(k) = \left\{\begin{array} {ll} 
\left\lfloor \frac{k}{q} \right\rfloor, & i=0, \\
k, & 1\le i \le \lfloor \frac{n-1}{2} \rfloor, \\ 
(U_i p^i)^{\gamma_0}, & \lfloor \frac{n+1}{2} \rfloor \le i\le n-1.  
\end{array} \right. 
$$

Suppose the pipeline length satisfies condition
$$
p \ge R-1.
$$

Then such a choice of traversal order and bank distribution function ensures
absence of conflicts when operating the self-sorting algorithm for the architecture
of a streaming FFT accelerator with $R$ banks of 1r1w memory when executing one
butterfly of size $R$ or $q$ butterflies of size $r$ per cycle. 
\end{theorem}

\proof
Initial stage $G_0=E_{0,\alpha}$ is the same as in theorem \ref{theorem:fft_1r1w}. 
Therefore it does not cause conflicts even without a pipeline. 
Butterflies of the remaining FFT stages have size $R$, and therefore all elements of the input of one butterfly are in different memory banks. 
It is required to prove that at each stage, writing results of butterflies to each memory cell occurs after reading from this cell at this stage. 

In stages $G_i=Q_i E_{i,\alpha}$ for $1\le i\le \lfloor (n{-}1)/2 \rfloor $, permutation $G_k$ changes only index $k_i$, 
which does not enter the butterfly number. Therefore, these stages are also executed in place and do not cause a conflict in memory access even without a pipeline.

Let $(n+1)/2 \le i\le n-1$. Divide all butterflies of the $i$-th stage into consecutive groups of $R$ pieces in accordance with ordinal numbers $T_i(k)$. Then in each group, all butterflies are processed, in whose numbers $k$ components $\bar{k}_{n-i}$, $\bar\bar{k}_{n-1-i}$ run through all possible values, and the remaining components of the digital representation are fixed. 
By definition of permutation $Q_i$, exactly to this set of memory cells are written the results of processing a group of $R$ butterflies. 
For pipeline length $p\ge R-1$, writing occurs after reading.

If number $n$ is even and $i=n/2$, then a group of $R$ butterflies is determined by components $\bar{k}_{n-i-1}$, 
$\bar\bar{k}_{n-1-i}$, which run through all possible values. 
It is important that portions of $r$ butterflies, determined by component $\bar\bar{k}_{n-1-i}$, 
are processed completely for each value of $\bar{k}_{n-i-1}$. 
Exactly to the memory read in butterflies of these small portions, writing of the result occurs in accordance with the definition of operation $Q_{n/2}$. \square  

\section{Prototyping Results}

Prototyping showed that when using a register file with width
16 bits per real value and a computational block with butterfly size 4,
memory area at size 128 Kibit constitutes about 80\% of accelerator
area. Changing address arithmetic to use dual-port
memory does not substantially change computational block area. Thus,
reducing memory area by 2 times by using 
static dual-port memory
allows reducing accelerator area by 40\%, and reduction by 6 times by
using embedded dynamic memory by 67\%, with proportional
reduction in static energy consumption. Synthesis of virtual
topology was performed for a low-power semiconductor
fabrication process with geometric norms 22 nm. 

\clearpage			
\chapter{Accelerating the Solution of the Yule--Walker Equations}
\label{chapt4}

This chapter considers hardware acceleration of the solution of Yule--Walker equations
based on the fast Schur algorithm using a hardware
FFT computation block.

First, an applied problem is described in which fast inversion of Toeplitz matrices is of great importance.

\section{Far Echo Suppression Using a Linear Filter with a Long Impulse Response}

In conference systems, a terminal has both a microphone and a speaker; the speaker reproduces speech from a remote speaker, 
while the microphone receives both the local speaker's speech and the signal from the speaker. This creates positive 
feedback.

The speaker's environment usually creates acoustic echo. Near echo, created by the local 
environment of the speaker, and far echo, arising from the reproduction of the speaker's speech at a remote terminal, are distinguished. 
Near echo is acceptable, while far echo participates in feedback formation and must be compensated.

For far echo compensation, a multiband adaptive equalizer is usually used. Since the equalizer does not 
take into account the signal phase, a half-duplex effect arises, when the speaker's 
speech cannot be interrupted, since the local speaker's voice is suppressed by the equalizer.

To avoid this effect, it is necessary to accurately estimate the echo transfer function to ensure
transmission of the local speaker's voice without distortion.

\begin{figure}[htbp]
\begin{center}
    \includegraphics[width=0.8\textwidth]{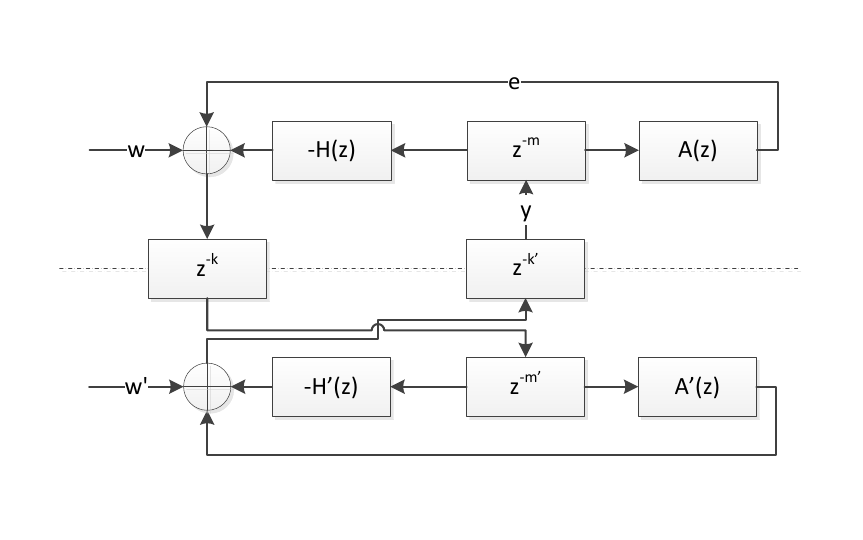}
\caption{Echo suppression model}
\label{f:Model}
\end{center}
\end{figure}

This work uses an echo suppression model using internal negative feedback, as shown in
Fig.~\ref{f:Model}.

Here $w$ is the local speaker's signal, $e$ is the echo signal, $y$ is the signal from the remote terminal. $H(z)$ is the transfer function
of the echo suppression filter, $A(z)$ is the transfer function of the filter modeling the acoustic feedback.  The transfer function
$A(z)$ is unknown, $z^{-m}$ is the delay of sound output by $m$ samples, $z^{-k}$ is the delay of the data transmission channel. 

It is assumed that the signals $w$ and $y$ are uncorrelated. In this case, the optimal filter $H$ equals $A$ and thus compensates for the feedback.

Let $x=e + w$ denote the microphone signal.

In the standard problem of optimal filter synthesis, given sequences $x=(x(t))_{t=0}^{N-1}$ and $y=(y(t))_{t=0}^{N-1}$, 
as well as a number $n\ll N$, it is required to find a sequence of numbers $h = (h_i)_{i=0}^{n-1}$ that minimizes the quality functional
$$
J(h) = \sum_{t=0}^{N+n-1} |x(t) - (h*y)(t)|^2,
$$
where $h*y$ denotes convolution, and the values $y(t)$ and $x(t)$ for $t<0$ or $t\ge N$ are considered zero. 

If $x$ and $y$ are sample realizations of a stationary random process, then the optimal sequence $h$ is close for large $N$ to the impulse response of the optimal linear filter.

The problem is solved by the least squares method. The vector $h$ is determined from the equation
$$
Th = c, 
$$
where the matrix $T = (T_{ij})_{i,j=0}^{n-1}$ and the vector $c=(c_i)_{i=0}^{n-1}$ have components
$$
T_{ij} = \frac{1}{N}\sum_{t=0}^{N-1} y(t)y(t-i+j), \qquad c_i = \frac{1}{N}\sum_{t=0}^{N-1} x(t)y(t-i).
$$
For ergodic processes, the sample matrix $T$ tends to the autocorrelation matrix, and the vector $c$ tends to the cross-correlation values between $x$ and $y$. 
In this case, the synthesized filter can be applied to estimate the process $x$ from $y$ at subsequent time moments, that is, for $t>N$.

For small $n$, the problem is easily solved by inverting the matrix $T$. 
However, for large values of $n$, the complexity of finding the filter coefficient vector $h$ and computing 
the convolution $h*y$ becomes determining for the possibility of real-time computation.

To invert the matrix $T$, the fast Schur algorithm~\cite{Ammar87thegeneralized} is used for factorization of $T^{-1}$ based on FFT.
For convolution, Gardner's algorithm~\cite{gardner1994efficient} is used, also based on FFT.

\section{Inversion of a Toeplitz Matrix Using Szego and Schur Polynomials}

This section presents the main mathematical results on inversion of Toeplitz matrices \cite{VoevTyrt,Ammar87thegeneralized}, which will be needed later to calculate the complexity of algorithms and their acceleration.

\subsection{Szego Polynomials and the Factorization of the Inverse of a Toeplitz Matrix}

It is required to quickly solve the linear equation $Tx=e_0$ for $x=(x_k)_{k=0}^n\in {\sf R}^{n+1}$, where $e_0=(1, 0, \ldots, 0)^*$ is 
the first unit vector, and the square matrix $T$ is Toeplitz,
$$
T = \begin{pmatrix} t_0 & t_1 & t_2 & \cdots & t_n \\
t_1 & t_0 & t_1 & \ldots & t_{n-1} \\
t_2 & t_1 & t_0 & \ldots & t_{n-2} \\
\vdots & & \ddots & \ddots & \vdots \\
t_n & t_{n-1} & t_{n-2} & \cdots & t_0
\end{pmatrix}.
$$

In applications, this matrix is usually nonsingular and positive definite. In particular, this occurs when the numbers $(t_k)_{k=0}^n$ 
are values of the correlation function of some regular stationary process. In this case, $x$ is the solution of the
Yule--Walker equation. For small dimensions $n$, this solution can be computed by the usual Levinson--Durbin method, which is done in
speech coding in mobile phones.  In particular, $n=10$ in the GSM standard. 

In what follows, for brevity of formulas, we will assume that $t_0=1$. Any Toeplitz matrix reduces to this case by multiplication by $t_0^{-1}$.
 
When reducing the dimension $n$ in the equation, it is sufficient to perform
truncation of the matrix $T$ and reduce the dimension of $e_0$. The truncated matrix of dimension $m+1$ will be denoted by $T_m$, and the truncated right-hand 
side by the vector $e_{0,m}$. Suppose that $x_0\ne 0$ for all $m$, and denote $d_m = x_0$. Introduce the normalized solution 
$\bar{\psi}_m = (\psi_{m,k})_{k=0}^m = x/x_0$. Thus, $T_m\bar{\psi}_m = e_{0,m} d_m$ for $0\le m\le n$.

Form the matrices
$$
R = \begin{pmatrix} 1 & \psi_{1,1} & \psi_{2,2} & \cdots & \psi_{n,n} \\
0 & 1 & \psi_{2,1} & \cdots & \psi_{n,n-1} \\
0 & 0 & 1 & \cdots & \psi_{n,n-2} \\
\vdots & \vdots & \vdots & \ddots & \vdots \\
0 & 0 & 0 & \cdots & 1 \end{pmatrix}, \qquad
D = \begin{pmatrix} d_0 & 0 & 0 & \cdots 0 \\
0 & d_1 & 0 & \cdots & 0 \\
0 & 0 & d_2 & \cdots & 0 \\
\vdots & \vdots & \vdots & \ddots & \vdots \\
0 & 0 & 0 & \cdots & d_n
\end{pmatrix}.
$$
Then, by properties of the Yule--Walker equation, $R^*TR = D$. 

For each solution $\psi_m$ of the Yule--Walker equation of order $m+1$, define a polynomial of degree $m$ 
$$
\psi_m(z) = 1 + \psi_{m,1} z + \cdots + \psi_{m m}z^m, \qquad m\ge 0,
$$
and a polynomial in inverse powers 
$$
\widetilde{\psi}_m(z) = \psi_{m m} + \psi_{m,m-1} z + \cdots + \psi_{m 1} z^{m-1} + z^m, \qquad m\ge 0.
$$

From the elements of the first column of the Toeplitz matrix $T$, define a function having the meaning of spectral density
$$
t(z) = 1 + t_1 (z+z^{-1}) + t_2(z^2 + z^{-2}) + \cdots.
$$
In this notation, the Yule--Walker equation can be written as
$$
\int_{|z|=1} t(z) \psi_m (z) z^{-k}\,dm(z) = 0, \qquad 0\le k\le m-1,
$$
where $dm(z)=dz/(2\pi i z)$ is the normalized Lebesgue measure on the unit circle. 

Since the polynomial $\psi_j$ has degree at most $j$, we have
$$
\int_{|z|=1} t(z) \widetilde{\psi}_m (z) (\widetilde{\psi}_k(z))^*\,dm(z) = 0, \qquad \forall \, k\ne m,
$$
which can be interpreted as orthogonality of the polynomials $\psi_j$ in the quasimetric generated by the kernel $t(\cdot)$. The norm of the polynomial equals
$$
\int_{|z|=1} t(z) |\widetilde{\psi}_k (z)|^2\,dm(z) = \int_{|z|=1} t(z) |\psi_k (z)|^2\,dm(z) = d_k.
$$

\begin{definition}
The polynomial $\psi_k(z)$ is called the Szego polynomial of order $k$ for the Toeplitz matrix $T$. 
\end{definition}

\vspace{3mm}
Szego polynomials have the following properties:
\begin{enumerate}
\item Orthogonality in the metric generated by the function $t(\cdot)$:
$$
\int_{|z|=1} t(z) \widetilde{\psi}_m (z) (\widetilde{\psi}_k(z))^*\,dm(z) = \delta_{k,m} d_k, \qquad k,m = 0,1, \ldots,
$$
where $\delta_{k,m}$ is the Kronecker symbol. 
\item Factorization of the inverse matrix.
The columns of coefficients of Szego polynomials generate an upper triangular matrix $R$ with ones on
the main diagonal, which factorizes the inverse matrix by Cholesky:
$$
T^{-1} = R D^{-1} R^*,
$$
where the matrix $D=\diag\{d_0, d_1, \ldots\}$ is diagonal. 
\item Recurrence formulas.
$$
\begin{pmatrix} \psi_m(z) \\ \widetilde{\psi}_m(z) \end{pmatrix} = 
\begin{pmatrix} 1 & \gamma_m z \\ \gamma_m & z \end{pmatrix}
\begin{pmatrix} \psi_{m-1}(z) \\ \widetilde{\psi}_{m-1}(z) \end{pmatrix}, \qquad m=1,2,\ldots
$$
with initial data 
$$
\begin{pmatrix} \psi_0(z) \\ \widetilde{\psi}_0(z) \end{pmatrix} = \begin{pmatrix} 1\\ 1 \end{pmatrix}.
$$
The quantities $\gamma_m$ are determined from the equations
\begin{eqnarray*}
\gamma_m & = & -\frac{1}{d_{m-1}}\left(t_m + \sum_{k=1}^{m-1} t_{m-k}\psi_{m-1,k}\right), \\
d_m & = & d_{m-1} (1 - |\gamma_m|^2).
\end{eqnarray*}
\item The quantities $d_k$ are explicitly expressed through the coefficients $\gamma_j$:
$$
d_k = \prod_{j=1}^k (1-|\gamma_j|^2).
$$
\end{enumerate}

\begin{definition}
The quantities $\gamma_k$ are called reflection coefficients, and the quantities $d_k$ are called residual variances for the sequence $(t_k)_{k=0}^{\infty}$.
\end{definition}

\subsection{The Schur Coefficient Problem}

A Schur function is an analytic function in the open unit disk ${\sf D}=\{z \, :\, |z|<1\,\}$ that takes values in
the closed disk $\bar{\sf D}$. 
Schur posed and solved the following problem (the coefficient problem for bounded analytic functions): given 
a set of complex numbers $c=(c_0, c_1,\ldots, c_{n-1})$, it is required to determine whether there exists a Schur function whose set of first 
Taylor coefficients coincides with the given set $c$. 

The solution is based on properties of the fractional-linear Mobius transformation
$$
z \rightarrow M_{\gamma}(z) = \frac{z+\gamma}{1 + \gamma^* z},
$$
where $\gamma\in {\sf D}$ is a given number. Here and below, $\gamma^*$ denotes the complex conjugate of $\gamma$. 

The Mobius transformation is invariant on the set of Schur functions: if $\phi$ is a Schur function,
then $M_\gamma (\phi)$ is also a Schur function, and vice versa. The transformation $M_\gamma$ has an inverse
$$
M_\gamma^{-1}(z) = \frac{z-\gamma}{1 - \gamma^* z},
$$
which is also a Mobius transformation. 

Let $\phi$ be a Schur function. Define $\phi_0=\phi$ and then by induction 
$$
\gamma_k=\phi_{k-1}(0), \qquad \phi_k(z) = \frac{1}{z} (M^{-1}_{\gamma_k} (\phi_{k-1}))(z), \qquad k=1,2,\ldots
$$
Obviously, $\phi_k$ is a Schur function for all $k$. If for some $k=N$ the function $\phi_N$ is a constant whose modulus equals $1$,
then further transformations are not applied. This is only possible when $\phi$ is a Blaschke product of order $N$.
For all other Schur functions, $|\gamma_k|<1$ for all $k$. The numbers $\gamma_k$ are called {\it Schur parameters} of the function $\phi$.

The inverse transformation is defined by the formula
$$
\phi_{k-1}(z) = \frac{\gamma_k + z\phi_k(z)} {1 + \gamma_k^* z \phi_k(z)}, \qquad |z|<1.
$$
Since $|\gamma_k|<1$ and $|\phi_k(z)|\le 1$, the Taylor series converges in the disk ${\sf D}$
$$
\phi_{k-1}(z) = (\gamma_k + z\phi_k(z)) \sum_{j=0}^\infty (-\gamma_k^* z \phi_k(z))^j =
\gamma_k + z (1-|\gamma_k|^2)\phi_k(z) + \cdots,
$$
each term of which is a polynomial in $z$ and $\phi_k(z)$. From this, by induction, it follows that the first $n$ Taylor coefficients 
$(c_k)_{k=0}^{n-1}$ of the function $\phi$ do not depend on $\phi_n$ and are completely determined by the Schur parameters $(\gamma_k)_{k=1}^n$. Conversely, 
for all $k\ge 1$, the number $\gamma_k$ is completely determined by the set of Taylor coefficients $(c_0, \ldots, c_{k-1})$. 

Thus, the condition $|\gamma_k|\le 1$ for $1\le k\le n$ is necessary and sufficient for the solvability of the posed coefficient problem.
Suppose it is satisfied. 

Substituting for the function $\phi_{j-1}$ its expression through $\phi_j$ for $j=1,2,\ldots,k$, we obtain that for any $k> 0$ 
$$
\phi(z) = \frac{\xi_k(z) + \widetilde{\eta}_k(z) \phi_k(z) } {\eta_k(z) + \widetilde{\xi}_k(z) \phi_k(z)},
$$
where $\xi_k$, $\eta_k$ are polynomials of degree at most $k-1$ and the polynomials $\widetilde{\xi}_k$, $\widetilde{\eta}_k$, as above, are written in inverse powers:
$$
\widetilde{\eta}_k(z) = z^k \eta_k(z^{-1}), \qquad \widetilde{\xi}_k(z) = z^k \xi_k(z^{-1}).
$$
Since 
$$
\phi(z) = \frac{\xi_k(z) + \widetilde{\eta}_k(z) \phi_k(z) } {\eta_k(z) + \widetilde{\xi}_k(z) \phi_k(z)} = 
\frac{\xi_{k-1}(z) + \widetilde{\eta}_{k-1}(z) \phi_{k-1}(z) } {\eta_{k-1}(z) + \widetilde{\xi}_{k-1}(z) \phi_{k-1}(z)},
$$
$$
\phi_{k-1}(z) = \frac{\gamma_k + z\phi_k(z)} {1 + \gamma_k^* z \phi_k(z)},
$$
these polynomials satisfy the recurrence equation
$$
\begin{pmatrix} \eta_k & \xi_k \\ \widetilde{\xi}_k & \widetilde{\eta}_k \end{pmatrix} = 
\begin{pmatrix} 1 & \gamma_k \\ \gamma_k^* z & z \end{pmatrix}
\begin{pmatrix} \eta_{k-1} & \xi_{k-1} \\ \widetilde{\xi}_{k-1} & \widetilde{\eta}_{k-1} \end{pmatrix}
$$
with initial data 
$$
\begin{pmatrix} \eta_0 & \xi_0 \\ \widetilde{\xi}_0 & \widetilde{\eta}_0 \end{pmatrix} = \begin{pmatrix} 1 & 0 \\ 0 & 1 \end{pmatrix}.
$$

\begin{theorem} \label{Coef_Schur}
For the coefficient problem for a set of numbers $(c_k)_{k=0}^{n-1}$ to be solvable, it is necessary and sufficient that all 
Schur parameters $(\gamma_k)_{k=1}^n$ of the polynomial $c(z)=c_0+c_1 z + \cdots + c_{n-1} z^{n-1}$ be at most $1$. 

Suppose this condition is satisfied. Then there exists a pair of polynomials $(\xi_n, \eta_n)$ of degree at most $n-1$ such that the set of all solutions
of the coefficient problem coincides with the set of functions of the form
$$
\phi(z) = \frac{\xi_n(z) + \widetilde{\eta}_n(z) \tau(z) } {\eta_n(z) + \widetilde{\xi}_n(z) \tau(z)},
$$
where $\tau(\cdot)$ is an arbitrary Schur function, $\xi_n(z)=z^n\xi(z^{-1})$, $\eta_n(z)=z^n\eta(z^{-1})$.
\end{theorem}

{\sf Definitions.} {\it

1. The polynomials $\xi_k$, $\eta_k$ are called Schur polynomials. They are completely determined by the Schur parameters $\gamma_j$
for $1\le j\le k$ or by the first $k$ Taylor coefficients of the function $\phi$.

2. Let $\phi$ be a Schur function and let $(\xi_n, \eta_n)$ be its Schur polynomials of order $n$,
$$
\phi(z) = \frac{\xi_n(z) + \widetilde{\eta}_n(z) \phi_n(z) } {\eta_n(z) + \widetilde{\xi}_n(z) \phi_n(z)}.
$$
Then the function $\phi_n$ is called the $n$th residual term of the function $\phi$, and the fraction $T_n \phi = \xi_n/\eta_n$ is called the $n$th 
approximation of the function $\phi$.
}

\vspace{3mm}
From the equality of determinants in the recurrence equation, it follows that
$$
\eta_n(z) \widetilde{\eta}_n(z) - \xi_n(z) \widetilde{\xi}_n(z) = z^n \delta_n, \qquad \delta_n = \prod_{k=1}^n (1-|\gamma_k|^2).
$$
This equality can also be rewritten as
$$
\eta_n(z) \eta_n(z^{-1}) - \xi_n(z) \xi_n(z^{-1}) = \delta_n.
$$
The independence of the first $n$ Taylor coefficients of the function $\phi$ from its residual term $\phi_n$ follows directly from the formula
$$
\phi(z) - (T_n\phi)(z) = \frac{z^n \delta_n \phi_n(z)}{\eta_n(z)(\eta_n(z) + \widetilde{\xi}_n(z) \phi_n(z))}.
$$

\subsection{Spectral Densities and Schur Functions}

Let $(r_k)_{k=0}^\infty$ be the correlation function of some stationary regular random process. The spectral density
of this process $S(z)$ is defined as the analytic continuation from the unit circle of the function
$$
S(z) = r_0 + r_1z + \bar{r}_1z^{-1} + r_2 z^2 + \bar{r}_2 z^{-2} + \cdots
$$
Obviously, if $|z|=1$, then $S(z)$ is a real number, which is nonnegative by properties of the correlation function. 

\begin{lemma} \label{Corr}
Let $(r_k)_{k=0}^\infty$ be a sequence of complex numbers and let the number $r_0$ be real and positive. Suppose that in the closed unit
disk the series
$$
R(z) = r_0 + r_1z + r_2 z^2 + \cdots
$$
converges.
Then for the function
$$
S(z) = r_0 + r_1z + \bar{r}_1z^{-1} + r_2 z^2 + \bar{r}_2 z^{-2} + \cdots = 2\Re R(z) - r_0
$$
to be nonnegatively definite on the unit circle, it is necessary and sufficient that the function
$$
\phi(z) = 1 - \frac{r_0}{R(z)}
$$
be a Schur function.
\end{lemma}

{\sf Proof.} If $\phi$ is a Schur function, then $R(z)\ne 0$ in the closed unit disk $\bar{\sf D}$. Conversely, if $S(z)\ge 0$
for $|z|=1$, then the set of values of the function $R(z)$ for $|z|=1$ has real part not less than $r_0/2>0$. By the maximum principle,
an analytic function $R$ in the unit disk takes values in the convex hull of values on the unit circle. Therefore,
$\Re R(z)\ge r_0/2 >0$ for $|z|\le 1$.

Further, we use the following simple statement: if $u$ is a complex number and $a>0$ is a real number, then the statements
$\Re u \ge 0$ and $|u-a|\le |u+a|$ are equivalent. Choose $a=r_0$ and $u = 2R(z)-r_0$ for $|z|\le 1$. Then the inequality $\Re u=S(z)\ge 0$ 
is equivalent to the inequality
$$
1\ge \left|\frac{2 R(z)-2r_0}{2R(z)}\right| = \left|1 - \frac{r_0}{R(z)}\right| = |\phi(z)|,
$$
which means that $\phi$ is a Schur function. \square

\begin{lemma} \label{z_phi}
Let the function $\phi(z)$ be analytic at zero. Then the functions $\phi(z)$ and $z\phi(z)$ can be Schur functions only simultaneously.
\end{lemma}

\proof Obviously, if $\phi(z)$ is a Schur function, then $z\phi(z)$ is also a Schur function, since $|z\phi(z)|\le |\phi(z)|\le 1$ for $|z|\le 1$.

Conversely, let $\psi(z)=z\phi(z)$ be a Schur function. Then the function $\phi(z)=\psi(z)/z$ is analytic in the unit disk. Consequently, $|\psi(z)|$ on the unit disk attains its maximum on the boundary, where $|z|=1$ and, therefore, $|\phi(z)| = |\psi(z)|\le 1$. \square

\vspace{3mm}
Thus, if $(r_k)_{k=0}^\infty$ is the correlation function of some stationary random process, then 
the calculation of Schur parameters can be performed for the function
$$
\phi(z) = \frac{\phantom{r_0 + {}} r_1\phantom{z} + r_2 z + \cdots}{r_0 + r_1 z + r_2 z^2 + \cdots}\; .
$$

\subsection{Relationship between Szego and Schur Polynomials}

\begin{theorem} \label{Schur_Sego}
Let the sequence of real numbers $(t_k)_{k=-\infty}^{\infty}$ be positive, i.e., $t_k=t_{-k}$ for all $k$ and the function
$$
S(z) = \sum_{k=-\infty}^{\infty} t_k z^k
$$
is correctly defined and nonnegative on the unit circle, $|z|=1$. 

From the sequence $(t_k)_{k=0}^{\infty}$, construct Toeplitz matrices $T_n$ and for them define Szego polynomials $\psi_n(z)$ and the sequence of reflection coefficients $(\gamma_n^{YW})_{n=1}^{\infty}$. 

Then the function
$$
\phi(z) = - \frac{\phantom{1 + {}} t_1 \phantom{z} + t_2 z\phantom{^2} + \ldots}{1 + t_1 z + t_2 z^2 + \ldots}
$$
is a Schur function, and the corresponding sequence of pairs of Schur polynomials $(\xi_n(z), \eta_n(z))_{n=0}^{\infty}$ and the sequence of Schur parameters $\gamma_n^{\phi}(z)$ satisfy the equations
$$
\psi_n(z) = \eta_n(z) + z \xi_n(z), \qquad \gamma_n^{YW} = \gamma_n^{\phi}.
$$
for all $n\ge 0$. 
\end{theorem}

\proof We proceed by induction on $n$. By definition, $\psi_0(z)=1$, $\eta_0(z)=1$, $\xi_0(z)=0$. Moreover, 
$$
\gamma_1^{\phi} = \phi(0) = - t_1, \qquad
\gamma_1^{YW} = - t_1.
$$
Hence $\psi_0(z) = \eta_0(z) + z \xi_0(z)$ and $\gamma_1^{\phi}=\gamma_1^{YW}$, which constitutes the statement for $n=0$.

Suppose the statement is proved for $n{-}1$, i.e.,
$$
\psi_{n-1}(z) = \eta_{n-1}(z) + z \xi_{n-1}(z), \qquad
\gamma_n^{\phi} = \gamma_n^{YW}.
$$
We prove this statement for $n$.

From the recurrence equation
\begin{eqnarray*}
\eta_n(z) & = & \eta_{n-1}(z) + \gamma_n^{\phi} \widetilde{\xi}_{n-1}(z), \\
\xi_n(z) & = & \xi_{n-1}(z) + \gamma_n^{\phi} \widetilde{\eta}_{n-1}(z)
\end{eqnarray*}
it follows that 
\begin{eqnarray*}
\eta_n(z) + \xi_n(z) & = & \eta_{n-1}(z) + \xi_{n-1}(z) + \gamma_n^{\phi} (\widetilde{\eta}_{n-1}(z) + \widetilde{\xi}_{n-1}(z)) \\
& = & 
\psi_{n-1}(z) + \gamma_n^{\phi}\widetilde{\psi}_{n-1}(z) = 
\psi_{n-1}(z) + \gamma_n^{YW}\widetilde{\psi}_{n-1}(z) = \psi_n(z),
\end{eqnarray*}
where substitutions were made according to the induction hypothesis. This proves the first statement: $\eta_n(z) + \xi_n(z) = \psi_n(z)$.

To prove the second property $\gamma_{n+1}^{YW}= \gamma_{n+1}^{\phi}$, we apply the formula proved above
$$
\phi(z) - \frac{\xi_n(z)}{\eta_n(z)} = z^n \delta_n \frac{\phi_n(z)}{\eta_n(z)(\eta_n(z) + \widetilde{\xi}_n(z) \phi_n(z))}, \qquad
\delta_n = \prod_{k=1}^n (1-|\gamma_k^{\phi}|^2),
$$
where $\phi_n(z)$ is the Schur residual term of order $n$. 

Introduce the notation 
$$
s(z) = t_1 + t_2 z + \ldots,
$$
so that $\phi(z) = - s(z)/(1+zs(z))$. Multiply the previous equation by $\eta_n(z)(1+zs(z))$:
$$
- s(z) \eta_n(z) - (1+zs(z))\xi_n(z) = z^n \delta_n \frac{(1+ zs(z))\phi_n(z)}{\eta_n(z) + \widetilde{\xi}_n(z) \phi_n(z)}.
$$
Write this equality as
$$
s(z)(\eta_n(z) + \xi_n(z)) + \xi_n(z) = -z^n \delta_n \frac{ (1+ zs(z))\phi_n(z)}{\eta_n(z) + \widetilde{\xi}_n(z) \phi_n(z)}.
$$
By the proved property, transform the left-hand side: 
$$
s(z) \psi_n(z) + \xi_n(z) = -z^n \delta_n \frac{ (1+ zs(z))\phi_n(z)}{\eta_n(z) + \widetilde{\xi}_n(z) \phi_n(z)}.
$$

The function on the right-hand side expands into a Taylor series in the closed unit disk. Terms of the series up to degree $n{-}1$ equal zero, and the coefficient at $z^n$ equals
$$
-\delta_n \frac{ \phi_n(0)}{\eta_n(0) + \widetilde{\xi}_n(0) \phi_n(0)} = -\delta_n  \gamma_{n+1}^{\phi},
$$
since $\eta_n(0)=1$, $\widetilde{\xi}_n(0)=0$. 

On the left-hand side, the degree of the polynomial $\xi_n(z)$ is at most $n{-}1$, therefore the coefficient at $z^n$ equals
$$
t_{n+1} + \sum_{k=1}^n t_{n+1-k} \psi_{n,k} = -\delta_n  \gamma_{n+1}^{YW},
$$
where notations have been introduced for the coefficients $\psi_n(z)$ and the residual variance $d_n$:
$$
\psi_n(z) = 1 + \psi_{n,1} z + \cdots + \psi_{n,n} z^n, \qquad 
d_n = \prod_{k=1}^n (1-|\gamma_k^{YW}|^2)
$$
and the definition of the reflection coefficient has been substituted
$$
\gamma_{n+1}^{YW} = -\frac{1}{d_n} \left(t_{n+1} + \sum_{k=1}^n t_{n+1-k} \psi_{n,k}\right).
$$

Equating coefficients at $z^n$, we obtain the required statement $\gamma_{n+1}^{YW} = \gamma_{n+1}^{\phi}$, which completes the proof of the induction step. \square

\begin{corollary} \label{d=delta}
Under the conditions of Lemma \ref{Schur_Sego}, the determinants of matrices consisting of Schur polynomials coincide with the residual variances in the Yule--Walker equation:
$$
d_n = \delta_n = \prod_{k=1}^n (1-|\gamma_k|^2) = z^{-n} \det \begin{pmatrix} \eta_n(z) & \widetilde{\xi}_n(z) \\ \xi_n(z) & \widetilde{\eta}_n(z) \end{pmatrix}.
$$
\end{corollary}

\section{Fast Schur Algorithm}

This section formulates a fast algorithm for computing coefficients of Schur polynomials. 
The general idea of the algorithm was outlined in~\cite{Ammar87thegeneralized}. 
To optimize the volume of computations and memory, this algorithm is presented in the form of recurrent computation on a binary tree. 
All auxiliary statements are formulated and proved, with the help of which the complexity is further calculated. 

\subsection{Transitivity of Schur Polynomials}

The fast algorithm is based on the following observation. Let $\phi_0=\phi$ be the original Schur function, $n>0$ and let $\phi_n$ be the $n$th 
residual term of the function $\phi_0$. Then the Schur parameters $(\gamma_{0,k})_{k=0}^\infty$ for the function $\phi_0$ and the Schur parameters 
$(\gamma_{m,n})_{m=0}^\infty$ of the Schur function $\phi_n$ are related by the equality $\gamma_{0,k+n} = \gamma_{k,n}$ for all $k\ge 0$. 
Indeed, the parameters $\gamma_{0,k+n}$ by definition are computed directly as Schur parameters $\gamma_{n,k}$ 
for the function $\phi_n$. 

A consequence of this observation is the following statement.
For any nonnegative numbers $m<n$, denote the Schur polynomials of order $n-m$ for the $m$th residual term $\phi_m$ of the function $\phi$ by 
$(\xi_{(m,n)}, \eta_{(m,n)})$. Associate with them the matrix
$$
\zeta_{(m,n)} = \begin{pmatrix} \eta_{(m,n)}, & \widetilde{\xi}_{(m,n)} \\ \xi_{(m,n)}, & \widetilde{\eta}_{(m,n)} \end{pmatrix}.
$$
The degree of the polynomials $\xi_{(m,n)}$ and $\eta_{(m,n)}$ is at most $n-m-1$. The polynomials in inverse powers have an additional factor $z$, 
since
$$
\widetilde{\xi}_{(m,n)}(z) = z^{n-m}\xi(z^{-1}), \qquad
\widetilde{\eta}_{(m,n)}(z) = z^{n-m}\eta(z^{-1}).
$$

\begin{lemma} \label{recursive_ksi_eta}
For any $k<m<n$ 
$$
\zeta_{(k,n)} = \zeta_{(k,m)} \zeta_{(m,n)}.
$$
\end{lemma}

{\sf Proof.} Let $k<m<n$ and let the functions $\phi_k$, $\phi_m$, $\phi_n$ be residual terms of the function $\phi$ of corresponding
orders. Schur polynomials are completely determined by the corresponding Schur parameters and relate residual terms as follows:
\begin{eqnarray*}
\phi_k & = & \frac{\xi_{(k,m)} + \widetilde{\eta}_{(k,m)} \phi_m}{\eta_{(k,m)} + \widetilde{\xi}_{(k,m)} \phi_m}, \\
\phi_m & = & \frac{\xi_{(m,n)} + \widetilde{\eta}_{(m,n)} \phi_n}{\eta_{(m,n)} + \widetilde{\xi}_{(m,n)} \phi_n}, \\
\phi_k & = & \frac{\xi_{(k,n)} + \widetilde{\eta}_{(k,n)} \phi_n}{\eta_{(k,n)} + \widetilde{\xi}_{(k,n)} \phi_n}.
\end{eqnarray*}
Eliminating $\phi_m$ from the first equation by substituting the second equation, we obtain a representation of $\phi_k$ as 
a fractional-linear function of $\phi_n$, as in the third equation. 
Schur polynomials do not depend on $\phi_n$, since they are determined by parameters $\gamma_j$ with $j\le n$. Therefore, the corresponding 
coefficients of the fractional-linear representation are proportional. By definition, $\eta_j(0)=1$ for all $j$. Therefore, 
the corresponding coefficients are equal, which is equivalent to the equation in the conclusion of the lemma. \square

\subsection{Transformation of the Coefficients of Schur Functions}

Let the Schur function $\phi$ be rational:
$$
\phi(z) = \frac{b_0(z)}{a_0(z)} = \frac{b_{0,0} + b_{0,1} z + b_{0,2} z^2 + \ldots} 
{a_{0,0} + a_{0,1} z + a_{0,2} z^2 + \ldots}.
$$
From the Mobius transformation, it follows that all residual terms in the Schur algorithm are also rational:
$$
\phi_n(z) = \frac{b_n(z)}{a_n(z)} = \frac{b_{n,0} + b_{n,1} z + b_{n,2} z^2 + \ldots} 
{a_{n,0} + a_{n,1} z + a_{n,2} z^2 + \ldots}, \qquad n=0, 1, 2, \ldots
$$
The coefficients of the numerator and denominator are given up to a common factor. We choose this factor from the condition $a_{n,0}=1$ for all $n$.

\begin{lemma} \label{ab_recursive}
Let $n>m\ge 0$. Then
\begin{eqnarray*}
\begin{pmatrix} a_m (z) \\ b_m(z) \end{pmatrix}
& = & \begin{pmatrix} \eta_{m,n}(z), & \widetilde{\xi}_{m,n}(z) \\ \xi_{m,n}(z) & \widetilde{\eta}_{m,n}(z) \end{pmatrix}
\begin{pmatrix} a_n (z) \\ b_n(z) \end{pmatrix}.
\end{eqnarray*}
\end{lemma}

\proof
From the definition of Schur polynomials, it follows that
$$
\frac{b_m(z)}{a_m(z)} = \phi_m(z) = \frac{\xi_{m,n}(z) + \widetilde{\eta}_{m,n}(z)\phi_n(z)}{\eta_{m,n}(z) + \widetilde{\xi}_{m,n}(z) \phi_n(z)} = \frac{\xi_{m,n}(z) + \widetilde{\eta}_{m,n}(z) \frac{b_n(z)}{a_n(z)}}{\eta_{m,n}(z) + \widetilde{\xi}_{m,n}(z) \frac{b_n(z)}{a_n(z)}}.
$$
After multiplying the numerator and denominator by $a_n(z)$, we obtain the statement of the lemma. The normalization of the polynomial $a_m(z)$ is preserved, since $\widetilde{\xi}_{m,n}(0)=0$, and therefore $a_m(0) = a_n(0) = 1$. \square

\subsection{Structure of the Binary Tree in Computing Schur Parameters}

Let $n=2^p$. Perform the operation of dichotomizing the integer interval $[0, n-1]$. To do this, first divide it into two parts 
$[0,2^{p-1}-1]$ and $[2^{p-1}, 2^p-1]$ and mark them with indices $i_p=0$ and $i_p=1$, respectively. Then divide each of these parts again 
in half and mark with index $i_{p-1}$, for example, the second part into segments $[2^{p-1}, 2^{p-1}+2^{p-2}-1]$ and $[2^{p-1}+2^{p-2}-1, 2^p-1]$. Continuing 
in this way, we obtain for each level of the binary tree $k=p, p-1,\ldots,0$ a partition of the integer interval $[0, n-1]$ into $2^{p-k}$ parts, and each 
of these parts is marked with a multi-index $\alpha=(i_{p-k}, \ldots, i_1)$ of zeros or ones ($i_j\in\{0, 1\}$).

Let $0\le k< p$. The length of each interval with multi-index $\alpha=(i_{p-k}, i_{p-k-1},\ldots, i_1)$ equals $2^k$. Denote its
boundaries $[c_\alpha, d_\alpha]$. Then
$$
c_\alpha = \sum_{j=1}^{p-k} 2^{j-1} i_j, \qquad
d_\alpha = c_\alpha + 2^k-1.
$$
To each interval $[c_\alpha, d_\alpha]$, assign the matrix 
$$
D_\alpha = \zeta_{(c_\alpha, d_\alpha+1)}.
$$
Then, by Lemma \ref{recursive_ksi_eta}, these matrices satisfy the recurrence equation
$$
D_{\alpha} = D_{(\alpha, 0)} D_{(\alpha, 1)}.
$$

Multi-indices form a natural structure of a binary tree. The root corresponds to the empty multi-index at $k=p$ and, accordingly,
the pair of Schur polynomials $(\xi_n, \eta_n)$, which is required to be calculated.
Vertices of the binary tree at $k=0$ correspond to a Schur transformation, which is completely determined by the corresponding 
Schur parameter $\gamma_j$. 

To compute the pair of Schur polynomials at the root of the binary tree, it is sufficient to start from the vertices and, applying the recurrence equation,
sequentially compute the pair of Schur polynomials from level $k$ to level $k+1$ for $k=0, 1, \ldots, p-1$. 
The total number of transformations equals
$$
N = \sum_{k=0}^{p-1} 2^{p-k-1} = 2^p-1 = n-1.
$$
The complexity of this algorithm consists of the complexity of determining all Schur parameters and the complexity of multiplying 
polynomials in the recurrent calculation of elements of the matrices $D$. 

\subsection{Computation of Schur Polynomials from Schur Parameters} \label{Schur_polynom_calc}

Let Schur parameters $(\gamma_j)_{j=1}^n$ be given. It is required to compute Schur polynomials $(\xi_n, \eta_n)$. We will assume that $n=2^p$.
With direct multiplication of polynomials, at each step it is required to perform multiplications in a number proportional to the step
number. Therefore, the total number of multiplications is proportional to $n^2$. Using Fourier transforms, this complexity can be 
significantly reduced.

Substitute Schur parameters into the vertices of the binary tree and perform multiplications, moving toward the root. At level $k$, 
$2^{p-k-1}$ identical operations will be performed. 

Consider a vertex at level $k>0$. It corresponds to a multi-index $\alpha$ of length $k<p$. Suppose the polynomials entering the matrices $D_{(\alpha, 0)}$ and $D_{(\alpha, 1)}$ attached to vertices of level $k-1$ connected to the selected vertex are computed. Denote the corresponding Schur polynomials $(\xi^{(\alpha,0)}, \eta^{(\alpha,0)})$ and $(\xi^{(\alpha,1)}, \eta^{(\alpha,1)})$. The degrees of these polynomials are at most $2^k-1$. It is required to compute polynomials $(\xi^\alpha, \eta^\alpha)$ of degree at most $2^{k+1}-1$ by the formula
\begin{eqnarray*}
\xi^\alpha & = & \xi^{(\alpha,0)} \eta^{(\alpha,1)} + \widetilde{\eta}^{(\alpha,0)} \xi^{(\alpha,1)}, \\
\eta^\alpha & = & \eta^{(\alpha,0)} \eta^{(\alpha,1)} + \widetilde{\xi}^{(\alpha,0)} \xi^{(\alpha,1)}.
\end{eqnarray*}

FFT is used to implement multiplication. The original polynomials have $m=2^k$ coefficients and degree $m-1$, and the result
of multiplying a pair of such polynomials has degree $2m-2$ and, accordingly, $2m-1$ coefficients. Therefore, the result value 
must be computed at $2m$ points on the unit circle.

For an arbitrary polynomial $r$ of degree at most $\ell-1$, denote the set of its values at $\ell $ equally spaced 
points on the unit circle by $F_\ell r$. This is the result of FFT from the set of coefficients of the polynomial $r$ on $\ell $ points. If the degree of $r$ is less than $\ell -1$, then the set of coefficients $\bar{r}$ is padded with zeros to dimension $\ell $, after which FFT is applied.

We assume that in memory are stored not the coefficients of polynomials $\xi^{(\alpha,0)}$, $\xi^{(\alpha,1)}$, $\eta^{(\alpha,0)}$, $\eta^{(\alpha,1)}$, 
but the values of Fourier transforms $F_m\xi^{(\alpha,0)}$, $F_m\xi^{(\alpha,1)}$, $F_m\eta^{(\alpha,0)}$, $F_m\eta^{(\alpha,1)}$, 
each of length $m$. Since the result of polynomial multiplication has length $2m$, these Fourier transforms must first 
be interpolated to $2m$ samples. 

\begin{lemma} \label{edge1}
Let $\alpha$ be a multi-index corresponding to some vertex of level $k>0$. Connected to it are vertices of level $k-1$ having multi-indices $(\alpha,0)$ and $(\alpha,1)$. Denote Schur polynomials at these vertices by $(\eta^{\alpha}, \xi^{\alpha})$, $(\eta^{(\alpha,0)}, \xi^{(\alpha,0)})$ and $(\eta^{(\alpha,1)}, \xi^{(\alpha,1)})$, respectively. 

Let $m=2^k$. Denote the results of DFT on $m$ points
\begin{eqnarray*}
&&(X^{\alpha}_n)_{n=0}^{n-1} = F_m \eta^{\alpha}, \qquad 
(Y^{\alpha}_n)_{n=0}^{n-1} = F_m \xi^{\alpha}, \qquad
(X^{(\alpha,0)}_n)_{n=0}^{n-1} = F_m \eta^{(\alpha,0)}, \\ 
&&
(Y^{(\alpha,0)}_n)_{n=0}^{n-1} = F_m \xi^{(\alpha,0)}, \qquad
(X^{(\alpha,1)}_n)_{n=0}^{n-1} = F_m \eta^{(\alpha,1)}, \qquad 
(Y^{(\alpha,1)}_n)_{n=0}^{n-1} = F_m \xi^{(\alpha,1)}.
\end{eqnarray*}

Then for $0\le n\le m-1$
\begin{eqnarray*}
X^{\alpha}_n & = & X^{(\alpha,0)}_n X^{(\alpha,1)}_n + (-1)^n (Y^{(\alpha,0)}_n)^* Y^{(\alpha,1)}_n, \\
Y^{\alpha}_n & = & Y^{(\alpha,0)}_n X^{(\alpha,1)}_n + (-1)^n (X^{(\alpha,0)}_n)^* Y^{(\alpha,1)}_n.
\end{eqnarray*}
\end{lemma}

\proof
By Lemma \ref{recursive_ksi_eta}
$$
\begin{pmatrix} \eta^{\alpha}(z) & \widetilde{\xi}^{\alpha}(z) \\ \xi^{\alpha}(z) & \widetilde{\eta}^{\alpha}(z) \end{pmatrix} = 
\begin{pmatrix} \eta^{(\alpha,0)}(z) & \widetilde{\xi}^{(\alpha,0)}(z) \\ \xi^{(\alpha,0)}(z) & \widetilde{\eta}^{(\alpha,0)}(z) \end{pmatrix}
\begin{pmatrix} \eta^{(\alpha,1)}(z) & \widetilde{\xi}^{(\alpha,1)}(z) \\ \xi^{(\alpha,1)}(z) & \widetilde{\eta}^{(\alpha,1)}(z) \end{pmatrix}.
$$
The value of DFT on $m$ samples from the set of coefficients of an arbitrary polynomial $r(z)$ of degree less than $m$ is the vector $(r(z_n)_{n=0}^{m-1}$, where $z_n=e^{-\frac{2\pi i}{m}n}$. Choose the first column in the last equation and substitute $z=z_n$:
\begin{eqnarray*}
X^{\alpha}_n & = & X^{(\alpha,0)}_n X^{(\alpha,1)}_n + \widetilde{\xi}^{(\alpha,0)}(z_n) Y^{(\alpha,1)}_n, \\
Y^{\alpha}_n & = & Y^{(\alpha,0)}_n X^{(\alpha,1)}_n + \widetilde{\eta}^{(\alpha,0)}(z_n) Y^{(\alpha,1)}_n.
\end{eqnarray*}
Vertices with multi-indices $(\alpha,0)$ and $(\alpha,1)$ are at level $k-1$, therefore $\widetilde{\eta}^{(\alpha,0)}(z)=z^{k-1} \eta^{(\alpha,0)}(z^{-1})$ and $\widetilde{\xi}^{(\alpha,0)}(z)=z^{k-1} \xi^{(\alpha,0)}(z^{-1})$. Substitution
$$
z_n^{k-1} = e^{-\pi i n} = (-1)^n, \qquad z_n^{-1} = z_n^*
$$
leads to the conclusion of the lemma. \square



\subsection{Computation of Residual Terms}

Schur parameters $\gamma_j$ are not known in advance. In the sequential calculation of these parameters from residual terms 
$\phi_1, \phi_2, \ldots$, it is required to compute all coefficients of numerators and denominators of these functions up to degree $n$. The number
of coefficients at each step is proportional to $n$, and the number of steps equals $n$. Therefore, the complexity is proportional to $n^2$. 
When using a binary tree of Schur transformations, the length of each residual term decreases, which significantly reduces
the volume of computations. 

Let $\alpha$ be a multi-index of length $p-k$, determining a vertex in the binary tree at level $k$. Associated with this vertex is an integer 
segment $[m_1, m_2]$ of length $2^k$ on the segment $[0,n-1]$. We will compute the first $m_2-m_1$ coefficients of the Schur function $\phi_{m_1}$,
which is the residual term of order $m_1$ of the original function $\phi$. 

Arcs of the binary tree leaving from one vertex are labeled with digits $0$ or $1$. We will call them, respectively, left and right 
branches. When moving only along left branches toward terminal vertices, the initial index of the interval $m_1$ does not change,
therefore the number of the residual term also does not change, and the length of the set of coefficients of the numerator and denominator decreases. 
Consequently, in such movement it is not required to recompute these coefficients. 
When shifting along the right branch, it is necessary to recalculate the parameters of the residual term.

At the initial moment, the set of coefficients in the root vertex is known and, consequently, in all vertices reachable from it
along left branches. The traversal of the binary tree will be performed in lexicographic order. At each moment in time, in all vertices
located above the analyzed one, computed coefficients of residual terms will be stored. Simultaneously, in each vertex
of the already traversed subtree, i.e., in each vertex lying to the left of the path from the current vertex to the root vertex, the 
computed pair of Schur polynomials will be stored. 

When moving along an arc labeled zero, Schur residual terms do not change, but Schur polynomials change, since the order of polynomials changes. When moving along an arc labeled one, Schur residual terms also change. The method of calculating numerators and denominators of Schur residual terms is based on the following statement. 

\begin{lemma} \label{edge0}
Let $\alpha$ be a multi-index of some vertex of level $k+1>0$. Connected to it are vertices of level $k$ with multi-indices $(\alpha, 0)$ and $(\alpha,1)$. At vertex $(\alpha,0)$, denote Schur polynomials by $(\xi^{(\alpha, 0)}, \eta^{(\alpha,0)})$, and the residual variance by $\delta^{(\alpha, 0)}$. At vertices $(\alpha, 0)$ and $(\alpha,1)$, denote Schur residual terms by $\phi^{\alpha}(z)=b^{\alpha}(z)/a^{\alpha}(z)$ and $\phi^{(\alpha,1)}(z)=b^{(\alpha,1)}(z)/a^{(\alpha,1)}(z)$, respectively. Then
$$
\begin{pmatrix} a^{(\alpha,1)} \\ b^{(\alpha,1)} \end{pmatrix} = \frac{1}{z^N \delta^{(\alpha, 0)}}
\begin{pmatrix} \widetilde{\eta}^{(\alpha, 0)}, & -\widetilde{\xi}^{(\alpha, 0)} \\ -\xi^{(\alpha, 0)}, & \eta^{(\alpha, 0)} \end{pmatrix}
\begin{pmatrix} a^\alpha \\ b^\alpha \end{pmatrix},
$$
where $N=2^k$.
\end{lemma}

\proof Let $j$ be a natural number whose binary representation is the multi-index $\alpha$. Then vertices with multi-indices $\alpha$ and $(\alpha, 0)$ correspond to the Schur residual term $\phi^{\alpha} = \phi_{2N j}$, and the vertex with multi-index $(\alpha, 1)$ corresponds to the residual term $\phi_{2Nj + N}$. Denote $m=2Nj$, $n=2Nj+N$. Then from the definition of polynomials $(\eta_{m,n}(z), \xi_{m,n}(z))$, it follows that $\xi^{(\alpha, 0)} = \xi_{m,n}$ and $\eta^{(\alpha, 0)} = \eta_{m,n}$. Apply Lemma \ref{ab_recursive}:
$$
\begin{pmatrix} a^\alpha \\ b^\alpha \end{pmatrix} = 
\begin{pmatrix} \eta^{(\alpha, 0)}, & \widetilde{\xi}^{(\alpha, 0)} \\ \xi^{(\alpha, 0)}, & \widetilde{\eta}^{(\alpha, 0)} \end{pmatrix}
\begin{pmatrix} a^{(\alpha,1)} \\ b^{(\alpha,1)} \end{pmatrix}.
$$
From recurrence equations for Schur polynomials, it follows that
$$
\det \begin{pmatrix} \eta^{(\alpha, 0)}, & \widetilde{\xi}^{(\alpha, 0)} \\ \xi^{(\alpha, 0)}, & \widetilde{\eta}^{(\alpha, 0)} \end{pmatrix} = \eta_{m,n} \widetilde{\eta}_{m,n} - \xi_{m,n} \widetilde{\xi}_{m,n} = \delta_{m,n} z^{n-m} = \delta^{(\alpha, 0)} z^N.
$$
Multiplying the previous equation by the inverse matrix leads to the statement of the lemma. \square

\vspace{3mm}
Suppose that at vertex $\alpha$, the FFT results on $2m$ samples $F_{2m}a^\alpha$ and $F_{2m}b^\alpha$ are known. The values 
$F_{2m}\xi^{(\alpha,0)}$ and $F_{2m}\eta^{(\alpha,0)}$ are also computed in advance during interpolation of Schur polynomials, as described in Section
\ref{Schur_polynom_calc}. As a result of multiplication, polynomials of degree $2m-1$ are obtained. It remains to select their part, up to degree 
$m-1$, which is formulated in the following statement. In it, $P_n r$ denotes the vector of coefficients of the polynomial $r(z)$ at powers from $0$ to $n-1$.

\begin{corollary} \label{edge0_FFT}
Let, under the conditions of Lemma \ref{edge0}, $(B^{\alpha}_j)_{j=0}^{2N-1}$ and $(A^{\alpha}_j)_{j=0}^{2N-1}$ be DFT on $2N$ points from $P_{2N}b^{\alpha}$ and $P_{2N}a^{\alpha}$. Let $(X_j)_{j=0}^{2N-1}$ and $(Y_j)_{j=0}^{2N-1}$ be DFT on $2N$ points from sets of coefficients $\eta^{(\alpha,0)}(z)$ and $\xi^{(\alpha,0)}(z)$, respectively. Define
$$
\begin{pmatrix} \widehat{A}_j \\ \widehat{B}_j \end{pmatrix} = \frac{1}{\delta^{(\alpha,0)}} \begin{pmatrix} X_j^* & -Y_j^* \\ -Y_j & X_j \end{pmatrix} \begin{pmatrix} A^{\alpha}_j \\ B^{\alpha}_j \end{pmatrix}, \qquad 0\le j\le 2N-1.
$$
Let $(\widehat{b}_k)_{k=0}^{2N-1}$ and $(\widehat{a}_k)_{k=0}^{2N-1}$ be the result of IDFT from vectors $((-1)^j\widehat{B}_j)_{j=0}^{2N-1}$ and $(\widehat{A}_j)_{j=0}^{2N-1}$, respectively.

Then $P_N(a^{(\alpha,1)}) = (\widehat{a}_k)_{k=0}^{N-1}$ and $P_N(b^{(\alpha,1)}) = (\widehat{b}_k)_{k=0}^{N-1}$.
\end{corollary}

\proof
If the degrees of polynomials $a^{\alpha}$ and $b^{\alpha}$ are greater than $2N-1$, then by Lemma \ref{edge0}, their truncation to degree $2N-1$ affects only coefficients of polynomials $a^{(\alpha,1)}$, $b^{(\alpha,1)}$ at degrees greater than $N-1$. The product 
$$
H(z) = \begin{pmatrix} \widetilde{\eta}^{(\alpha, 0)}, & -\widetilde{\xi}^{(\alpha, 0)} \\ -\xi^{(\alpha, 0)}, & \eta^{(\alpha, 0)} \end{pmatrix}
\begin{pmatrix} P_{2N} a^\alpha \\ P_{2N} b^\alpha \end{pmatrix}
$$
has degree at most $3N-1$, since polynomials $\xi^{(\alpha,0)}$, $\eta^{(\alpha,0)}$, $\widetilde{\eta}^{(\alpha,0)}$, $\widetilde{\xi}^{(\alpha,0)}$ have degrees at most $N$. At the same time, the first $N$ coefficients of the polynomial $H(z)$ equal zero, since after division by $z^N$ a polynomial is obtained by Lemma \ref{edge0}. Consequently, the polynomial $H(z)$ can have nonzero coefficients only at degrees from $N$ to $3N-1$ and 
$$
H(z)=z^N G(z), \qquad \deg G\le 2N-1. 
$$

The first $N$ coefficients of the polynomial $G(z)$ by Lemma \ref{edge0} coincide with vectors $\delta^{(\alpha,0)} \col(a^{(\alpha,1)}_n, b^{(\alpha,1)}_n)$ for $0\le n\le N-1$.  The DFT vector from $G(z)$ on $2N$ points coincides by definition with the vector $(G(z_j))_{j=0}^{2N-1}$, where $z_j=e^{-\frac{2\pi i}{2N}j}$. Therefore, the first $N$ coefficients in IDFT from the vector $(G(z_j))_{j=0}^{2N-1}/\delta^{(\alpha,0)}$ coincide with $P_N(a^{(\alpha,1)}_n, b^{(\alpha,1)}_n)$. It remains to prove that
$G(z_j)/\delta^{(\alpha,0)} = \col(\widehat{A}_j, (-1)^j \widehat{B}_j)$ or that
$$
\frac{1}{z_j^N} H(z_j) = \begin{pmatrix} X_j^* & -Y_j^* \\ -(-1)^jY_j & (-1)^j X_j \end{pmatrix} \begin{pmatrix} A^{\alpha}_j \\ B^{\alpha}_j \end{pmatrix}, \qquad 0\le j\le 2N-1.
$$

Obviously, $(z_j)^N=(-1)^j$. Further, the vertex with multi-index $(\alpha, 0)$ is at level $k$, therefore $\widetilde{\eta}^{(\alpha,0)}(z)=z^N \eta^{(\alpha,0)}(z^{-1})$ and $\widetilde{\eta}^{(\alpha,0)}(z)=z^N \eta^{(\alpha,0)}(z^{-1})$. Consequently, for $0\le j\le 2N-1$
$$
\widetilde{\eta}^{(\alpha,0)}(z_j) = z_j^N \eta^{(\alpha,0)}(z_j^*) = (-1)^j X_j^*, \qquad
\widetilde{\xi}^{(\alpha,0)}(z_j) = z_j^N \xi^{(\alpha,0)}(z_j^*) = (-1)^j Y_j^*,
$$
which proves the statement of the corollary. \square

\subsection{Formulation of the Fast Schur Algorithm}

This section formulates the fast Schur algorithm, based on properties established in previous sections.

Let $n=2^p$ and let the first column $(t_i)_{i=0}^n$ of a self-adjoint Toeplitz matrix $T$ of order $n+1$ be given and $t_0=1$. It is required to find the Szego polynomial $\psi(z)$ of degree at most $n$, generated by the Toeplitz matrix $T$. Suppose that the matrix $T$ is positive definite.

Construct a binary tree of height $p$, whose leaves are numbered by numbers $i=0,1,\ldots, n-1$. Leaves form level zero. Let $1\le k\le p$. Level with number $k$ consists of $2^{p-k}$ vertices. A vertex of level $k$ with number $j$ is marked with a multi-index $\alpha$ of zeros and ones, which is the binary representation of the number $j$. This vertex is connected to exactly two vertices of level $(k-1)$, whose binary representations are $(\alpha, 0)$ and $(\alpha, 1)$.  

Define the Schur function
$$
\phi_0(z) = - \frac{ \sum_{i=1}^n t_i z^{i-1}} { \sum_{i=0}^{n-1} t_i z^i}.
$$
The residual term of the function $\phi_0$ of order $k$ will be denoted by $\phi_k(z)$. 

Each vertex of the binary graph is loaded with the following data: parameters of polynomials of the numerator and denominator of the Schur residual term and parameters of Schur polynomials. Consider a vertex of level $k\ge 0$ with number $j$. Introduce the notation $m=2^k j$. Associated with this vertex are the following data:
\begin{itemize}
\item the first ${2^k}$ coefficients of the normalized numerator $b_m(z)$ and normalized denominator $a_m(z)$ of the residual term $\phi_m(z)$;
\item DFT results $F_{2^k} \eta_{m, m+2^k}$ and $F_{2^k} \xi_{m, m+2^k}$ on ${2^k}$ points from Schur polynomials of order $2^k$, calculated from the initial Schur function $\phi_m$;
\item the quantity $\delta_{m,m+2^k}^{-1}$, inverse to the residual variance over $2^k$ steps from the initial Schur function $\phi_m(z)$.
\end{itemize}

Before the start of the algorithm for traversing this tree, all Schur polynomials are unknown, and among polynomials of numerators and denominators of residual terms, only $(b_{0,k}, a_{0,k})$ are known. These are the initial polynomials of the numerator and denominator of $\phi_0(z)$. In memory are stored DFTs from coefficients of polynomials
$$
b_{0,k} = \sum_{i=1}^{2^k} t_i z^{i-1}, \qquad a_{0,k} = \sum_{i=0}^{2^k-1} t_i z^i, \qquad 0\le k\le m.
$$
The goal of the algorithm is to compute Schur polynomials at the root of the tree: $(\eta_{0,p}(z), \xi_{0.p}(z))$, for which it will be necessary to compute all data at all vertices. 

Vertices are traversed in lexicographic order. In traversed vertices, all loaded data are calculated. Traversed vertices are determined as follows.

With each vertex, a unique path to the root vertex is associated.

Let the current vertex be at level $k$ and have number $j<2^{p-k}$. There exists a unique path from it to the root vertex of the tree. At level $\ell > k$, this path passes through a vertex with number $i_{\ell}=[j/2^{\ell-k}]$, where $[\cdot ]$ is the integer part of the number. Then traversed are all vertices of level $\ell\le k$ with number $q<2^{k-\ell}(j+1)$, as well as all vertices of level $\ell > k$ with number $q > i_{\ell}$. For $\ell>k$, at vertex $i_{\ell}$, values of the numerator and denominator $(B_{i_{\ell},\ell}, A_{i_{\ell},\ell})$ are already computed, but not the values of Schur polynomials $(X_{i_{\ell}, \ell}, Y_{i_{\ell}, \ell})$.

\vspace{3mm}
{\bf Tree Traversal Algorithm.}

The initial vertex is the first leaf with multi-index $\alpha=(0, \ldots, 0)$. At this vertex, values $\eta^{\alpha}=1$, $\xi^{\alpha}=-t_1$ and $(\delta^{\alpha})^{-1}=1/(1-|t_1|^2)$ are given. 

The binary tree is traversed in lexicographic order. Let the current vertex have number $j$ at level $k<p$. The binary representation of the number $j$ forms the multi-index $\alpha$. Also introduce notations for the order $m=2^k$ of Schur polynomials at this vertex and for the number of the Schur residual term denote $\ell=2^k j$. At this vertex, vectors $F_m \eta_{\ell, \ell+m}$, $F_m \xi_{\ell, \ell+m}$ of dimension $m$ are already calculated, as well as the inverse of the residual variance $\delta$.

Processing of the current vertex consists of the following operations. First, DFT values on $m$ samples $F_m \eta_{\ell, \ell+m}$ and $F_m \xi_{\ell, \ell+m}$ are interpolated to $2m$ samples $F_{2m} \eta_{\ell, \ell+m}=(X_q)_{q=0}^{2m-1}$ and $F_{2m} \xi_{\ell, \ell+m}=(Y_q)_{q=0}^{2m-1}$. Further processing depends on the parity of the number $j$.

{\bf 1.} Let the current vertex be even, $j = 2j_0$. Then its multi-index can be represented as $(\alpha, 0)$, where $\alpha$ is the multi-index of a vertex of level $k+1$ connected to the current vertex and to which the Schur residual term $\phi_{\ell}$ is attached. The following operations are performed.
\begin{enumerate}
\item Calculation of DFT $F_{2m} b^{\alpha}=(B^{\alpha}_q)_{q=0}^{2m-1}$, $F_{2m} a^{\alpha}=(A^{\alpha}_q)_{q=0}^{2m-1}$ on $2m$ points from the first $2m$ coefficients of the numerator and denominator of the residual term $\phi_{\ell}(z)$.
\item Calculation of parameters of the new Schur residual term $\phi_{\ell+m}$, attached to the adjacent vertex of level $k$ with number $j+1$, having multi-index $(\alpha,1)$. For this, by Corollary \ref{edge0_FFT}, quantities are computed
$$
\begin{pmatrix} \widehat{A}_q \\ \widehat{B}_q \end{pmatrix} = \begin{pmatrix} X_q^* & -Y_q^* \\ -Y_q & X_q \end{pmatrix} \begin{pmatrix} A^{\alpha}_q \\ B^{\alpha}_q \end{pmatrix}, \qquad 0\le q\le 2m-1.
$$
Then IDFTs are computed 
$$
(\widehat{b}_q)_{q=0}^{2m-1}=F^{-1}_{2m} ((-1)^q \widehat{B}^{\alpha}_q)_{q=0}^{2m-1}, \qquad (\widehat{a}_q)_{q=0}^{2m-1}=F^{-1}_{2m} (\widehat{A}^{\alpha}_q)_{q=0}^{2m-1}
$$ 
on $2m$ samples. By Corollary \ref{edge0_FFT}, the first $m$ coefficients of the numerator and denominator of the function $\phi_{\ell+m}(z)$ form vectors $(\delta^{-1} \widehat{b}_q)_{q=0}^{m-1}$ and $(\delta^{-1} \widehat{a}_q)_{q=0}^{m-1}$, respectively. These data are attached to vertex $(\alpha,1)$ of level $k$ and from it along zero branches to all vertices of levels less than $k$. 
\item At vertex of level zero with multi-index $\beta=(\alpha, 1, 0, \ldots, 0)$, quantities are determined: $\eta^{\beta}=1$, $\xi^{\beta} = \widehat{b}_0$ and $(\delta^{\beta})^{-1} = 1/(1 - |\widehat{b}_0|^2)$. This vertex becomes the current one at the next step of the algorithm.
\end{enumerate}

{\bf 2.} Let the current vertex be odd, $j=2j_0+1$. Then its multi-index can be represented as $(\alpha, 1)$, where $\alpha$ is the multi-index of a vertex of level $k+1$ connected to the current vertex. At the same time, the vertex with multi-index $(\alpha, 0)$ has already been processed earlier, and in it DFT results on $2m$ samples $(X^{(\alpha, 0)}_q)_{q=0}^{2m-1}$ from $\eta^{(\alpha,0)}$ and $(Y^{(\alpha, 0)}_q)_{q=0}^{2m-1}$ from $\xi^{(\alpha,0)}$ are computed, as well as the residual variance $\delta^{(\alpha, 0)}$.

By Lemma \ref{edge1}, DFT values from coefficients of Schur polynomials $(X^{\alpha}_q)_{q=0}^{2m-1}=F_{2m} \eta^{\alpha}$, $(Y^{\alpha}_q)_{q=0}^{2m-1}=F_{2m} \xi^{\alpha}$ at vertex with multi-index $\alpha$ are determined by the formulas
\begin{eqnarray*}
X^{\alpha}_q & = & X^{(\alpha,0)}_q X_q + (-1)^q (Y^{(\alpha,0)}_q)^* Y_q, \\
Y^{\alpha}_q & = & Y^{(\alpha,0)}_q X_q + (-1)^q (X^{(\alpha,0)}_q)^* Y_q, \qquad 0\le q\le 2m-1.
\end{eqnarray*}
Moreover, the inverse residual variance is computed by the rule 
$$
(\delta^{\alpha})^{-1} = (\delta^{(\alpha,0)})^{-1} \delta^{-1}.
$$ 
At the next step of the algorithm, the vertex with multi-index $\alpha$ becomes the current vertex.

This ends the body of the loop in the algorithm for calculating data in the loaded binary tree. Processing of a new current vertex begins. The algorithm finishes when the current vertex is at level $p$. 

\begin{theorem} \label{tree}
Let the Toeplitz matrix $T$ be positive definite. Then the formulated tree traversal algorithm finishes and at vertex of level $p$ are found Schur polynomials for the function $\phi_0(z)$.
\end{theorem}

\proof From the positive definiteness of $T$, it follows that $|\gamma_m|<1$ for all $0\le m<n=2^p$, where $\gamma_m$ are reflection coefficients. 
By Lemma \ref{Schur_Sego}, reflection coefficients coincide with Schur parameters for the function $\phi_0$. 
At vertices of level zero, quantities $\delta^{\alpha}$ were defined as $1-|\gamma_m|^2$, where $m$ is the number of the vertex of level $0$. 
Consequently, for all $m$, the inversion $(\delta^{\alpha})^{-1}$ is performed correctly. 

The method of calculating Schur polynomials is based on Lemma \ref{edge1} and Corollary \ref{edge0_FFT}. \square

\section{Complexity of Computing Schur Polynomials}

Consider the general complexity of the fast Schur algorithm on an abstract machine. For this, we will need information
on special implementations of FFT, given in Appendix~\ref{Appendix:fft}.

\subsection{General Estimate of the Number of Operations}

Ammar and Gregg ~\cite{Ammar-Complexity} give an estimate of the complexity of the Schur algorithm for
polynomials with real coefficients in software implementation and
using FFT with split-radix, which amounts to $(8/3)n\log_2^2n +
O(n\log_2n)$ real multiplication operations and $(16/3)n\log_2^2n + O(n\log_2n)$ addition operations.

The complexity of the binary tree traversal algorithm is completely determined by its size: height $p$ or dimension $n=2^p$. 
As in Appendix~\ref{Appendix:fft}, denote the complexity of complex FFT on $n$ samples by $\phi(n)$, 
real FFT by $\tau(n)$, and double real interpolation by $I(n)$. 

\begin{lemma} \label{complexity_recurrent}
The complexity $T(n)$ of implementation of the fast Schur algorithm for dimension $n=2^p$ satisfies the equation
$$
T(2n) = 2T(n) + 4I(n) + 4\tau(2n) +(34n+2)\mr + (24n -2)\ar.
$$
\end{lemma}

\proof
We prove the lemma by induction on the height $p$ of the binary tree. 
The algorithm for traversing a binary tree of height $p+1$ consists of traversing two binary trees of height $p$, 
with root vertices having multi-indices $0$ and $1$, 
as well as processing these two vertices. Therefore, the quantity $T(2n)-2T(n)=\Delta T$ equals the sum of the number of operations in the even and odd vertices of the tree.

Interpolation of DFT of one real polynomial from $n$ to $2n$ samples requires $I(n)$ operations. 
For two Schur polynomials and for two vertices, we get $4I(n)$ operations. 

Consider processing of an even vertex. 
In item 1, execution of two DFTs on $2n$ real samples requires $2\tau(2n)$ operations.

In item 2, vectors $(\widehat{A}_q, \widehat{B}_q)_{q=0}^{2n-1}$ are computed, 
then two real IDFTs on $2n$ samples are performed and the result is divided by a real number $\delta$. 

All complex-valued vectors in the algorithm are DFTs from real arrays. Therefore, it is sufficient to store only half of their values in memory. 
Values $(\widehat{A}_q, \widehat{B}_q)$, as well as $(A^{\alpha}_q, B^{\alpha}_q)$ and $(X_q, Y_q)$ are real for $q=0$ and for $q=n$. 
In addition to them, values at $1\le q\le n-1$ are stored in memory, which are complex. Implementation of multiplication at $q=0$ or at $q=n$ requires $4$ real multiplications and $2$ real additions. Implementation of multiplication at $1\le q\le n-1$ requires $4$ complex multiplications and $2$ complex additions. 

Finally, multiplication of real vectors $(\widehat{b}_q)_{q=0}^{n-1}$ and $(\widehat{b}_q)_{q=0}^{n-1}$ by $\delta^{-1}$ requires $2n$ real multiplications. 

In item 3, one real multiplication and one real addition are performed, as well as one operation of inversion of a real number.

In total, the number of operations in items 2 and 3 of processing an even vertex is
$$
\Delta_2 = (n-1)(4\mc+2\ac)+2(4\mr+2\ar) + 2\tau(2n) + 2n\mr + 2\mr + \ar.
$$

Consider processing of an odd vertex. Computation of residual variance $\delta^{\alpha}$ consists of one real multiplication. 
Computation of $(X^{\alpha}_q, Y^{\alpha}_q)$ at $q=0$ or at $q=n$ requires $4$ real multiplications and $2$ real additions. 
At $1\le q\le n-1$, this operation is complex. The complexity of implementation of processing an odd vertex equals
$$
\Delta_3 = (n-1)(4\mc + 2 \ac) + 2(4\mr+2\ar) + \ar.
$$
In total, the total number of operations is 
$$
\Delta T = 4I(n) + 4\tau(2n) + 2(n-1)(4\mc+2\ac)+4(4\mr+2\ar) + (2n+2)\mr + 2\ar,
$$
which corresponds to the statement of the lemma. \square

For $p=0$, the number $n=2^p$ is not even, and therefore Lemma \ref{complexity_recurrent} is not applicable. 

\begin{lemma} \label{init_recurrent}
The number of operations for finding Schur polynomials of order $n=2$ equals
$$
T(2) = 20 \mr + 15 \ar.
$$
\end{lemma}

\proof
Let Schur polynomials of order $1$ be given at two adjacent vertices of level zero:  $(\eta_0(z), \xi_0(z)) = (1, \gamma_0)$ and $(\eta_1(z), \xi_1(z)) = (1, \gamma_2)$. Consider the work of the tree traversal algorithm in this case. 

Interpolation from $m=1$ values of $(\eta_0(z), \xi_0(z))$ and $(\eta_0(z), \xi_0(z))$ to $2m$ samples does not require computations, since these are constants.

Processing of the zero vertex consists of three items. In item 1, it is required to compute values of two first-degree polynomials at points $z_0=1$ and $z_1=-1$. This requires $4$ real additions. 

In item 2, matrix multiplication for $q=0$ and $q=1$ requires a total of $8$ multiplications and $4$ additions. In the following two IDFTs on $2$ samples, it is sufficient to compute only $\widehat{b}_0$ and $\widehat{a}_0$, for which $2$ additions are needed. To them are added two multiplications by $\delta^{-1}$. 

In item 3, one multiplication, one addition, and one inversion of a real positive number are performed. 

When processing an odd vertex, $4$ multiplications and $2$ additions are performed at $q=0$ and at $q=1$. In addition, values of $\delta^{-1}$ are multiplied. 

In total, the total number of operations corresponds to the conclusion of the lemma. \square

\vspace{3mm}
To calculate values $T(n)$, we apply the following auxiliary statement, in which $p=\log_2 n$.

\begin{lemma}[\cite{Ammar-Complexity}] \label{recursive_twice}
Let a sequence of positive numbers $(a(p))_{p=1}^{\infty}$ satisfy the equation
$$
a(p+1) = 2 a(p) + 2^p p c_0 + 2^p c_1 + p c_2 + c_3 + c_4 (-1)^p.
$$
Then
\begin{eqnarray*}
a(p) & = & 2^{p-2} p^2 c_0 + 2^{p-2} p(2c_1-c_0) + 2^{p-1}(2c_2+c_3-c_1-c_4/3+a(1)) \\
&&
- c_2 p - c_2 + c_4 (-1)^p/3.
\end{eqnarray*}
\end{lemma}

\proof
The solution of the linear difference equation 
$$
a(p+1) = 2 a(p) + b(p)
$$ 
with an arbitrary input sequence $b$ has the form
$$
a(p) = 2^p a(1) + \sum_{k=1}^{p-1} 2^{p-k-1} b(k).
$$
It remains to compute the sums
\begin{eqnarray*}
\sum_{k=1}^{p-1} 2^{p-k-1} k 2^k & = & 2^{p-1} \sum_{k=1}^{p-1} k = 2^{p-1} \frac{p(p-1)}{2}, \\
\sum_{k=1}^{p-1} 2^{p-k-1} 2^k & = & \sum_{k=1}^{p-1} 2^{p-1}  = 2^{p-1} (p-1), \\
\sum_{k=1}^{p-1} 2^{p-k-1} k & = & \sum_{k=1}^{p-1} \sum_{j=1}^k 2^{p-k-1} = \sum_{j=1}^{p-1} \sum_{k=j}^{p-1} 2^{p-k-1}\\
& = &
\sum_{j=1}^{p-1} (2^{p-j}-1) = 2^p-p-1, \\
\sum_{k=1}^{p-1} 2^{p-k-1} & = & 2^{p-1} - 1, \\
\sum_{k=1}^{p-1} 2^{p-k-1} (-1)^k & = & -\frac{1}{3}(2^{p-1} +(-1)^p).
\end{eqnarray*} 
Substitution of terms leads to the conclusion of the lemma. \square

\vspace{3mm}
Using Lemma \ref{recursive_twice}, one can calculate the exact complexity of the fast Schur algorithm, 
in which one of the methods of FFT implementation is chosen. 
For any implementation method, real FFT has half the complexity in the main term compared to complex, 
and double interpolation has the same order:
\begin{eqnarray*}
\tau(n) & \sim & \phi(n/2), \\
I(n) & \sim & \tau(n) + \phi(n/2).
\end{eqnarray*}

\begin{corollary} \label{complexity_Schur}
Let the implementation of complex FFT have complexity 
$$
\phi(n) = (C\ar + D \mr) n \log_2n + {\mathcal O}(n).
$$
Then the complexity of implementation of the Schur algorithm has order
$$
T(n) = 2(C\ar + D_n \mr) n \log_2^2n + {\mathcal O}(n \log_2n).
$$
\end{corollary}

\proof
We use Lemmas \ref{complexity_recurrent} and \ref{recursive_twice}. The main term in the sum from the recurrence equation in Lemma \ref{complexity_recurrent}
$$
4I(n) + 4\tau(2n) \sim 4(\tau(n) + \phi(n/2)) + 4\phi(n) \sim 8 \phi(n).
$$ 
has order $n \log_2n$ and multiplier at it
$$
c_0 = 8(C\ar + D \mr).
$$
By Lemma \ref{recursive_twice}, the main term of the solution $T(n)$ has order $n \log_2^2 n$ and multiplier $c_0/4$, which is stated in the corollary. \square

\vspace{3mm}
As discussed in ~\ref{chapt2}, since FFT is used for computing
convolution, the order of data in intermediate vectors is not important. Thus,
one can use for FFT addressing of decimation-in-frequency type and for IFFT decimation-in-time type. 
This allows eliminating the permutation operation.
As discussed in Section~\ref{arithmetic_opt}, adders should be taken into account
only when using floating-point numbers. For fixed-point
computations, if the number of multiplication and addition operations
does not differ much, then the cost of addition operations can be neglected,
since it does not qualitatively affect the total cost of computations. Since
computation of complex multiplication in hardware implementation does not reduce to
real multiplications, one should separately account for the complexity of the algorithm in
terms of complex multiplication and addition. 

For hardware implementation of FFT, the radix-4 algorithm is more preferable,
having 6\% more multiplication operations compared to
split-radix algorithms, but a regular data structure. Variants
of parallelization of this algorithm for single-port and dual-port random access
memory are given in Chapter~\ref{chapt2}.

Let us compute the number of complex operations when using complex FFT with
radix 4. The number of complex multiplications in FFT is $(3/8)n\log_2n +
O(n\log_2n)$, and additions $n\log_2n + O(n\log_2n)$. Substituting values for $C$ and $D$ into the formula from Corollary 
\ref{complexity_Schur},
we obtain $(3/2)n\log_2^2n + O(n\log_2n)$ complex multiplications and $4n\log_2^2n +
O(n\log_2n)$ complex additions. These values correspond to the results of \cite{Ammar-Complexity}.

It is also necessary to estimate the number of operations for accessing addressable memory and
memory or a block for computing complex exponentials. The number of complex
exponentials exactly equals the number of multiplications; for simplicity, we will not
consider grouping butterflies for their reuse.

Under the assumption of storing intermediate computation results in non-addressable
registers of size $O(1)$, for each stage of FFT with radix 4, $n$
reads and writes of complex data are required.
In this case, we obtain $2n\log_2^{2}n+O(n\log_{2}n)$
reads for the Schur algorithm and the same number of writes of complex data. For
the real Schur algorithm, we obtain $n\log_2^{2}n+O(n\log_{2}n)$ operations
of access to addressable memory of each type.

For the algorithm to work, besides FFT, it is required to perform vector operations of addition and
multiplication with total complexity $O(n\log_2n)$. In addition, it is required to perform
division operations. Division operations can be replaced by computing $1/x$ and
multiplication.

\subsection{General Estimate of the Amount of Addressable Memory}

Let us estimate the minimum necessary number of memory cells. Memory is required for
storing input and output data and intermediate values. In addition, memory
may be required for storing complex exponentials. If for computing
complex exponentials a specialized block for computing
elementary functions from Chapter~\ref{chapt3} is used, then memory for them is not required.

Ammar's implementation of the Schur algorithm requires $6n$ complex or real memory cells
of data depending on the data type of the Schur algorithm without accounting for complex
exponentials. 

\begin{lemma}\label{lemma:memory}
For implementation of the fast Schur algorithm of length $n=2^p$, $M(n)=4n$
memory cells are sufficient, real or complex depending on the type of input data.
\end{lemma}

\proof
The binary tree representing the fast Schur algorithm is traversed in lexicographic order.
The algorithm is recursive, so each vertex corresponds to the same task, but with different
numerical parameters and different dimensions. Input data $(a, b)$ for a vertex of level $k$ have dimension $2^k$.
Output data are DFTs from a pair of Schur polynomials $(\xi, \eta)$ of size $2^k$.

From the root vertex, arcs with indices $0$ and $1$ lead to vertices of binary subtrees, which due to the recursiveness of the algorithm 
are processed independently. First, the subtree with the vertex into which arc $0$ leads is processed. Input data are 
half of the components of vectors $a$ and $b$, therefore for calculation $M(n/2)$ memory cells are required. The result is vectors 
$X$ and $Y$ of dimension $n=2^k$, equal to DFTs from the corresponding pair of Schur polynomials. They occupy $2n$ cells. 

Processing of the second subtree, into which arc $1$ leads, begins with calculation of initial data in accordance with item 1 of the algorithm. At this step,
DFTs from input data $a$, $b$ of length $n$ are performed. They are assumed to be performed in place, without involving additional memory. Finally, 
DFTs from Schur polynomials obtained from both subtrees are multiplied componentwise.

Thus,
$$
M(n) = 2n + \max\{ 2n, M(n/2)\},
$$
$$
M(1)=4.
$$
Therefore $M(n)=4n$. 

\square

\vspace{3mm}
The obtained estimate is 33\% better than Ammar's result. 
It is possible to achieve it in practice when using in-place FFT computation.

\section{Estimate of Optimal Parallelism and Computation Time of the Fast Schur Algorithm on a Device with
Hardware-Accelerated FFT}

We will consider a device with computation saturation based on one memory access operation,
i.e., having a sufficient set of computational blocks to launch processing of one read
datum in the pipeline each cycle. This corresponds to modern engineering practice of developing 
digital semiconductor devices with small lithographic norms.

Then the performance of such a device is determined by the number of reads in the algorithm and the pipeline length; accounting for
arithmetic operations is not required. Write operations also do not need to be accounted for, 
if one read operation generates at most one write operation, 
otherwise write operations should be accounted for.

We will consider the Schur algorithm in terms of vector operations over complex data, for real data - over 
pairs of real numbers. 
Pseudocode for implementation of the considered algorithm is given in Table~\ref{table:FFTSchur}.

\begin{table} [htbp] 
\centering 
\caption{Pseudocode of the fast Schur algorithm.} 
\label{table:FFTSchur}
\begin{tabular}{l}
$(\zeta_{0,n-1}(z),\eta_{0,n-1}(z))=Schur(\widehat{b}_{0}(z),\widehat{a}_{0}(z), n)$\\
$if\quad n>1$\\
\quad \quad $(\zeta_{0,n/2-1}(z),\eta_{0,n/2-1}(z))=Schur(\widehat{b}_{0}(z),\widehat{a}_{0}(z), n/2)$\\
\quad \quad $B=F_{n}\widehat{b}_{0}(z),A=F_{n}\widehat{a}_{0}(z)$\\
\quad \quad $P=F_{n}\tilde{\zeta}_{0,n/2-1}(z)\cdot B+F_{n}\eta_{0,n/2-1}(z)\cdot A$\\
\quad \quad $Q=F_{n}\tilde{\eta}_{0,n/2-1}(z)\cdot B+F_{n}\zeta_{0,n/2-1}(z)\cdot A$\\
\quad \quad $p(z)=W_{n}P, q(z)=W_{n}Q$\\
\quad \quad $(\zeta_{n/2,n-1}(z),\eta_{n/2,n-1}(z))=Schur((z^{-n/2}p(z))_{+},(z^{-n/2}q(z))_{+}, n/2)$\\
\quad \quad $F_{n}\zeta_{0,n-1}(z)=F_{n}\zeta_{0,n/2-1}(z)\cdot
            F_{n}\zeta_{n/2,n-1}(z) + F_{n}\eta_{0,n/2-1}(z)\cdot
            F_{n}\tilde{\eta}_{n/2,n-1}(z)$\\
\quad \quad $F_{n}\eta_{0,n-1}(z)=F_{n}\zeta_{0,n/2-1}(z)\cdot
F_{n}\eta_{n/2,n-1}(z) + F_{n}\eta_{0,n/2-1}(z)\cdot
F_{n}\tilde{\zeta}_{n/2,n-1}(z)$\\
\quad \quad $\eta_{0,n-1}(z)=W_{n}F_{n}\eta_{0,n-1}(z)$\\
\quad \quad $\zeta_{0,n-1}(z)=W_{n}F_{n}\zeta_{0,n-1}(z)$\\
$else$\\
\quad \quad $\gamma=-\widehat{a}(0)$\\
\quad \quad $\zeta_{0,0}(z) = 1/(1-|\gamma|^2)$ \\
\quad \quad $\eta_{0,0}(z) = \bar{\gamma}/(1-|\gamma|^2)$ \\
$end$\\
\end{tabular}
\end{table}

Here
$\widehat{b}_{k}(z)=z^{-1}(b_{k}(z)-b(0))$,
$\widehat{a}_{k}(z)=z^{-1}a_{k}$, i.e., we cut off
$b_{k}(0)\equiv1$ and $a_{k}(0)\equiv0$. $F_{n}$ is the Fourier
transform of length $n$, $W_{n}$ is the inverse Fourier transform,
$\cdot$ is elementwise multiplication.
Denote $\zeta_{l,m}(z)=\tilde{\xi}_{l,m}(z)$. $p(z)_{+}$ means the part of a generalized polynomial with
nonnegative powers.

We will not use double interpolation, computing it by definition using FFT. 
This leads to an increase in arithmetic complexity by 25\%, however 
it allows using an efficient implementation of FFT and vector operations
based on reconfigurable computational blocks, which allows reducing 
the total complexity of the block by 2 times or more compared to a microprocessor-based block 
due to reduction of program memory, instruction decoder, and overhead
for organizing loops.

At the same time, one vector operation corresponds to
one instruction of the architectural template of the accelerator (Definition~\ref{defn:instruction}). Elementary operations will be: 
stage of complex FFT/IFFT with radix 4, stage of complex FFT/IFFT with radix 2,
last stage of real FFT with radix 2, last stage of real FFT with radix 4, 
first stage of real IFFT with radix 2, first stage of real IFFT with radix 4,
leaf stage of the Schur algorithm, vector complex multiplication, vector complex addition,
vector real multiplication, vector real addition.

The number of read operations for complex operations equals the total length of input data vectors, 
and for real - half the length.

The most complex in terms of computational operations is the FFT stage with radix 4. 
Per one complex datum, for its execution, 1 complex multiplier, 2 complex adders
and a block for approximating complex exponentials are required. 
Complex computation blocks can be used for real computations.
The block for approximating complex exponentials can be used for division by approximating $1/x$.

Let us estimate the energy consumption of the device by the method proposed in Chapter~\ref{chapt1}.
For this, we need to estimate the length of the critical path in cycles and the algorithm runtime in cycles.
That is, we need to estimate the number of reads in the entire algorithm and on the critical path with the above
choice of elementary operations.

\subsection{Estimate of the Number of Complex Reads in the Real Schur Algorithm}

Consider a depth-sorted data flow graph corresponding to the pseudocode in Table~\ref{table:FFTSchur}
with additional constraints. At one depth, only operations of the same type can be located and the maximum 
width of vector operations of complex reading at one depth does not exceed $n$, with input vector length $n$.

\begin{figure}[htbp]
	\center
    \includegraphics[width=0.8\textwidth]{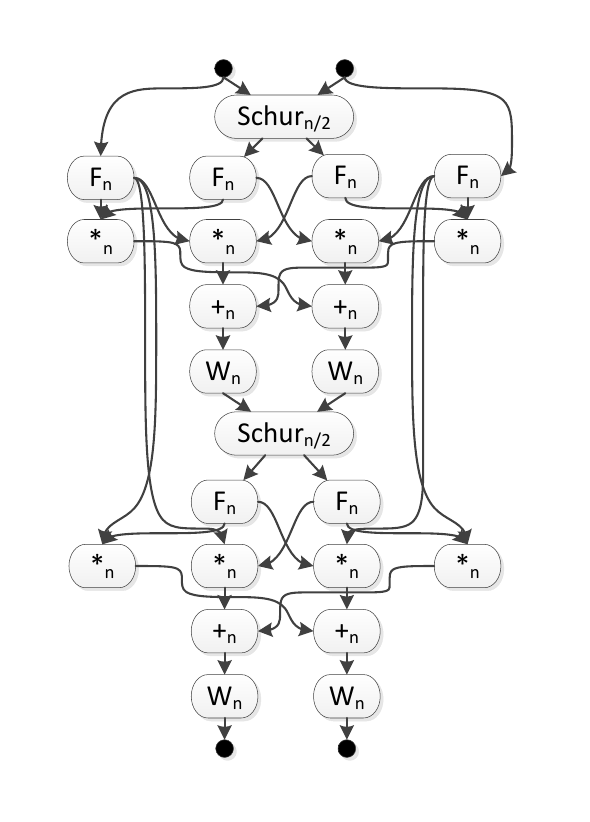}
    \caption{Data flow graph of the fast Schur algorithm for implementation on a reconfigurable computational block.}
    \label{fig:schur-df}
\end{figure}

The data flow graph is shown in Figure~\ref{fig:schur-df}.

Let $n=2^k$. The number of stages for real FFT with radix 4 and length $n$ 
$$
s=\lceil\log_4 n\rceil = \lceil k/2 \rceil.
$$ 

Results of counting read operations for processing in one node with real input data length $n=2^k$ 
are given in Table~\ref{table:op-count}.

\begin{table} [htbp] \centering 
\caption{Number of read operations in the fast Schur algorithm.} 
\label{table:op-count} 
\begin{tabular}{|l|r|r|r|r|} 
\hline 
Operation     &Length&Number&Stages&Reads\\ 
              &      &per node&     &per stage\\ 
\hline 
Real FFT      &$n$  &6&$\lceil\log_4 n\rceil$&$n/2$\\ 
Real IFFT     &$n$  &4&$\lceil\log_4 n\rceil$&$n/2$\\ 
Complex mult. &$n$  &8&1&$n/2$\\ 
Complex add.  &$n$  &4&1&$n/2$\\ 
Recursive call&$n/2$&2&&$T_1(n/2)$\\
\hline 
Leaf op.      &1&1&1&1\\
\hline 
\end{tabular} 
\end{table}

Let us prove an auxiliary statement.

\begin{lemma}\label{lemma:sum}
Let $Q(2^m)$, $m\in \mathbb{Z}_+$ be described by the following recurrence formula
\begin{eqnarray*}
Q(1) = a\\
Q(2^m)=b2^mm+c2^m+dm+e+2Q(2^{m-1}).
\end{eqnarray*}

Then 
$$
Q(2^m)=b2^{m-1}m^2+(b+2c)2^{m-1}+(a+2d+2e)2^m-dm-2d.
$$
\end{lemma}
\proof
Group terms and sum the resulting series.
$$
Q(2^m)=2^ma+\sum_{k=1}^{m}2^{m-k}(b2^kk+c2^k+dk+e)
$$
$$
=2^m\left(a+\sum_{k=1}^{m}bk+c+d2^{-k}k+e2^{-k}\right)
$$
$$
=a2^m+(bm+b+2c)2^{m-1}m+d(2^{m+1}-m-2)+e2^{m-1}
$$
$$
=b2^{m-1}m^2+(b+2c)2^{m-1}+(a+2d+2e)2^m-dm-2d.
$$
\square

\vspace{3mm}

\begin{lemma} \label{T1_bound}
The number of read operations $T_1(2^m)$ in implementation of the fast Schur algorithm satisfies inequalities $T_1(2^m)^-\le T_1(2^m)\le T_1(2^m)^+$, where
\begin{eqnarray*}
T_1(2^m)^- & = & 1.25\cdot2^mm^2+7.25\cdot2^m, \\
T_1(2^m)^+ & = & 1.25\cdot2^mm^2+9.75\cdot2^m.
\end{eqnarray*}
\end{lemma}
\proof
Using Table~\ref{table:op-count}, we compose a recurrence formula for the number of read operations in implementation of the fast Schur algorithm.
\begin{eqnarray*}
T_1(1) = 1\\
T_1(2^m)=10\lceil m/2\rceil 2^{m-1}+12\cdot2^{m-1}+2T_1(2^{m-1}).
\end{eqnarray*}

We use the estimate 
\begin{equation}\label{eq:ceil-est}
m/2\leq \lceil m/2 \rceil\leq m/2+0.5, \qquad m \in \mathbb{N}
\end{equation}
$$
T_1(1)^-=T_1(1)=T_1(1)^+=1
$$
$$
T_1(2^m)^-=2.5\cdot2^mm+6\cdot2^m+2T_1^-(2^{m-1})
$$
$$
T_1(2^m)^+=2.5\cdot2^mm+8.5\cdot2^m+2T_1^+(2^{m-1}).
$$

Then by Lemma~\ref{lemma:sum}
$$
T_1(2^m)^-=2.5\cdot2^{m-1}m^2+(2.5+12)2^{m-1}+2^m=1.25\cdot2^mm^2+7.25\cdot2^m
$$
$$
T_1(2^m)^+=2.5\cdot2^{m-1}m^2+(2.5+17)2^{m-1}+2^m=1.25\cdot2^mm^2+9.75\cdot2^m.
$$
\square

\subsection{Estimate of the Critical Path Length}

The algorithm cannot be parallelized more than the length of its critical path. It
is a limiting factor of parallelism, since all operations on
the critical path are executed sequentially. 

Results of counting read operations for processing in one node with real 
input data length $n=2^k$ are given in Table~\ref{table:crit}. Table~\ref{table:crit}
differs from Table~\ref{table:op-count} since all operations, except addition,
are performed separately over each generator or Schur polynomial and can be 
parallelized by a factor of two.

\begin{table} [htbp] \centering 
\caption{Number of read operations on the critical path of the fast Schur algorithm.} 
\label{table:crit} 
\begin{tabular}{|l|r|r|r|r|} 
\hline 
Operation     &Length&Number&Stages&Width\\ 
              &      &per node&     &\\ 
\hline 
Real FFT      &$n$  &2&$\lceil\log_4 n\rceil$&$2n$\\ 
Real IFFT     &$n$  &2&$\lceil\log_4 n\rceil$&$n$\\ 
Complex mult. &$n$  &2&1&$2n$\\ 
Complex add.  &$n$  &2&1&$n$\\ 
Recursive call&$n/2$&2&$T_{max}(n/2)$&$n$\\
\hline 
Leaf op.      &1&1&1&1\\
\hline 
\end{tabular} 
\end{table}

Similarly to the previous, using Table~\ref{table:crit}, we compose a recurrence formula for the number of read operations on the critical path 
in implementation of the fast Schur algorithm.

\begin{lemma} \label{T_max}
The number of read operations $T_{max}(2^m)$ on the critical path in implementation of the fast Schur algorithm satisfies inequalities $T_{max}(2^m)^-\le T_{max}(2^m)\le T_{max}(2^m)^+$, where
\begin{eqnarray*}
T_{max}(2^m)^- & = & 13\cdot2^m-2m-4, \\
T_{max}(2^m)^+ & = & 17\cdot2^m-2m-4.
\end{eqnarray*}
\end{lemma}
\proof
We use estimate~\ref{eq:ceil-est} to compute upper and lower estimates of $T_{max}(2^m)$.
$$
T_{max}(2^m)^-\leq T_{max}(2^m) \leq T_{max}(2^m)^+
$$
$$
T_{max}(1)^-=T_{max}(1)=T_{max}(1)^+=1
$$
$$
T_{max}(2^m)^-=2m+4+2T_{max}^-(2^{m-1})
$$
$$
T_{max}(2^m)^+=2m+6+2T_{max}^+(2^{m-1}).
$$

Then by Lemma~\ref{lemma:sum}
$$
T_{max}(2^m)^-=(1+4+8)2^m-3m-6=13\cdot2^m-2m-4
$$
$$
T_{max}(2^m)^+=(1+4+12)2^m-3m-6=17\cdot2^m-2m-4.
$$
\square

\vspace{3mm}
Thus, we can
conclude that the maximum speedup from parallelization of the Schur algorithm is at most $C\log^2_2n$ times.

Let us make an additional assumption about the absence of reordering of scalar
operations between vector operations, which follows from Definition~\ref{defn:instruction} of the instruction of the architectural template. 

Under this assumption, the execution time equals $lT_{max}$, where $l$ is the pipeline length, 
if the number of computational blocks is greater than or equal to the maximum
parallelism of the widest vector operation. The widest vector operations
are located at the top level of recursion of the Schur algorithm.

When performing one read per cycle, the pipeline length cannot be less than the size of the FFT butterfly,
since computing the result of a butterfly requires reading all input data.
                                                                   
\subsection{Estimate of Computation Time of the Fast Schur Algorithm for
Real Data on Parallel Pipelined Processors}

Consider $2^k$ computational blocks that can
execute in one cycle 2 complex addition operations, one complex multiplication operation and two
operations of computing real-valued elementary functions, one write operation and one
read operation of complex data. As before, we make an assumption
about the absence of reordering of scalar operations between vector operations.

In accordance with the "Width" column of Table~\ref{table:crit}, this number of blocks corresponds to maximum parallelization for
the Schur algorithm of length $2^{k-1}$ for real data. The limiting factor with such a choice of processor
element is the number of reads from memory.

\begin{lemma} \label{Tl1}
Let $n=2^m$ and let the pipeline have length $l$. Then the execution time of the fast Schur algorithm $T_{2k}(n)$ for 
real data on $2^k$ processors satisfies the inequality $T_{2k}^-(n)\le T_{2k}(n)\le T_{2k}^+(n)$, where
\begin{eqnarray*}
T_{2k}^-(n) &=& 2^{m-k}(13\cdot2^k-1.5k+1.25m^2-1.25k^2-7.75)\\
            &&+(l-1)(13\cdot2^m-2m-4), \\
T_{2k}^+(n) &=& 2^{m-k}(17\cdot2^k-1.5k+1.25m^2-1.25k^2-2.75)\\
            &&+(l-1)(17\cdot2^m-2m-4).
\end{eqnarray*} 
\end{lemma}
\proof
The execution time of the fast Schur algorithm for real data on $2^k$ processors can be represented as 
$$T_{2^k}(n)=T_{2^k}(2^m)=\hat{T}_{2^k}(2^m)+(l-1)T_{max}(2^m),$$
where $\hat{T}_{2^k}$ is the execution time at pipeline length $l=1$. 
Since for starting dependent computations, pipeline clearing is required.

Let us compute $\hat{T}_{2^k}$. The fast Schur algorithm contains $2^{m-k+1}$ sequentially executed subtasks of size $2^{k-1}$,
each of them is executed in critical path length $T_{max}(2^{k-1})$ at $l=1$

The remaining $T_1(2^m)-2^{m-k+1}T_1(2^{k-1})$ operations at higher recursion levels are ideally
parallelized on $2^k$ processors without idle time.

\begin{eqnarray*}
T_{2^k}(2^m)=2^{m-k+1}T_{max}(2^{k-1})+\frac{T_1(2^m)-2^{m-k+1}T_1(2^{k-1})}{2^k}+(l-1)T_{max}(2^m)
\\
T_{max}(1)=T_1(1)
\\
T_{2^{m+1}}(2^m)=T_{max}(2^m)
\end{eqnarray*}

Let us compute upper and lower estimates for $T_{2^k}(2^m)$ using upper and lower estimates for $T_1$ and $T_{max}$.
$$
T_{2^k}^-(2^m)=2^{m-k+1}T_{max}^-(2^{k-1})+\frac{T_1^-(2^m)-2^{m-k+1}T_1^+(2^{k-1})}{2^k}+(l-1)T_{max}^-(2^m)
$$
$$
=2^{m-k}(13\cdot2^k-1.5k+1.25m^2-1.25k^2-7.75)+(l-1)(13\cdot2^m-2m-4)
$$

$$
T_{2^k}^+(2^m)=2^{m-k+1}T_{max}^+(2^{k-1})+\frac{T_1^+(2^m)-2^{m-k+1}T_1^-(2^{k-1})}{2^k}+(l-1)T_{max}^+(2^m)
$$
$$
=2^{m-k}(17\cdot2^k-1.5k+1.25m^2-1.25k^2-2.75)+(l-1)(17\cdot2^m-2m-4)
$$
\square

\vspace{3mm}
Let us compute the average estimate 
\begin{equation}\label{eq:amdal-est0}
T_{2^k}^*(2^m) = \frac{T_{2^k}^-(2^m)+T_{2^k}^+(2^m)}{2}
\end{equation}
$$
=2^{m-k}(15\cdot2^k-1.5k+1.25m^2-1.25k^2-5.25)+(l-1)(15\cdot2^m-2m-4)
$$

\subsection{Estimate of Optimal Parallelism}
Suppose it is required to develop a hardware acceleration block for computing
the fast Schur algorithm for the acoustic echo suppression problem. Sampling
frequency 16 kHz, impulse response length 4096 samples, which
corresponds to duration 0.25 sec. All computations will be performed with floating
point FP32 with rounding of denormalized numbers. Adaptation frequency $f^*=1$ Hz.

As a basic computational block, consider a block containing one
complex multiplier, two complex adders and a module for computing
complex exponentials and $1/x$. 

Let the block have a pipeline of length $l=5$. For sequential computation of FFT with
radix 4, $l\geq4$ is required. In addition, by Theorem~\ref{theorem:fft_1rw}, for applying
single-port 1rw memory, $l$ must be odd. 

Let the block not
store data in memory during idle time; loading and unloading are performed before and
immediately after completion of computations. Consequently, the number of gates in
the part of the circuit not disconnected from power during idle time is $N_0 = 0$. It should be noted
that this block is a universal accelerator for FFT computation 
and vector computations and is not specific to the Schur algorithm. 
When adding to the module for computing elementary functions additional tables for
computing $\log_2$ and $\sqrt{x}$, which slightly increases its size, this
block covers a wide class of digital signal processing algorithms.

By Lemma~\ref{lemma:memory}, for implementation of the fast Schur algorithm of length 4096 with 
FP32 data, the required memory size is 64 Kibibytes. 

Let us use Tables~\ref{table:mem-area} and~\ref{table:mem-vs-fp32}.
When using single-port memory, the number of memory
gates can be estimated as $N_1=48M$, where $M$ is the number of gates of an FP32 multiplier
without denormalization.

We will consider the simplest implementations of a complex multiplier in the form of 
4 real multipliers and 2 real adders and a complex adder 
in the form of 2 real adders. Additional tables for $1/x$ we estimate as $0.5$ of 
the area of the block. Let us use Tables~\ref{table:fp-vs-int} and~\ref{table:interp-comp}.

\begin{table} [htbp] \centering 
\caption{Size of the computational part of the block in real FP32 multipliers w/o denorm} 
\label{table:ops} 
\begin{tabular}{|l|r|r|} 
\hline 
Operation&Quantity&Size\\ 
\hline 
Complex multiplication&1&\\ 
\quad Multiplication &4&M\\ 
\quad Addition &2&0.64M\\ 
Complex addition&2&\\ 
\quad Addition &2&0.64M\\
Complex exponential and $1/x$&2.5&1.5M\\ 
\hline 
Total&&10.31M\\
\hline 
\end{tabular} 
\end{table}

In Table~\ref{table:ops}, a calculation of the number of gates in the computational part of the block is given,
the total number is $\hat{N}_2=10.31M$.

By formula~\ref{eq:power}, without parallelization, power equals 
$$
P_1=B (c_0 / f + c_1)(\hat{N}_2 + N_1)=58.31C,
$$ 
where $C$ is some constant independent of the algorithm.

Minimum power at infinite parallelization equals $P_0=10.31C$, which
corresponds to a reduction in energy consumption by a factor of 6 or by 82\%.

Suppose that the external interface of the block has width 64 bits.
Let data loading and unloading operations occur without overlap through
the external interface of the block, i.e., per cycle a read or write of a memory
word occurs.

We will consider only parallelization by a power of 2, $p=2^k$. Such
a choice is related to the complexity of splitting memory into banks and organizing their addressing
in other cases.

Let us estimate the total runtime in cycles, using the estimate of execution time on $2^k$ processors $T_{2^k}^*(2^m)$.
$$
\bar{T}_{2^k}(2^m)=T_{2^k}^*(2^m) + T_{io}(2^m)
$$

Here $T_{io}(n)=2n$ is the input-output time. Per cycle, a
read or write of a pair of real data occurs.

Using estimate~\ref{eq:amdal-est0}, we define 
$$
\bar{T}_{2^k}(2^m) = 2^{m-k}(15\cdot2^k-1.5k+1.25m^2-1.25k^2-5.25)+4(15\cdot2^m-2m-4)+2^{m+1}
$$
$$
=77\cdot2^m+2^{m-k}(1.25m^2-1.25k^2-1.5k-5.25)-8m-16.
$$

Total algorithm runtime without parallelization
$$
\bar{T}_{1}(2^m) = 77\cdot2^m+2^m(1.25m^2-5.25)-8m-16.
$$

For $m=12, k=0$ and duty cycle $S=1$, the clock frequency $f=f^*\bar{T}_1(4096)\simeq 1.03$ MHz,
which is in the interval of linear growth of power with frequency for
low-micron lithographic processes.

For parallelizing FFT by a factor of $p$ on single-port memory using the algorithm
from Theorem~\ref{theorem:fft_1rw}, $2p$ memory banks are required. However, in practice,
the minimum size of a single-port static memory bank in libraries
is 8 Kibibytes, which corresponds to maximum parallelization by a factor of 4.

Let us use formula~\ref{eq:opt-par} for optimal parallelism based on
Amdahl's law.

Critical path 
$$
\bar{T}_{max}(2^m)=\bar{T}_{2^{m+1}}(2^m) = 77\cdot2^m-10m-20.
$$

Thus, for $m = 12$, the non-parallelizable part of the algorithm for Amdahl's
law $K = \frac{\bar{T}_{max}(4096)}{\bar{T}_{1}(4096)}=\simeq0.31$
$$
p_0=\arg \min_p{P}=\sqrt\frac{48(1 - K)}{10.31K} \simeq 3.25
$$

The nearest admissible parallelization equals 4. This estimate does not account for the computation
graph but only the critical path length.

Let us compute the power estimate using formula~\ref{eq:power}
$$
P(p)=C\frac{p \hat{N}_2 + N_1}{l^*(p)}, \qquad l^*(p)=\frac{p}{1+K(p-1)}.
$$

At this, power $P(4)\simeq 42.77C$, which corresponds to a reduction in power by 27\%.

Let us check the adequacy of this estimate by performing more accurate computations based on
algorithm runtime depending on parallelization.
 
Replace the estimate of the speedup function by Amdahl's law with its definition in the power estimate~\ref{eq:power}
$$
l(p)=\frac{\bar{T}_1(2^m)}{\bar{T}_p(2^m)}
$$
$$
P(p)=C(\hat{N}_2p+N_1)\frac{77\cdot2^m+2^m(1.25m^2-5.25)-8m-16}{77\cdot2^m+2^{m-k}(1.25m^2-1.25k^2-1.5k-5.25)-8m-16}, \qquad p=2^k.
$$

For $p=2$ and $p=4$, power equals respectively
$$P(2)\simeq 44.43C$$
$$P(4)\simeq 42.07C.$$

Thus, optimal parallelism equals $p=4$, which leads to
a reduction in power by 28\%.

The estimate of optimal parallelism and power based on Amdahl's law provides
adequate qualitative results, close to the power estimate based on
analysis of algorithm runtime.

We can use register dual-port memory, whose blocks have
smaller size but larger area of one cell. The minimum block size
is 1 Kibibyte. Admissible parallelization is $p=1..64$.
At this, the relative area of 64K memory is $N_1=96M$, 
and $P(1)=106.31C$.

Again, let us use formula~\ref{eq:opt-par} for optimal parallelism based on
Amdahl's law.
$$
p_0=\arg \min_p{P}=\sqrt\frac{96(1 - K)}{10.31K} \simeq 4.6
$$

Let us compute the power estimate using formula~\ref{eq:power} and the speedup function
by Amdahl for $\lfloor p_0 \rfloor = 4$ and $\lceil p_0 \rceil = 8$.

At this, power $P(8) = 70.06 C$, and $P(4) = 65.78 C$, which corresponds to
a reduction in power by 38\%.

Let us use the speedup function based on algorithm analysis. For such
an architecture, compute values $l(p)$ and $P(p)$ for powers of 2 in the parallelism interval
$p=1..64$.

\begin{table} [htbp]
\centering
\caption{Dependence of power on parallelism when using register memory.}
\label{table:powrer-schur-reg-mem}
\begin{tabular}{|r|r|r|}
\hline
$p$&$l(p)$&$P(p)/C$\\
\hline
    1 &1 &106.31\\
    2 &1.54&75.5\\
    4 &2.1&64.69\\
    8 &2.6&68.66\\
    16&2.9&89.44\\
    32&3.1&137.43\\
    64&3.19&236.78\\
\hline
\end{tabular}
\end{table}

From Table~\ref{table:powrer-schur-reg-mem}, we see that the minimum
power $P(4) = 64.69C$ is achieved at $p=4$, which coincides with the optimum
when using Amdahl's law.

Both power estimates are greater than power estimates when
using single-port memory and $p=4$. 

We can conclude that from the point of view of energy efficiency,
for implementation of the fast Schur algorithm under the given conditions, the optimal
architecture is one with 4 parallel computational blocks and 8 banks of single-port
static memory. The reduction in static power from using this
architecture is 27\%. We can also conclude that the fast Schur algorithm is quite
poorly amenable to parallelization on a pipelined computational block.
The main factors preventing parallelization are the long critical path
of the algorithm and the long pipeline of the block. 

In addition, the estimate of power and parallelism using Amdahl's law provides
the same qualitative results compared to the significantly more labor-intensive
estimate based on algorithm analysis.

\section{Hybrid Filtering Algorithm with Low
Delay}\label{s:Gardner} To solve the filtering problem, it is required to
compute the linear convolution of signal $y$ and filter coefficients $h$:

$$(h*y)(t)=\sum_{i=0}^{n-1}h(i)y(t-i)$$

The impulse response $h$ is split into blocks of unequal length
$h_i=[h_{k_i},\ldots,h_{k_{i+1}}]$, where $k_0=0$, $k_i<k_{i+1}$,
$k_p=n$. Denote $l_i=k_{i+1}-k_i$ the length of block with number $i$.

$$(h*y)(t)=\sum_{i=0}^{p-1}\sum_{j=0}^{l_i-1}h_i(j)y(t-k_i-j)=\sum_{i=0}^{p-1}(h_i*y)(t-k_i)$$

Denote the convolution
$c_i(t)=(h_i*y)(t-k_i)=\sum_{j=0}^{l_i-1}h_i(j)y(t-k_i-j)$.

Convolution with the first block $c_{0}$ is implemented by definition.

Values $c_i, 0<i<p$ are computed blockwise using DFT of length
$2l_i$.

Denote the delay of block $d_i=k_i-l_i$. $d_i$ is the distance in samples
between the last value $y(t)$ necessary for computing $c_i$, and
the moment of using the computation result. Thus, $d_i/F$
is the maximum time available for computing $c_i$.

Consider splitting into blocks of length

\begin{equation}\label{e:Split}
\{l\}_0^{p-1}=\{2m, m, m, 2m, 2m, 4m, 4m, ..\}
\end{equation}

\begin{lemma}
Splitting~(\ref{e:Split}) ensures $d_i\geq l_i, i>0$. That is,
the time for computing convolution is not less than the time for collecting data.
\end{lemma}
{\sf Proof.}
For even $i=2j$, $d_{2j}=2^{j}+2^{j-1}-2{j-1}=2^{j}\geq 2^{j-1}$;
for odd $i=2j-1$, $d_{2j-1}=2^{j}-2{j-1}=2^{j-1}$.
\square

Thus, all block convolutions can be executed simultaneously and
uniform processor loading is ensured.

We can estimate the maximum computational costs of the algorithm per
one sample $w=\sum_{i=0}^{p-1}w_i/d_i$, where $w_i$ is the computational
complexity of finding convolution with one block.

\begin{lemma}
The complexity of the algorithm with splitting~(\ref{e:Split}) is of order
$4m+2.5\log_{2}^{2}n$ per sample.
\end{lemma}
{\sf Proof.}
For the zero block $w_0/m=4m$. Let us split the following blocks into pairs,
since blocks in a pair have the same size, addition can be
performed in the frequency domain. Let $F_{2l_i}h_i$ be computed in advance,
then for computing convolutions of a pair of blocks, we need one forward and
one inverse DFT.

Therefore
$$
(w_{2k+1}+w_{2k+2})/(2^k m)=2w(F_{2^{k+1} m})/(2^k m)=5(k+1) + O(1).
$$

Introduce the notation
$$
p=2\log_2 n-2\log_2 m-1.
$$
Summing the terms, we obtain:
$$w=4m +
5\sum_{k=0}^{(p-1)/2-1}(k+O(1))=4m+\frac{5}{8}p^2+O(p)=4m+2.5\log_2^2n+O(\log
n).
$$
\square

Since the signal is reproduced through a speaker with a small delay $m$, we can set
$h_{0}\equiv0$ and exclude the linear term from the complexity estimate.

For computations, the FFT computation block used for computing
the optimal filter can be reused.
\chapter*{Conclusion}						
\addcontentsline{toc}{chapter}{Conclusion}	

Energy efficiency is a key parameter of modern wireless low-power devices. With
the transition to semiconductor fabrication technologies with small nodes (45 nm and below),
leakage currents proportional to the area of components connected to the supply start to exert a decisive influence on
power consumption. This work proposes an effective method for assessing and selecting the optimal parallelism of a hardware computing architecture based on the characteristics of the computational algorithm. The problem addressed is the
development of energy-efficient computing devices for a wide class of FFT-based
digital signal processing algorithms. The foundation is a universal vector computing block for FFT computation and a block for evaluating elementary functions. Improvements in energy efficiency are also demonstrated for each component by reducing the number of computational operations and
using more energy-efficient memory compared with published results.
The application of this block to a more complex algorithm for superfast factorization of Toeplitz matrices is demonstrated, and the analysis of energy efficiency
and optimal parallelism is provided.

The main results of the work are as follows.
\begin{enumerate}
\item A power consumption model has been developed for a low-power digital circuit executing a known computational algorithm. The model includes dynamic and static energy losses and the possibility of power-down during idle periods. The problem of selecting the optimal parallelism in this model has been solved.

\item The problem of minimizing power consumption when computing values of standard functions has been reduced to minimizing table lengths. New quasi-spline approximation methods with nonuniform tabulation, convenient for hardware implementation, have been developed. Using these methods, the table lengths of all major standard functions have been reduced compared with known counterparts, which has led to a substantial decrease in energy consumption and an increase in the speed of hardware computing blocks.

\item A theorem has been proved on placing FFT data in multi-bank memory when computing with arbitrary mixed radices, ensuring homogeneity of the synchronous dataflow graph for computations, which provides maximum computation speed for a given parallelism and zero buffer memory size. Explicit FFT formulas in the form of Kronecker products by stages of arbitrary orders have been obtained.

\item A theorem has been proved on a self-sorting FFT modification in multi-bank memory with mixed radices, as well as an analogous theorem for a computing device with single-port memory.

\item For the fast Schur algorithm, the minimum memory size has been found, the length of the critical path has been computed, and an estimate of the optimal parallelism has been obtained.
\end{enumerate}

The proposed methods and computational algorithms were tested during
the practical development of low-power semiconductor circuits with small
geometrical nodes.

\clearpage		
\listoffigures									
\addcontentsline{toc}{chapter}{\listfigurename}	
\clearpage

\listoftables									
\addcontentsline{toc}{chapter}{\listtablename}	
\clearpage			

\addcontentsline{toc}{chapter}{\bibname}	
\selectlanguage{English}					
\bibliography{biblio_eng}						

\appendix                       
\chapter{Methods of Power Consumption Optimization}\label{Appendix:optim}

We provide a brief overview of power consumption optimizations at various levels and then a detailed description of several architectural and system-level techniques.

\section{Supply Voltage Reduction} According to~\ref{eq:d_pow}, dynamic power is proportional to the square of the voltage. The power loss from leakage currents depends linearly on the supply voltage. Thus, reducing the supply voltage leads to a reduction in device power. Lowering the supply voltage is the most significant mechanism for reducing energy consumption. The downside of lowering the supply voltage is the decreased reliability and gate switching delay due to process variations in the fabrication, leading to a reduced frequency. Various parallelization methods are used to maintain the required computational performance. Effective parallelization requires modifying or developing new algorithms that allow parallel execution.

\section{Multiple Supply Voltages} In some cases the supply voltage of individual parts of the circuit cannot be reduced due to constraints on the minimum performance or the need to interface with external circuits outside the die. Another approach is to reduce the supply voltage in individual circuit paths. In this case, paths with lower delay can have a lower supply voltage. Voltage level partitioning can occur at the module level when different modules have different performance requirements. For example, in the case of a circuit consisting of a controller and a compute accelerator, if the controller has a shorter critical path at the same supply voltage, its supply voltage can be reduced to align the critical paths. Level shifters must be inserted to match voltage levels between interacting circuits. In addition, extra supply voltages are required in the power delivery network. The presence of additional components reduces the benefit of using multiple supply voltages and significantly complicates the design of the semiconductor die. Multiple supply voltages are usually used together with multiple clock frequencies.

\section{Transistor Scaling} The input capacitance of a gate is proportional to the width of the transistor gates connected to the input. The optimal gate width is proportional to the required switching speed of the driven load. To reduce input capacitance, transistors can be scaled on paths with small delay. Since the size of the transistors in a CMOS gate is tied to balancing different paths inside the gate, this leads to scaling of the entire gate. When the gate width is reduced, the propagation delay increases, which changes the critical paths in the circuit. This optimization is usually performed by automatic physical synthesis tools.

\section{Body Biasing} Body biasing is used to reduce leakage current in standby. The goal is to increase the gate supply voltage $V$. To do this, a voltage of the opposite polarity is applied to the transistor body. Experiments with 65 nm CMOS transistors have shown that body biasing reduces leakage current by 20\%-30\% at the nominal supply voltage. The drawback of this approach is the need for an additional supply voltage, which complicates the device power delivery network.

\section{Multiple Clock Signals} The design of application-specific integrated circuits often includes integrating off-the-shelf components in addition to components tailored to the performance requirements of the target circuit. Off-the-shelf blocks may have excessive or insufficient performance. Achieving the required performance requires raising the frequency and supply voltage of critical components. Other components may use a reduced clock frequency. Additionally, a reduced supply voltage may be used. Having multiple clock generators complicates the circuit and requires synchronization between modules with different clock frequencies. A special case is an integral clock frequency. For example, memory can operate at double or half the frequency compared to the computational logic. An integral frequency ratio does not require complex synchronization.

\section{Clock Gating} Clock gating is often used in practice for energy savings. It is applied when a portion of the circuit remains idle for a long time. Energy savings are achieved by disabling the clock distribution network and eliminating switching activity. Clock gating does not affect leakage currents and does not reduce static power.

\section{Reduction of Switching Activity} Before the output signal of a gate is established, transient processes occur due to delays between the establishment of input signals. The transients end when the input signals reach stable levels for the gate or group of gates. These transients lead to recharging of capacitances and an increase in the active power of the circuit. If the input signals arrive simultaneously, the transient activity is reduced. For this, the path lengths in the circuit must be balanced. Tree structures have shorter and more balanced paths compared to chains. Using pipelining also balances and shortens paths because spurious values do not propagate through buffer registers.

\section{Transistor Stacking} In this method, transistors are connected in series between the supply and ground. This increases the voltage at the junction points between transistors, which raises the threshold voltage of the transistors that are not directly connected to ground. Raising the threshold voltage reduces leakage current and, therefore, static power consumption.

\section{Gates with Different Threshold Voltages} Gates with different threshold voltages can be used to reduce power. Standard cell libraries usually provide three options: standard, high, and low threshold voltage. Gates with a high threshold voltage have leakage currents an order of magnitude lower than gates with a low threshold voltage, but their switching delay is significantly higher. Thus, gates with standard or low threshold voltage can be used on the critical path, while the rest of the circuit can use high-threshold gates. This method mainly reduces the static power of the device because gates not on the critical path have lower delay. In practice, 98\% of all gates in a low-power circuit can have a high threshold voltage. A small reduction in active power may also be observed due to the reduced switched capacitance and signal swing at internal nodes of the gate.

\section{Block Duplication} One method of hardware-level parallelization is duplicating hardware blocks, such as multipliers, adders, or processors connected in parallel. In this case, the blocks can operate at a lower frequency while maintaining the required performance, which allows the supply voltage to be reduced compared to sequential computations. As the number of blocks increases, the switched capacitance and leakage currents grow faster than linearly. Additional overhead arises from the need to multiplex data, synchronize, and increase interconnect lengths. Block duplication increases area and can potentially raise the cost of manufacturing the die. For deep-submicron process nodes, the area of logic circuits is not a limiting factor for circuits of moderate complexity, since the minimum die area is determined by the density of the I/O pads. Block duplication only makes sense when using vector and parallel algorithms that can exploit it.

\section{Pipelining} Another approach to parallelizing computations is pipelining. The goal of pipelining is to balance and limit critical paths, which makes it possible to reduce the clock frequency and supply voltage. Using a pipeline introduces additional costs for registers to buffer data between stages.

\section{Resource Allocation} Resource allocation can lead to a significant reduction in the switching frequency of transistors, which reduces active power. Resource multiplexing reduces area, but the switching frequency may increase due to the uncorrelated nature of multiplexed data, causing higher power consumption. Using independent processing circuits for correlated data can reduce active power at the cost of increased static power due to the larger circuit area.

\section{Data Representation Optimization} Numerical data can be represented using fixed-point and floating-point formats. The bit width of floating-point numbers is typically a multiple of 8. The width of fixed-point numbers can be arbitrary. Intermediate data are usually stored in fixed-point format. Reducing data width saves energy, while increasing it improves accuracy. The accuracy of input and output data is usually defined by external requirements. The precision of intermediate data should be minimized to reduce circuit complexity under constraints on the output accuracy. Because rounding and saturation are nonlinear operations, they should be avoided when possible to enable linear arithmetic optimizations and simplify accuracy analysis. For example, an adder needs $N+1$ result bits to add two $N$-bit numbers without overflow in two's complement arithmetic. Consider the direct-form implementation of an FIR filter of length $M$. It requires $M-1$ adders connected in a chain. The first adder has two $N$-bit inputs. Without width optimization, the output of the last adder in the chain will have a width of $M+N$. Rewriting the FIR filter formula in a tree form shows that the output width can be reduced to $N + \lceil \log_2 M\rceil$ without overflow and loss of accuracy. In many cases, the probability of overflow or underflow on real data is very low. In such cases, saturation and rounding can be used to reduce the width of intermediate data while maintaining average accuracy. Reducing data width simplifies the circuit and saves energy.

\section{Arithmetic Optimizations} Arithmetic optimizations include changing the architecture to reduce the switching frequency of transistors required to compute the result. They also include reducing the cost of the elementary operations used. Elementary integer operations can be ordered by decreasing complexity as division, multiplication, addition, shifts, and bitwise operations. As an example, consider an FIR filter with known constant coefficients. In this case, multiplication can be replaced by a set of shifts and additions that are less expensive to compute than a multiplier. Such a replacement reduces area, delay, and power consumption of the circuit.

\section{Time-Multiplexed Resource Sharing} Leakage currents that define static power depend on the number of gates in the circuit. Thus, reducing the number of gates in the circuit reduces static power. To achieve this, different functional blocks inside the circuit can be reused in the computation algorithm. If the required resource is busy, intermediate data are stored in memory until the resource becomes available.

Compared to directly mapping the data flow graph of the algorithm onto the hardware, resource multiplexing requires additional control logic and a register file or addressable memory for intermediate data. If resources remain idle, multiplexing yields substantial energy savings despite the overhead. Another way to reuse resources is pipelining.

\section{Power Gating} In standby mode, leakage currents are the main cause of power consumption. Disconnecting the circuit from the power supply eliminates leakage and saves energy. This is achieved using large transistors that disconnect power from the rest of the circuit in standby mode. These transistors are placed on the power rail, and sometimes on the ground rail. The transistors are controlled by a high-level ``sleep'' signal that is inactive during normal operation. During standby, the control logic asserts the ``sleep'' signal, which disconnects power from the rest of the circuit.

The limited resistance of the gating transistors leads to noise on the power rail of the circuits they feed. To reduce power rail noise, these transistors must have low resistance and therefore large size and long switching time. The large size of the gating transistors increases the die area. The long switching time leads to a prolonged transient exceeding one clock cycle, which complicates the control circuitry.

Unlike clock gating, power gating leads to loss of data. Therefore, it can only be applied to circuits that support this capability. Non-volatile or always-on memory must be used to preserve data needed for future operation. Using additional memory introduces extra overhead for data copying and storage.

Another option is to connect buffer registers in the circuit to an always-on power supply. In this case, the register state is not lost. Having two parallel power distribution networks within a single module complicates the wiring on the die and introduces substantial overhead.

\section{Dynamic Voltage and Frequency Scaling} Dynamic voltage and frequency scaling simultaneously adjusts both parameters depending on the computational load to minimize power consumption. This method is typically used in multicore processors. The operating system monitors the load on the cores and, based on the load and task deadlines, raises or lowers the voltage and frequency to meet performance requirements. Since the method is applicable only beyond the region where power scales linearly with frequency, its use in low-power circuits is limited.

\section{Use of Parallel Algorithms} In many cases, applying different algorithms that are more amenable to parallelization is a more effective way to parallelize than modifying the original algorithm. Parallelization can be achieved at the level of primitive operations. For example, replacing bitwise computation in a CORDIC scheme with a polynomial approximation substitutes iterative summation with a parallel multiplication implementation. Parallelization can also be achieved by changing the order of computations to remove false dependencies between operations through memory or intermediate data. Using parallel algorithms can increase computational complexity compared to the sequential algorithm.		
\chapter{Polynomial Interpolation} \label{Appendix:poly}
We briefly present the interpolation theory results that we need.

\section{Interpolation Polynomials}\label{sect3_1_1}

Assume that distinct numbers $x_{0}, x_{1}, \ldots$ and the corresponding function values $f(x_{0})$, $f(x_{1})$, $\ldots$ are fixed.

\begin{definition}
First-order divided differences are the numbers
$$
f(x_{k}, x_{k+1}) = \frac{f(x_{k+1})-f(x_{k})}{x_{k+1}-x_{k}} , 
\qquad k = 0, 1,  \ldots  
$$

Divided differences of arbitrary order $n>1$ are defined iteratively:
$$
f(x_{k}, x_{k+1}, \ldots, x_{k+n}) = \frac{f(x_{k+1}, \ldots, x_{k+n}) - 
f(x_{k}, \ldots, x_{k+n-1})}{x_{k+n}-x_{k}}. 
$$
\end{definition}

\begin{lemma}\label{lemma:interp_newton}
$$
f(x_{n}) = f(x_{0}) + (x_{n}-x_{0})f(x_{0}, x_{1}) + \ldots   + 
(x_{n}-x_{0})\ldots  (x_{n}-x_{n-1}) f(x_{0}, x_{1}, \ldots  , x_{n}).
$$
\end{lemma} 

{\sf Proof.} 
This equality can be transformed as follows:
\begin{eqnarray*}
&&(\ldots  ((\frac{f(x_{n})-f(x_{0})}{x_{n}-x_{0}} - f(x_{0}, x_{1}))
\frac{1}{ x_{n}-x_{1}} - f(x_{0}, x_{1}, x_{2}))\frac{1}{ x_{n}-x_{2}} - \ldots  \\
&&- f(x_{0}, x_{1}, \ldots  , x_{n-1}))\frac{1}{ x_{n}-x_{n-1}} = 
f(x_{0}, x_{1}, \ldots  , x_{n}),
\end{eqnarray*}
which follows from the definition of divided differences and from the possibility of permuting their arguments. \square

\begin{definition}\label{definition:fundamental}
We call the following polynomials the fundamental polynomials of the node system $(x_{0},\ldots, x_{n})$:
$$ l_{k}(x) = \frac{\omega (x)}{ (x-x_{k})\omega '(x_{k})}, 
\qquad \omega (x) = \prod^{n}_{j=0} (x-x_{j}).
$$
\end{definition}

\begin{theorem}

There exists a unique polynomial of degree at most $n$ that satisfies the interpolation data $(x_{0}, f(x_{0}))$, $\ldots$, $(x_{n}, f(x_{n}))$. It can be given by the Lagrange formula
$$
P(x) = \sum^{n}_{k=0} {l_{k}(x) f(x_{k})},  
$$
or by the Newton formula
$$
P(x) = f(x_{0}) + (x-x_{0})f(x_{0}, x_{1}) + \ldots   + (x-x_{0})\ldots  
(x-x_{n-1})f(x_{0}, x_{1}, \ldots  , x_{n}).
$$
\end{theorem}

{\sf Proof.} 

1. We show that the Lagrange polynomial is an interpolation polynomial. Since
$$
\frac{\omega (x)}{ x-x_{i}} = \prod^{n}_{j=0,  j\neq i} (x-x_{j}), 
$$
as $x\to x_{i}$ this value tends to $\omega '(x_{i})$. Moreover, this function has zeros at all nodes $x_{j}$, $0\le j\le n$, except the node $x_{i}$. Therefore, in the Lagrange formula at $x=x_{i}$ only one term remains in the sum, namely, the term with $k=i$. This term coincides with $f(x_{i})$, which proves the required interpolation property.

In the Newton formula, at $x=x_{k}$ we obtain the value $f(x_{k})$ by Lemma~\ref{lemma:interp_newton}.

The uniqueness of the interpolation polynomial is proved by contradiction. Suppose $P_{1}(x)$ and $P_{2}(x)$ are two interpolation polynomials of degree at most $n$. Then the polynomial $P_{1}(x)-P_{2}(x)$ has degree at most $n$ and has at least $n+1$ zeros, namely, $x_{0}$, $\ldots$, $x_{n}$. Consequently, this polynomial is identically zero. \square

\section{Error of Interpolation Polynomials}

Let distinct numbers $x_{0}$, $x_{1}$, $\ldots$, $x_{n}$, the function values $f(x_{0})$, $f(x_{1})$, $\ldots$, $f(x_{n})$, and the corresponding interpolation polynomial $P(x)$ be given. We seek to estimate the interpolation error $R(x)=f(x)-P(x)$ for points $x$ different from the interpolation nodes.

\begin{lemma}\label{lemma:interp_error}
$$
R(x) = \omega (x) f(x_{0}, x_{1}, \ldots  , x_{n}, x), 
$$
where $\omega (x)=(x-x_{0})\ldots  (x-x_{n})$.
\end{lemma} 

{\sf Proof.} 
Choose a point $x$ different from all interpolation nodes. Add the point $x$ to the set of interpolation nodes together with the value $f(x)$. Apply Lemma~\ref{lemma:interp_newton} to the resulting set of $n+2$ initial data:
\begin{eqnarray*}
f(x) & = & f(x_{0}) + (x-x_{0})f(x_{0}, x_{1}) + \ldots   + 
(x-x_{0})\ldots  (x-x_{n-1})f(x_{0}, x_{1}, \ldots  , x_{n}) \\
&& + (x-x_{0})\ldots  (x-x_{n-1})(x-x_{n}) f(x_{0}, x_{1}, \ldots, x_{n}, x).
\end{eqnarray*}
But the sum of the terms on the right-hand side, except for the last one, coincides with the interpolation polynomial in Newton form. The claim of the lemma follows. \square

Thus, representations of the divided difference $f(x_{0}, x_{1}, \ldots  , x_{n}, x)$ can be used to estimate the interpolation error. 

\begin{lemma}\label{lemma:discrete_derivatives_rep}
Let the nodes $x_{0}$,  $x_{1}$, $\ldots$,  $x_{n}$ lie on the segment $[a,b]$, and let the function $f$ have a derivative of order $n$ on this segment. Then
\begin{eqnarray*}
f(x_{0}, x_{1}, \ldots, x_{n}) = \int^{1}_{0} dt_{1} \int^{t_{1}}_{0} dt_{2} 
\ldots\\
\int^{t_{n-1}}_{0} dt_{n} 
f^{(n)}\left(x_{0}+\sum^{n}_{i=1}t_{i}(x_{i}-x_{i-1})\right). 
\end{eqnarray*}
\end{lemma}

{\sf Proof.}
The argument of the function $f^{(n)}$ lies in the interval $[a,b]$, since it equals the following convex combination of the numbers $x_{i}$:
$$
x_{0}+\sum^{n}_{i=1}t_{i}(x_{i}-x_{i-1}) = 
(1-t_{1})x_{0} + (t_{2}-t_{1})x_{1} + \ldots   +
(t_{n-1}-t_{n})x_{n-1} + t_{n}x_{n}.
$$

Therefore the function $f^{(n)}$ is defined on the integration domain. 

For $n=1$ the statement is obvious. Assume it is proved for $n-1$. Integrating once the right-hand side in the theorem statement, we obtain
\begin{eqnarray*}
\frac{1}{ x_{n}-x_{n-1}} \int^{1}_{0}dt_{1}\int^{t_{1}}_{0}dt_{2} \ldots  
\int^{t_{n-2}}_{0}dt_{n-1} \\ 
\Bigg(f^{(n-1)} \left(x_{0}+ \sum^{n-1}_{i=1}t_{i}(x_{i}-x_{i-1}) +
t_{n-1}(x_{n}-x_{n-1})\right) \\
- f^{(n-1)}\left(x_{0}+\sum^{n-1}_{i=1}t_{i}(x_{i}-x_{i-1})\right)\Bigg) &&\\ 
= \frac{1}{ x_{n}-x_{n-1}} (f(x_{0}, x_{1}, \ldots  , x_{n-2}, x_{n}) - 
f(x_{0}, x_{1}, \ldots  , x_{n-2}, x_{n-1})) && \\
= \frac{1}{ x_{n}-x_{n-1}}(f(x_{0}, x_{1}, \ldots  , x_{n-2}, x_{n}) - 
f(x_{n-1}, x_{0}, x_{1}, \ldots  , x_{n-2})) &&\\ 
= f(x_{0}, x_{1}, \ldots, x_{n}).
\end{eqnarray*}

In the penultimate equality we used the symmetry of the function $f(x_{0}, x_{1}, \ldots, x_{n})$ with respect to its arguments, and the last equality is the definition of $f(x_{0}, x_{1}, \ldots, x_{n})$. 
\square

\begin{lemma}\label{lemma:discrete_derivatives_mid} 

Let the nodes $x_{0}$, $x_{1}$, $\ldots$, $x_{n}$ lie on the segment $[a,b]$, and let the function $f$ have a continuous derivative of order $n$ on this segment. Then there exists a point $\xi \in [a,b]$ such that
$$
f(x_{0}, x_{1}, \ldots, x_{n})  = \frac{1}{ n!} f^{(n)}(\xi ). 
$$
\end{lemma}

{\sf Proof.}
We use Lemma~\ref{lemma:discrete_derivatives_rep}. Since the function $f^{(n)}(x)$ is assumed to be continuous, the value of the integral equals the value of this function at some average point of the integration domain times the volume of that domain. The volume equals 
$$ 
\int^{1}_{0}dt_{1}\int^{t_{1}}_{0}dt_{2} \ldots  \int^{t_{n-1}}_{0}dt_{n} = 
\frac{1}{ n!}. \qquad \square  
$$

\begin{theorem}\label{theorem:interp_acc}
Suppose the function $f$ has a continuous derivative of order $n+1$ on the segment $[a,b]$ containing all interpolation nodes, which are assumed to be distinct. Then for $x\in [a, b]$
$$
R(x) = \omega (x) \int^{1}_{0}dt_{1}\int^{t_{1}}_{0}dt_{2} \ldots  
\int^{t_{n}}_{0}dt_{n+1} f^{(n+1)}[x_{0}+
\sum^{n+1}_{i=1}t_{i}(x_{i}-x_{i-1})], 
$$
where $x_{n+1}=x$, $\omega (x)=(x-x_{0})\ldots  (x-x_{n})$.

Moreover, there exists a point $\xi $ on $[a,b]$ such that 
$$
R(x) = \frac{\omega (x)}{ (n+1)!} f^{(n+1)}(\xi ). 
$$
\end{theorem} 

{\sf Proof.} 
Consider the expression for the interpolation error via divided differences from Lemma~\ref{lemma:interp_error} and use Theorems~\ref{lemma:discrete_derivatives_rep} and~\ref{lemma:discrete_derivatives_mid} to express the divided differences.
\square

Functions whose higher-order derivatives exist and are bounded are usually called smooth. For such functions the values between the interpolation nodes cannot be arbitrary, because the rate of change of the function is bounded. Therefore the interpolation error can be relatively small. 

Consider $W^n(M, [a,b])$—the class of all functions on the segment $[a,b]$ whose derivative of order $n$ is bounded by the given number:
$$
|f^{(n)}(x)| \leq M.
$$

This is a fairly wide class of functions admitting interpolation of any $n$ initial data (since the interpolation polynomial has degree $n - 1$, and therefore $f^{(n)}(x)=0$).

\section{Interpolation of Derivatives}\label{sect3_1_2}

Let the values $y_1$, $y_2$, $\ldots$, $y_n$ of the function $f$ be known at the interpolation nodes $x_1$, $x_2$, $\ldots$, $x_n$. We need to estimate the derivative of $f$ of order $m$. 

As an estimate of the derivative of the function $f$ we choose the corresponding derivative of the interpolation polynomial $P(x)$. Then the interpolation error equals 
$$
R^{(m)}(x) = f^{(m)}(x) - P^{(m)}(x).
$$

\begin{lemma}\label{lemma:zero_reduction}
Let $m \geq 1$ and $n$ be integers with $m \geq n \geq 0$, let the function $f$ be $m$ times differentiable on the segment $[a,b]$, and suppose $f$ has $k \geq m$ distinct zeros on $[a,b]$. Then $f^{(n)}$ has at least $k-n$ distinct zeros on $[a,b]$.
\end{lemma}

{\sf Proof.} 
The statement follows by induction on $n$ from Rolle's theorem. 
\square
  
\begin{theorem}\label{theorem:interp_acc_diff}
Let $x$ be some number, and let the function $f$ have $n$ continuous derivatives on the segment $[a,b]$ that contains all interpolation nodes and the point $x$. 


Then on the segment $[a,b]$ there exists a point $\xi$ such that 
$$
R^{(m)}(x) = \frac{\omega^*(x) }{ (n-m)!} f^{(n)}(\xi),
$$
$$
\omega^*(x)=\prod_{i=1}^{n-m}{(x-x^*_i)}, \{x^*_i | R^{(m)}(x^*_i) = 0\}.
$$
\end{theorem}

{\sf Proof.}

By Lemma~\ref{lemma:zero_reduction} the function $R^{(m)}(x)$ has at least $n - m$ distinct zeros on $[a,b]$. Consider the interpolation polynomial $Q(x)$ of degree $n - m - 1$ for the function $f^{m}$ at the zeros of $\omega^*(x)$. The algebraic polynomials $P(x)$ and $Q(x)$ of degree $n - m - 1$ coincide at $n - m$ distinct points. Hence they are equal. Apply Theorem~\ref{theorem:interp_acc} to the polynomial $Q(x)$ to complete the proof.
\square

\begin{corollary}\label{corollary:interp_acc_diff}
From Theorem~\ref{theorem:interp_acc_diff} it follows that 
$$
\max_{x} |R^{(m)}(x)| \leq  \frac{M}{ (n-m)!} \max_{x} |\omega^*(x)| \leq \frac{M (b-a)^{n-m}}{ (n-m)!}. 
$$
\end{corollary}
The given bound does not depend on the interpolated function and the interpolation nodes.

\section{Approximation Error of an Interpolation Polynomial}\label{sect3_1_3}

In the approximation problem the polynomial is not interpolation, that is, at the known points it takes values different from the values of the functions being approximated. We estimate the deviation of the approximation polynomial from the interpolation polynomial of the same degree when the errors at the interpolation nodes are known.

Consider the Lebesgue numbers for the system of interpolation nodes $(x_{1},\ldots,
x_{n})$, $x_i\in I_0$, where $I_0=[-1,1]$, through the fundamental polynomials from Definition~\ref{definition:fundamental}: 
$$
\lambda_{n,\nu}=\sup_{x\in[-1,1]}\sum_{k=1}^{n} |l_k^{(\nu)}(x) |, \qquad
0\leq\nu<r,\quad r \geq 1
$$

\begin{lemma}\label{lemma:lebesgue_norm}
Estimate the norm of the derivative of the interpolation polynomial constructed from the pairs ${(x_1, y_1),\ldots,(x_r,y_r)}$:  
$$
\|p(x,y)^{(\nu)}\| \leq \lambda_{r,\nu} \|y\|.
$$
\end{lemma}

{\sf Proof.}

Consider the interpolation polynomial in Lagrange form:
$$\|p(x,y)^{(\nu)}\| = \|\sum^{n}_{k=1} {l_{k}^{(\nu)}(x) y_{k}} \| \leq
\sum^{n}_{k=1} \|l_{k}^{(\nu)}(x)\| |y_{k}| \leq \lambda_{r,\nu}\|y\|
\square
$$ 

\begin{lemma}\label{lemma:lebesgue_invariant}
Lebesgue numbers under linear transformations of variables are related by
$$
\hat{\lambda}_{r,\nu}=\left(\frac{b-a}{2}\right)^{-\nu}\lambda_{r,\nu},
$$
where $\hat{\lambda}_{r,\nu}$ are the Lebesgue numbers for the result of the linear mapping of the original node system from $I_0$ to the interval $[a, b]$.
\end{lemma}

{\sf Proof.}
Let $x\in[-1,1]$, consider the change of variable
$$\hat{x}=\frac{b-a}{2}x+\frac{b+a}{2}$$

For the node system $(\hat{x_1}\ldots\hat{x_r})$ the fundamental polynomials are
$l_k(x)=\hat{l}_k(\hat{x})$. Differentiating $\nu$ times, we obtain
$$
l_{k}^{(\nu)}(x)=\left(\frac{b-a}{2}\right)^\nu\hat{l}_k^{(\nu)}(\hat{x}).
$$

Then the Lebesgue numbers are expressed by definition as
$$
\hat{\lambda}_{r,\nu}=
\sup_{\hat{x}\in[-a,b]}\sum_{k=1}^{r}|\hat{l}_k^{(\nu)}(\hat{x})|=
\left(\frac{b-a}{2}\right)^{-\nu}\lambda_{r,\nu}
$$ 
\square

One may pose the problem of choosing an optimal node system that minimizes the Lebesgue number. An explicit formula for the optimal nodes is currently unknown. Extended Chebyshev node systems~\cite{Smith_thelebesgue} are considered a good approximation; they are obtained by mapping the Chebyshev node system 
$$
T_k=\cos \frac{\pi (2k-1) }{ 2n}, \qquad k =1, \ldots, n,
$$
onto the interval $[cos 3 \pi / (2n), cos \pi /(2n)]$:
$$
\hat{T}_k=\cos \frac{\pi (2k - 1) }{ 2n} \cos^{-1} \frac{\pi }{ 2n}, \qquad k = 1, \ldots, n.
$$

In this case $\hat{T}_1=1$, $\hat{T}_n=-1$. The lemma implies that $\lambda(\hat{T}) < \lambda(T)$.

%
%

\begin{theorem}\label{theorem:approx_acc_diff}
Let $f\in W^r(M;[a,b]])$, $\xi=(\xi_1,\ldots,\xi_r)\in\mathbb{R}^r$ be an arbitrary vector, $x=(x_1,\ldots,x_r)$ an arbitrary interpolation node system on $I=[a,b]$, $p(x, \xi)$ a polynomial of degree $r-1$ such that $p(x_i, \xi) = \xi_i$, and $f_x=(f(x_1),\ldots,f(x_r))\in\mathbb{R}^r$. Then for any $0 \leq \nu < r$ the inequality 
$$
\|p(x,f_x)^{(\nu)}-p(x,\xi)^{(\nu)}\| \leq
\lambda_{r\nu}(I_0)\left(\frac{2}{b-a}\right)^{\nu}
\sup_{1\leq k \leq r}|f(x_k)-\xi_k|
$$
holds.
\end{theorem}

{\sf Proof.}
We use Lemmas~\ref{lemma:lebesgue_norm} and~\ref{lemma:lebesgue_invariant}:
\begin{eqnarray*}
\|p(x,f_x)^{(\nu)}-p(x,\xi)^{(\nu)}\| = 
\|p(x,f_x-\xi)^{(\nu)}\| \leq \lambda_{r,\nu}\|f_x-\xi\| \\
=\left(\frac{b-a}{2}\right)^\nu\lambda_{r,\nu}\|f_x-\xi\|
=\lambda_{r\nu}(I_0)\left(\frac{2}{b-a}\right)^{\nu}
\sup_{1\leq k \leq r}|f(x_k)-\xi_k|.
\end{eqnarray*}
\square

\begin{theorem}\label{theorem:approx_acc}
Let $f\in W^r(M;I)$, $\xi=(\xi_1,\ldots,\xi_r)\in\mathbb{R}^r$ be an arbitrary vector. Then for any $0 \leq \nu < r$ and $x\in I$ the inequality
$$
\|f^{(\nu)}(x)-p(x,\xi)^{(\nu)}\| \leq \frac{M (b-a) ^{n-\nu}}{ (n-\nu)!}
+\left(\frac{2}{b-a}\right)^{\nu}\lambda_{r \nu}(I_0)\sup_{1\leq k \leq r}|f(x_k)-\xi_k|
$$
holds.
\end{theorem}

{\sf Proof.}
We use the triangle inequality:
$$
\|f^{(\nu)}(x)-p(x,\xi)^{(\nu)}\| \leq \|f^{(\nu)}-p(x,f_x)^{(\nu)}\|
+ \|p(x,f_x)^{(\nu)}-p(x,\xi)^{(\nu)}\|.
$$
Estimate the first term by Theorem~\ref{theorem:interp_acc_diff} and the second term by Theorem~\ref{theorem:approx_acc_diff}.

\square

This inequality makes it possible to estimate the accuracy of approximating a function by a polynomial from the residuals at arbitrary interpolation nodes. In contrast to the error bound for polynomials of best uniform approximation, when non-optimal interpolation nodes are chosen it is coarse because of rounding errors, which leads to excessive refinement of the segment grid in piecewise approximation and excessive precision of the coefficients.
\chapter{Mixed Integer Linear Programming Problem}\label{Appendix:lp}

As is well known, the canonical form of the linear programming problem is defined as follows:
$$
\min_x \{f^T x\, : \,  Ax \leq b\}.
$$

A distinctive characteristic of the tabular optimization problem is that it involves only a small set of variables yet an extremely large number of constraints proportional to the number of grid nodes. Only the constraints near the points of maximum deviation of the interpolation polynomial are significant; all other constraints are satisfied automatically.

Another important aspect of the table reduction problem is that the mantissa length of the coefficients of the interpolation polynomials is fixed. To solve the linear programming problem with integrality constraints on some variables, known as the mixed integer linear programming problem, the branch-and-bound algorithm is employed.

\section{Choosing an Algorithm for Solving the Linear Programming Problem}\label{sect3_3_1}

In this work, the linear programming problem was solved using the function $linprog$ from the optimization toolbox of the $Matlab$ system~\cite{Matlab}. However, its use warrants clarification, because the function implements three different solution algorithms, only one of which, after certain modifications, is suitable for the class of problems under consideration.

The linear programming problem is often solved by the simplex method~\cite{Dantzig63}. The method is guaranteed to converge and yields the exact solution of the problem. Its convergence rate is exponential in the worst case~\cite{klee:1972}. For random matrices, the expected number of simplex steps grows linearly with the number of variables~\cite{Schrijver:1986:TLI:17634}. Slow convergence is observed for the problems considered here, which makes the method impractical for high-dimensional instances.

Another, more recent algorithm is the interior-point method proposed by Karmarkar~\cite{Karmarkar:1984:NPA:800057.808695}. The method relies on a barrier function that grows rapidly as the boundary of the feasible set is approached. The minimum of the sum of the barrier and objective functions converges to the solution as the growth rate of the barrier function decreases. Unlike the simplex method, the solutions are interior points of the feasible set. For the class of problems considered here, the method fails to converge because the auxiliary optimization problem becomes ill-conditioned due to the large number of redundant constraints.

The last method is a modification of the quadratic optimization method with an active set of constraints proposed by Gill~\cite{Gill:1984:POP:1271.1276} to reduce the effective dimension of the problem. At each step, an active subset of equality-type constraints $\hat{A}x=\hat{b}$ is selected from the constraint set, while the remaining constraints are inactive.

\begin{table}[htbp]\centering
\small
    \captionof{table}{Pseudocode of the active-set method.}
    \label{listing:active-set}
    \begin{alltt}
Find initial point
iter = 0
repeat until constraints are satisfied \(\vee\) iter > MaxIter
    iter = iter + 1
    solve the problem with the active equality constraints
    find the Lagrange multipliers for the active constraints
    remove constraints with negative multipliers
    add the violated constraint to the active set
end repeat
    \end{alltt}
\end{table}

The general outline of the method is shown above. When the iteration limit is exceeded, the method may terminate successfully at an infeasible point, which often happens in the implementation of the function $linprog$.

To reach the optimization goal, the function $linprog$ has to be restarted several times until the constraints are satisfied.

\section{Branch-and-Bound Method for Solving the Mixed Integer Linear Programming Problem}\label{sect3_3_2}
To solve the table minimization problem, a method is needed that preserves the integrality of the polynomial coefficients. Simple rounding of the solution to the continuous problem produces large errors. The branch-and-bound method is used to solve the mixed integer problem. The essence of the method is to split the solution space into two subspaces ${x_i \leq \lfloor \bar{x}_i \rfloor}$ and ${x_i \geq \lceil \bar{x}_i \rceil}$, where $\bar{x}_i$ is a component of the current solution. The desired optimum may therefore reside in either subspace. Here $v = f^Tx$, and $w$ is the current bound for the branch-and-bound method. LP is the function that solves the auxiliary linear programming problem. The mixed algorithm differs from the listing in that not all variables are rounded. The parameter $\varepsilon$ determines the discretization accuracy of the integer variables and strongly influences the convergence rate of the method. Consequently, the method is approximate, $\|x^*-x\| \leq  \varepsilon$, where $x^*$ is the true solution. This must be taken into account when applying the algorithm.

Note that any violation of the constraints added by the algorithm while solving the auxiliary linear programming problem causes the algorithm to loop.

\begin{table}[htbp]\centering
    \captionof{table}{Pseudocode of the branch-and-bound method for integer linear programming}
    \label{listing:branch-bound}
    \begin{alltt}
\(w=\infty\)
\((\hat{x}, \hat{v}, \hat{w})\)=ILP(\(f,A,b,x,\varepsilon,w\)) 
    \((\hat{x},\hat{v})\)=LP(\(f,A,b,x\))
    \(\hat{w}=w\)
    if \(\hat{v}\geq\hat{w}\)
        return
    end
    if \( \|\hat{x}-\lfloor\hat{x}+0.5\rfloor \|<\varepsilon \)
        \(\hat{w}=\hat{v}\)
        return
    end	
    \(\exists{i},|\hat{x}\sb{i}-\lfloor\hat{x}\sb{i}+0.5\rfloor|\geq\varepsilon\)	     
    \((A,b)=\{Ax\leq{b}\wedge{x}\sb{i}\leq\lfloor\hat{x}\sb{i}\rfloor\}\) 
    \((\bar{x},\bar{v},\bar{w})\)=ILP(\(f,A,b,x,\varepsilon,\hat{w})\)	
    if \(\bar{w}<\hat{w}\)
        \(\hat{x}=\bar{x},\hat{v}=\bar{v},\hat{w}=\bar{w}\) 
    end	
    \((A,b)=\{Ax\leq{b}\wedge{x}\sb{i}\geq\lceil\hat{x}\sb{i}\rceil\}\)	
    \((\bar{x},\bar{v},\bar{w})\)=ILP(\(f,A,b,x,\varepsilon,\hat{w})\)	
    if \(\bar{w}<\hat{w}\)
        \(\hat{x}=\bar{x},\hat{v}=\bar{v},\hat{w}=\bar{w}\) 
    end	
end
    \end{alltt}
\end{table}
The general scheme of the integer linear programming method is shown above.		
\chapter{Implementation of Special Types of FFT}\label{Appendix:fft}

FFT implementation techniques are well developed. This section collects the fastest algorithms for the real FFT and for double interpolation.

We introduce notation.
The FFT of length $n$ is defined by the matrix ${\mathcal F}_n=(\omega_n^{jk})_{j,k=0}^{n-1}$, where $\omega_n=e^{-\frac{2\pi i}{n}}$. The matrix ${\mathcal F}_n={\mathcal F}_n^T$ is symmetric. Its inverse matrix satisfies the properties $W_n={\mathcal F}_n^{-1}=n^{-1}\bar{{\mathcal F}}_n=n^{-1}(\bar{\omega}_n^{jk})$.

Let $e_0$, $\ldots$, $e_{n-1}$ be the standard basis vectors in ${\sf R}^n$. Introduce the reflection matrix $K_n=(e_0, e_{n-1}, \ldots, e_1)$, the inversion matrix $J_n=(e_{n-1}, \ldots, e_0)$, and the diagonal matrix $D_n=\diag\{(\omega_n^j)_{j=0}^{n-1}\}$. Then, obviously,
$$
K_n {\mathcal F}_n = \bar{\mathcal F}_n = n W_n, \qquad J_n {\mathcal F}_n = \bar{\mathcal F}_n \bar{D}_n.
$$
If the vector $x$ is real, then the vector $y={\mathcal F}_nx$ has conjugate-even symmetry $K_ny=\bar{y}$, which we denote by CE (conjugate-even symmetry). For even $n$ this means that the components $y_0$ and $y_{n/2}$ are real and $y_{n-k}=\bar{y}_k$ for $1\le k<n/2$.

Further $\ac$, $\mc$ denote the cost of complex addition and complex multiplication, respectively, and $\ar$, $\mr$ the cost of the analogous real operations. If the costs of real multiplication and addition coincide, then this cost is denoted by $\tau_{\mathbb R}$.

We assume that complex addition and multiplication are performed in the usual way, and therefore
$$
\ac = 2 \tau_{\mathbb R}, \qquad \mc = 2\ar + 4\mr = 6 \tau_{\mathbb R}.
$$
In some cases the architecture of the computing device makes it possible to account for the fact that multiplication by $1$ and by $i$ has no cost, and multiplication by $e^{\pi i/4}$ costs $2\ar + 2\mr = 4 \tau_{\mathbb R}$.

We denote the costs of the complex and real FFT of order $n$ by $\phi(n)$ and $\tau(n)$, respectively. Obviously, the cost of cyclic convolution equals $3\phi(n)+n \mc$ in the complex case and $3\tau(n)+(n/2-1) \mc + 2\mr$ in the real case. Further we assume that $n$ is a power of $2$, and therefore multiplication and division by $n$ reduce to a shift and have no cost.

\section{Complex Radix-2 Algorithm}

Throughout this section we use the following notation. If $x$ is a vector of length $n=2m$, then $x_0$ and $x_1$ denote the slices consisting of the first $m$ components and the last $m$ components of the vector $x$, respectively. The vector of length $m$ composed of the even components of $x$ is denoted by $x'_0$, and of the odd components by $x'_1$.

Introduce the permutation matrix $P_n$ of size $n$, for which $P_n^Tx=\col(x'_0, x'_1)$. We prove that for $n=2m$ one has
$$
P_n^T {\mathcal F}_n = \begin{pmatrix} {\mathcal F}_m & {\mathcal F}_m \\ {\mathcal F}_m D'_m & -{\mathcal F}_m D'_m \end{pmatrix},
$$
where $D'_m=\diag\{(\omega_n^j)_{j=0}^{m-1}\}$. Indeed, let $j$ be the row index and $k$ the column index. Then
\begin{eqnarray*}
\omega_n^{2jk} & = & \omega_m^{jk}, \qquad 0\le j<m, \quad 0\le k < n, \\
\omega_n^{(2j+1)k} & = & \omega_m^{jk} \omega_n^k, \qquad 0\le j<m, \quad 0\le k < m, \\
\omega_n^{(2j+1)(k+m)} & = & - \omega_n^{(2j+1)k}, \qquad 0\le j<m, \quad 0\le k < m.
\end{eqnarray*}
Since ${\mathcal F}_nx= (P_n^T{\mathcal F})^T P_n^Tx$, we have
$$
{\mathcal F}_n x = \begin{pmatrix} {\mathcal F}_m & D'_m {\mathcal F}_m \\ {\mathcal F}_m & -D'_m {\mathcal F}_m \end{pmatrix} \begin{pmatrix} x'_0 \\ x'_1 \end{pmatrix} = \begin{pmatrix} y_0 \\ y_1 \end{pmatrix} = \begin{pmatrix} u_0 + u_1 \\ u_0 - u_1 \end{pmatrix}, \qquad \begin{pmatrix} u_0 \\ u_1 \end{pmatrix} = \begin{pmatrix} {\mathcal F}_m x'_0 \\ D'_m {\mathcal F}_m x'_1 \end{pmatrix}.
$$
This equality underlies the standard radix-2 FFT algorithm in the time domain. Typically the algorithm begins with the bit-reversal permutation
$$
\Pi_n = P_n \diag\{P_{n/2}, P_{n/2}\} \cdots \diag\{ P_4, \ldots, P_4\}.
$$

The standard radix-2 algorithm in the frequency domain is based on the formula
$$
\begin{pmatrix} y'_0 \\ y'_1 \end{pmatrix} = P_n^T {\mathcal F}_n x = \begin{pmatrix} {\mathcal F}_m u_0 \\ {\mathcal F}_m D'_m u_1 \end{pmatrix}, \qquad \begin{pmatrix} u_0 \\ u_1 \end{pmatrix} = \begin{pmatrix} x_0 + x_1 \\ x_0 - x_1 \end{pmatrix}.
$$
In it the butterflies are performed first, and then the bit-reversal permutation.

\begin{lemma} \label{radix2}
The cost of the standard radix-2 algorithm in the frequency domain or in the time domain, in an implementation that does not take into account the specifics of multiplication by $1$, $i$, and $(1\pm i)/2$, equals
$$
\phi(n) = n p \ac + \frac{n}{2}(p - 1)\mc = ( 3np - n) \ar + (2np - 2n) \mr, \qquad n=2^p\ge 4.
$$

When the specifics of multiplication by $1$, $i$, and $(1\pm i)/2$ are taken into account, the cost of the standard radix-2 algorithm equals
$$
\phi(n) = (3 n p - 3n + 4)\ar + (2n p - 7n + 12)\mr, \qquad n=2^p\ge 4.
$$
Initial data: $\phi(1)=0$, $\phi(2) = 2\ac$.
\end{lemma}

\proof
The specifics of multiplication by $1$, $i$, and $(1\pm i)/2$ are accounted for when multiplying by $D'_m$. Without accounting for the specifics, the cost satisfies the recurrence
$$
\phi(n) = 2\phi(n/2) + n \ac + n/2\, \mc, \qquad \phi(2) = 2 \ac.
$$
The solution of this equation for $n=2^p\ge 2$ is
$$
\phi(n) = n p \ac + \frac{n}{2}(p - 1)\mc = ( 3np - n) \ar + (2np - 2n) \mr,
$$
which agrees with the lemma.

When the specifics of multiplication by $1$, $i$, and $(1\pm i)/2$ are taken into account, the recurrence has the form
$$
\phi(n) = 2\phi(n/2) + (3n-4) \ar + (2n-12) \mr, \qquad \phi(4) = 16 \ar,
$$
and the solution for $n=2^p\ge 4$ is
$$
\phi(n) = (3 n p - 3n + 4)\ar + (2n p - 7n + 12)\mr.
$$
\square

\section{Complex Split-Radix Algorithm in the Frequency Domain}

Let $n=2m=4\ell$. Alongside the diagonal matrices $D_n$ of size $n$ and $D'_m$ of size $m$, introduce the matrix $D''_{\ell} = \diag\{(\omega_n^j)_{j=0}^{\ell-1}\}$. It is clear that $(D''_{\ell})^4 = (D'_{\ell})^2 = D_{\ell}$ and that $D'_m=\diag\{D''_{\ell}, i D''_{\ell}\}$.

In the radix-2 algorithm in the frequency domain we further decompose the butterfly used to compute $y'_1$. Introduce the notation $v=y'_1$ and $t=u_1$. Then
$$
\begin{pmatrix} v'_0 \\ v'_1 \end{pmatrix} = P_m^T y'_1 = P_m^T {\mathcal F}_m D'_m u_1 = \begin{pmatrix} {\mathcal F}_{\ell} & {\mathcal F}_{\ell} \\ {\mathcal F}_{\ell} D'_{\ell} & -{\mathcal F}_{\ell} D'_{\ell} \end{pmatrix} \begin{pmatrix} D''_{\ell} t_0 \\ i D''_{\ell} t_1 \end{pmatrix} = \begin{pmatrix} {\mathcal F}_{\ell} D''_{\ell} z_0 \\ {\mathcal F}_{\ell} D'_{\ell} D''_{\ell} \bar{z}_1 \end{pmatrix},
$$
where $z_0 = t_0 + i t_1$, $z_1 = \bar{t}_0 + i \bar{t}_1$. In this case $D'_{\ell} D''_{\ell} = D_{\ell} \bar{D}''_{\ell}$ and $J_{\ell}\bar{{\mathcal F}}_{\ell}={\mathcal F}_{\ell}D_{\ell}$. Therefore $v'_1 = J_{\ell} \overline{{\mathcal F}_{\ell} D''_{\ell} z_1}$.

The resulting formula defines the recursive rule for computing the component $y'_1$ in the FFT algorithm in the frequency domain. The component $y'_0$ is still computed as ${\mathcal F}_m u_0$.

\begin{lemma} \label{split_complex}
The cost of the complex split-radix algorithm in the frequency domain equals
$$
\phi(n) = \left(\frac{8}{3}np - \frac{16}{9} n - \frac{2}{9} (-1)^p + 2\right) \ar + \left(\frac{4}{3} np - \frac{38}{9}n + \frac{2}{9}(-1)^p + 6\right)\mr,
$$
where $n=2^p\ge 4$, totaling $4np-6n+8$ real operations. Initial values: $\phi(1)=0$, $\phi(2) = 4\ar$, $\phi(4) = 16\ar$.
\end{lemma}

\proof
From the recurrence relation obtained it follows that the cost of the algorithm satisfies the equation
\begin{eqnarray*}
\phi(n) & = & \phi(n/2) + 2 \phi(n/4) + n \ac + 2\ell \ac + 2(\ell-2) \mc + 2(2\mr+2\ar) \\
& = &
\phi(n/2) + 2 \phi(n/4) + (4n-4)\ar + (2n-12)\mr.
\end{eqnarray*}
Denote $\psi_p = \phi(2^p)$. Then the sequence $\psi_p$ satisfies the linear difference equation
$$
\psi_p = \psi_{p-1} + 2 \psi_{p-2} + 2^p \ac + 2^{p-1} \ac + 2(2^{p-2}-2) \mc + 2(2\mr+2\ar).
$$
The solution is the function stated in the lemma. \square

\section{Real Radix-2 Algorithm in the Time Domain}

It is also called the Edson-Bergland algorithm.

Let the vector $x$ be real and let the radix-2 algorithm in the time domain be applied. The result of each stage is a CE vector. One vector butterfly contains two real additions for the zero and middle frequencies, and in the remaining $n/2-2$ butterflies only half of them are processed. Each butterfly includes two complex additions and one complex multiplication. Therefore the recurrence equation is
$$
\tau(n) = 2\tau(n/2) + 2\ar + (n/4-1)(2\ac + \mc) = 2\tau(n/2) + (3n/2-4)\ar + (n-6)\mr.
$$
for $n\ge 4$. Initial data: $\tau(1)=0$, $\tau(2)=2 \ar$, $\tau(4)=6\ar$. The solution of the equation for $n=2^p\ge 4$ is
$$
\tau(n) = \left(\frac{3}{2} n p - \frac{5}{2}n + 4\right)\ar + \left( np - \frac{7}{2}n + 6\right)\mr.
$$

The algorithm is inconvenient for implementing convolution, since it requires performing the bit permutation of the input real array before starting the work.

\section{Real Split-Radix Algorithm in the Frequency Domain}


Consider the split-radix algorithm in the frequency domain whose input is a real array. Then, in the notation of the complex version of the algorithm, the vector $t$ is real. Therefore $z_0=z_1$ and, consequently, $v'_1 = J\bar{v}'_0$. This vector need not be computed separately.

Thus, computing the real FFT $\col(y'_0, y'_1)$ of length $2m$ reduces to computing the real FFT $y'_0$ of length $m$ and computing the complex FFT $v'_0={\mathcal F}_{\ell}D''_{\ell}z_0$ of length $\ell=m/2$.

\begin{lemma} \label{split_real}
The cost of the real split-radix algorithm in the frequency domain equals
$$
\tau(n) = \left(\frac{4}{3} np - \frac{17}{9}n - \frac{1}{9}(-1)^p + 3\right)\ar + \left(\frac{2}{3} np - \frac{19}{9} n + \frac{1}{9} (-1)^p + 3\right) \mr,
$$
where $n=2^p\ge 2$, totaling $2np - 4n + 6$ real operations, provided the split-radix algorithm is also used to implement the complex FFT.
\end{lemma}

\proof
The recurrence equation for the cost is determined by the decomposition and by multiplication by $D''_{\ell}$:
\begin{eqnarray*}
\tau(n) & = & \tau(n/2) + \phi(n/4) + n \ar + (\ell-2) \mc + (2\mr + 2\ar) \\
& = &
\tau(n/2) + \phi(n/4) + (3n/2-2)\ar + (n-6)\mr.
\end{eqnarray*}
Substituting the value of $\phi(n)$ derived above and solving the resulting first-order linear equation yields the statement of the lemma. \square

\vspace{3mm}
The inverse transform from a CE vector to a real vector requires the same number of operations as stated above.

The convolution of real arrays of length $n=2^p$ can be computed using this split-radix algorithm in the frequency domain, multiplication, and the dual split-radix algorithm in the time domain. The FFT algorithms are in-place and require exactly $n$ cells of real memory.

\section{Double Real Interpolation}

Given a vector $x\in {\sf R}^m$. It is known that $x={\mathcal F}_m^{-1} u$, where the vector $u$ is real. We need to compute the vector $p={\mathcal F}_n \col(x, 0)$, where $n=2m$.

Split the target vector into even and odd components: $P_n p=\col(p'_0, p'_1)$. It is clear that $p'_0=u$ is the original vector and $p'_1 = {\mathcal F}_mD'_m x$ is the vector to be computed.

The real vector $x={\mathcal F}_m^{-1} u$ is computed by the algorithm in the frequency domain with cost $\tau(m)$.

To find $p'_1 = {\mathcal F}_mD'_m x$ we apply the split-radix algorithm in the frequency domain. Since the vector $x$ is real, we take the first $x_0$, $x_1$ as the vectors of the first and last $\ell=m/2$ components and form the vector $z_0=x_0 +i x_1$. In accordance with the split-radix algorithm the vector $p'_1$ is completely determined by the vector ${\mathcal F}_{\ell} D''_{\ell} z_0$.

\begin{lemma} \label{twice_interp}
The double real interpolation algorithm with an input vector of length $m$ has cost
$$
I(m) = \tau(m) + \phi(m/2) + (m/2-2) \mc + (2\mr + 2\ar).
$$
If the real and complex FFTs are implemented by split-radix algorithms, then the cost of double real interpolation equals
$$
I(m) = \left(\frac{8}{3} mp - \frac{28}{9} m + \frac{1}{9} (-1)^p + 3\right) \ar + \left(\frac{4}{3} mp - \frac{26}{9}m - \frac{1}{9}(-1)^p + 3\right)\mr.
$$
Here $m=2^p \ge 2$, totaling $4mp - 6m + 6$ real operations. Initial values: $I(2)=0$, $I(4) = 2 \ar$.
\end{lemma}

\proof
Multiplication by $D''{\ell}$, taking into account the specifics of $1$ and $(1+i)/\sqrt{2}$, requires $\ell-2$ complex multiplications, 2 real additions, and 2 real multiplications. To compute the vector $p_1$ one also needs one real FFT $x={\mathcal F}_m^{-1}u$ and one complex FFT ${\mathcal F}_{\ell} D''_{\ell} z_0$. This yields the first statement of the lemma.

The second statement follows from a direct substitution of the conclusions derived above.

Thus, instead of computing ${\mathcal F}_mx$ we perform ${\mathcal F}_mu$, and these operations have the same cost.

Compared with the real split-radix algorithm, the computation of the vectors $x_0{+}x_1$ and $x_0{-}x_1$ is absent, since $x_1=0$. Therefore the cost of the entire double interpolation operation for $m=2^p$ is
\begin{eqnarray*}
I(m) & = & \tau(m) + \phi(\ell) + (\ell-2) \mc + (2\mr + 2\ar) \\
& = &
\tau(m) + \phi(m/2) + (m-2) \ar + (2m - 6)\mr \\
& = &
\left(\frac{4}{3} mp - \frac{17}{9}m - \frac{1}{9}(-1)^p + 3\right)\ar + \left(\frac{2}{3} mp - \frac{19}{9} m + \frac{1}{9} (-1)^p + 3\right) \mr \\
&&
+ \left(\frac{4}{3}m(p{-}1) {-} \frac{8}{9} m {+} \frac{2}{9} (-1)^p {+} 2\right) \ar {+} \left(\frac{2}{3} m(p{-}1) {-} \frac{19}{9}m {-} \frac{2}{9}(-1)^p {+} 6\right)\mr \\
&&
+ (m-2) \ar + (2m - 6)\mr \\
& = &
\left(\frac{8}{3} mp - \frac{28}{9} m + \frac{1}{9} (-1)^p + 3\right) \ar + \left(\frac{4}{3} mp - \frac{26}{9}m - \frac{1}{9}(-1)^p + 3\right)\mr.
\end{eqnarray*}
totaling $4mp - 6m + 6$ real operations. Initial values: $I(2)=0$, $I(4) = 2 \ar$.

\square		

\end{document}